\documentclass[fleqn,usenatbib]{mnras}

\usepackage{newtxtext,newtxmath}
\usepackage[T1]{fontenc}

\DeclareRobustCommand{\VAN}[3]{#2}
\let\VANthebibliography\thebibliography
\def\thebibliography{\DeclareRobustCommand{\VAN}[3]{##3}\VANthebibliography}

\usepackage{graphicx}	% Including figure files
\usepackage{amsmath}	% Advanced maths commands
\usepackage{lipsum}
\usepackage{booktabs}
\usepackage{siunitx}
\usepackage{multirow,bigdelim} % for the multirow curly bracket thing
\usepackage{bm}

\DeclareSIUnit\Rjup{\ensuremath{\mathit{R}}_\mathrm{Jup}}
\DeclareSIUnit\Mjup{\ensuremath{\mathit{M}}_\mathrm{Jup}}

\DeclareSIUnit\Rsun{\ensuremath{\mathit{R}_{\sun}}}
\DeclareSIUnit\Msun{\ensuremath{\mathit{M}_{\sun}}}
\DeclareSIUnit\Lsun{\ensuremath{\mathit{L}_{\sun}}}
\DeclareSIUnit\rhosun{\ensuremath{\rho_{\sun}}}

\DeclareSIUnit\Rstar{\ensuremath{\mathit{R}_{\star}}}
\DeclareSIUnit\Mstar{\ensuremath{\mathit{M}_{\star}}}

\DeclareSIUnit \parsec {pc}
\DeclareSIUnit \au {au}
\DeclareSIUnit \ppt {ppt}
\DeclareSIUnit \ppm {ppm}
\DeclareSIUnit \hour {h}
\DeclareSIUnit \year {yr}
\DeclareSIUnit \muunit {\mu}
\DeclareSIUnit \perpix {\ensuremath{\mathrm{px}^{-1}}}
\DeclareSIUnit\gauss{G}
\DeclareSIUnit\angstrom{\text {Å}}

\newcommand{\software}[1]{\textsc{#1}}
\newcommand{\satellite}[1]{\textit{#1}}

\newcommand*{\teffA}{\SI{5685 \pm 65}{\kelvin}}
\newcommand*{\massA}{\SI{1.04 \pm 0.07}{\Msun}}
\newcommand*{\ProtA}{\SI{13.6 \pm 1.6}{\day}}
\newcommand*{\ProtAspot}{\SI{15.1 \pm 0.6}{\day}}

\newcommand*{\teffB}{\SI{5250 \pm 90}{\kelvin}}
\newcommand*{\massB}{\SI{0.88 \pm 0.07}{\Msun}}

\newcommand*{\radiusb}{\SI{1.270 \pm 0.030}{\Rjup}}
\newcommand*{\massb}{\SI{1.25 \pm 0.03}{\Mjup}}
\newcommand*{\periodb}{\SI[group-digits=false]{2.6556743 \pm 0.0000013}{\day}} % using only our photometry
\newcommand*{\periodbnounit}{\SI[round-mode=places, round-precision=6, group-digits=false]{2.6556743 \pm 0.0000013}{}} % using only our photometry (no cheops, vedad analysis)
\newcommand*{\tpribnounit}{\SI[round-mode=places, round-precision=4, group-digits=false]{2459258.824996 \pm 0.000097}{}} % using only our photometry

\newcommand*{\ra}[2][]{{
    \ang[
        angle-symbol-degree=\textsuperscript{h},
        angle-symbol-minute=\textsuperscript{m},
        angle-symbol-second=\textsuperscript{s},
        #1]{#2}%
}}

\newcommand*{\gspot}{\ensuremath{\mathcal{G}_\mathrm{spot}}}
\newcommand*{\gquiet}{\ensuremath{\mathcal{G}_\mathrm{quiet}}}

\newcommand*{\vsini}{\ensuremath{v\sin{i_\star}}}

\newcommand{\manualcite}[2]{\textcolor{blue}{\hyperlink{bib:#1}{[#2]}}}
\title[Resolving star spots on WASP-85\,A]{Resolving star spots on WASP-85\,A using high-resolution transit spectroscopy}

\author[V. Kunovac et al.]{%
Vedad Kunovac,$^{1,2}$\thanks{Royal Society Newton International Fellow}\thanks{Contact: vedad.kunovac@warwick.ac.uk}
Heather Cegla,$^{1,2}$
Hritam Chakraborty,$^3$
Cis Lagae,$^{1,2}$
David J.~A. Brown,$^{1,2}$
\newauthor
Alix Freckelton,$^4$
Samuel Gill,$^{1,2}$
Mercedes L\'{o}pez-Morales,$^5$
James McCormac,$^{1,2}$
Annelies Mortier,$^4$ % check with heather
\newauthor
Mathilde Timmermans,$^{4,6}$
Thomas G. Wilson,$^{1,2}$
Romain Allart,$^7$,
Edward M. Bryant,$^{1,2}$ % check with heather
\newauthor
Matthew R. Burleigh,$^8$
Lauren Doyle,$^{1,2}$ % check with heather
Edward Gillen,$^9$ % check with heather
James S. Jenkins$^{10,11}$,
Marina Lafarga,$^{1,2}$
\newauthor
Monika Lendl,$^3$
Mahmoud Oshagh,$^{12}$
Vatsal Panwar,$^{1,2}$
Peter P. Pedersen,$^{13,14}$
Amaury Triaud$^4$,
\newauthor
Richard G. West$^{1,2}$
and Peter J. Wheatley,$^{1,2}$
\\\\
$^{1}$Centre for Exoplanets and Habitability, University of Warwick, Coventry, CV4 7AL, UK \\
$^{2}$Department of Physics, University of Warwick, Coventry, CV4 7AL, UK \\
$^{3}$Geneva Observatory, University of Geneva, Chemin Pegasi 51, 1290 Versoix, Switzerland \\
$^{4}$School of Physics \& Astronomy, University of Birmingham, Edgbaston, Birmingham B15 2TT, United Kingdom \\
$^{5}$Space Telescope Science Institute, 3700 San Martin Drive, Baltimore MD 21218, USA \\
$^{6}$Astrobiology Research Unit, Université de Liège, 19C Allée du 6 Août, 4000 Liège, Belgium\\
$^{7}$D\'epartement de Physique, Institut Trottier de Recherche sur les Exoplan\`etes, Universit\'e de Montr\'eal, Montr\'eal, Qu\'ebec, H3T 1J4, Canada \\
$^{8}$School of Physics and Astronomy, University of Leicester, Leicester LE1 7RH, UK \\
$^{9}$Astronomy Unit, School of Physics and Astronomy, Queen Mary University of London, London E1 4NS, UK \\
$^{10}$Instituto de Estudios Astrof\'{\i}sicos, Facultad de Ingenier\'{\i}a y Ciencias, Universidad Diego Portales, Av. Ej\'ercito Libertador 441, Santiago, Chile \\
$^{11}$Centro de Excelencia en Astrof\'{\i}sica y Tecnolog\'{\i}as Afines (CATA), Camino El Observatorio 1515, Las Condes, Santiago, Chile \\
$^{12}$Autotrader UK, 1 Tony Wilson Place, Manchester M15 4FN UK \\
$^{13}$Cavendish Laboratory, JJ Thomson Avenue, Cambridge CB3 0HE, UK \\
$^{14}$Institute for Particle Physics and Astrophysics , ETH Z\"urich, Wolfgang-Pauli-Strasse 2, 8093 Z\"urich, Switzeland \\
}

\date{Accepted XXX. Received YYY; in original form ZZZ}

\pubyear{2015}

\begin{document}
\label{firstpage}
\pagerange{\pageref{firstpage}--\pageref{lastpage}}
\maketitle

\begin{abstract}
Stellar surface inhomogeneities such as spots and faculae introduce Doppler variations that challenge exoplanet detection via the radial velocity method. While their impact on disc-integrated spectra is well established, detailed studies of the underlying local line profiles have so far been limited to the Sun. We present an observational campaign targeting the active star WASP-85\,A during transits of its hot Jupiter companion. The transits span two stellar rotation periods, allowing us to probe the evolution of active regions. From ground-based photometry we identify seven active regions, six containing dark spots. Using simultaneous ESPRESSO transit spectroscopy, we spatially resolve these regions on the stellar surface by using the planet as a probe. We detect significant bisector shape changes, line broadening, and net redshifts during spot occultations, with velocity shifts of \qtyrange[range-units=single,range-phrase=\text{--}]{108}{333}{\metre\per\second} (mean uncertainty \qty{50}{\metre\per\second}). The observed broadening is consistent with the Zeeman splitting, implying magnetic field strengths (Stokes $I$) $B = \qtyrange[range-units=single,range-phrase=\text{--}]{2.7}{4.4}{\kilo\gauss}$ (mean uncertainty \qty{0.6}{\kilo\gauss}), comparable to solar umbrae. Combined with our photometric spot model, this yields lower limits to the disc-integrated field $Bf = \qty[separate-uncertainty,multi-part-units=single]{16 \pm 3}{\gauss}$ and \qty[separate-uncertainty,multi-part-units=single]{61 \pm 9}{\gauss} for the two hemispheres probed -- at least three times higher than Sun-as-a-star values. We also measure centre-to-limb variations in FWHM, line depth, equivalent width, and convective blueshift, which broadly agree with solar observations and 3D MHD models. This work demonstrates a new way to characterise the surfaces of exoplanet host stars, paving the way for future analyses incorporating synthetic line profiles from 3D MHD simulations.

\end{abstract}


%%%%%%%%%%%%%%%%%%%%%%%%%%%%%%%%%%%%%%%%%%%%%%%%%%

%%%%%%%%%%%%%%%%% BODY OF PAPER %%%%%%%%%%%%%%%%%%

\section{Introduction}
Approximately \qty{20}{\percent} of all confirmed exoplanets have been discovered using the Doppler (radial velocity) technique.\footnote{\url{https://exoplanetarchive.ipac.caltech.edu/docs/counts_detail.html}} The method has been tremendously successful in identifying multiple exoplanet demographic populations from hot Jupiters \citep{mayor1995,butler1997,udry2003,cumming2008} to super-Earths and mini-Neptunes \citep{howard2010,udry2007a,mayor2011}. However, the limiting factor in detecting the lowest-mass planets -- in particular those with a mass similar to Earth in the habitable zone of their stars -- is the activity caused by the stars themselves due to surface phenomena \citep{dumusque2011,lovis2011,fischer2016}. The disc-integrated radial velocity has contributions from not only the reflex motion due to orbiting planets, but also from dark spots \citep{saar1997,desort2007,lagrange2010,boisse2011,haywood2014,lisogorskyi2020}, bright faculae and plages \citep{meunier2010,haywood2016}, convective granulation \citep{cegla2013,meunier2015,cegla2019} and acoustic pulsations \citep{chaplin2019,kunovachodzic2021} -- all of which can mimic or obscure planetary signals at the sub-\unit{\metre\per\second} level. Based on resolved-disc measurements of the Sun, the dominant rms component is from the variable suppression of convective blueshift due to evolving magnetic regions such as spots and plage/faculae \citep{haywood2016}.

For most stars,\footnote{The exceptions are bright, nearby -- and often evolved -- stars that can be resolved with interferometry.} we only observe the integrated light from the visible stellar disc; resolved-disc measurements are generally limited to the Sun. Consequently, there remains uncertainty about how well stellar models reproduce surface physics on other stars. Advanced 3D magnetohydrodynamic (MHD) simulations of the solar convection zone and photosphere \citep[e.g.][]{vogler2005,stein2024,rodriguezdiaz2024} accurately capture much of the observed solar surface behaviour and represent a significant improvement over 1D models \citep{beeck2012,pereira2013}. Line synthesis from 3D models that include non-LTE effects successfully reproduces many centre-to-limb (CLV) trends in the solar atmosphere for several well-studied lines, including \ion{Fe}{i} \citep{lind2017}, \ion{Na}{i} and \ion{K}{i} \citep{canocchi2024}, \ion{O}{i} \citep{amarsi2018} and many others \citep{takeda2019}.

Transiting exoplanets present an opportunity to probe the local photosphere along the transit chord, particularly by using differential spectroscopy, i.e. comparing in-transit observations to those obtained outside transit \citep{cegla2016,bourrier2017}. These methods are able to retrieve the local line profiles along the transit chord, enabling studies of CLV effects in stars other than the Sun \citep{dravins2017a}. By using cross-correlation functions (CCFs) to stack thousands of lines together one can boost S/N of the underlying local line profiles, which comes at the expense of averaging lines with different properties and different centre-to-limb effects. Regardless, this technique has led to detections of differential rotation \citep{cegla2016,doyle2023} and centre-to-limb convective blueshift \citep{doyle2022} in several solar-type stars.

The occultation of star spots during exoplanet transits has long been used to infer the distribution of spot sizes, latitudes, and contrasts on other stars from high-quality photometric light curves obtained with missions such as \satellite{Kepler} \citep{morris2017,dai2018,breton2024}. These studies suggest that spotted stars share similar active latitudes with the Sun, but exhibit spot coverage fractions orders of magnitude larger than solar values at activity maximum \citep{morris2017}.
In spectroscopy, spot occultations have typically been treated as a nuisance as they may significantly bias the sky-projected stellar obliquity $\lambda$ \citep{oshagh2018}, which is relied on to infer the formation and dynamical histories of multiple planet populations \citep{triaud2018a,albrecht2021}. However, they may offer a complementary opportunity to probe the local, small-scale magnetic fields on stellar surfaces through Zeeman broadening of spectral lines -- an effect that has not previously been measured directly. 
Traditionally, magnetic field measurements have relied on either polarimetric Zeeman-Doppler imaging (ZDI; e.g. \citealt{see2019,bellotti2025}) to map large-scale fields, or Zeeman broadening and intensification in disc-integrated spectra to characterise smaller-scale fields \citep{reiners2012a,kochukhov2020}. 
% spots and active regions impact planet searches
% spectra of spots have not been directly observed in other stars
% magnetic fields are difficult to measure in other stars, some measurements exist with polarimetry, others measure zeeman splitting in individual lines on bright stars

The paper is organised as follows. In Section~\ref{sec:target_observations} we summarise the properties of the WASP-85 system and describe the photometric and spectroscopic data on which this paper is based on. In Section~\ref{sec:rotation_period} we analyse long-term photometry to derive a rotation period for WASP-85\,A. In Section~\ref{sec:transit_analysis} we analyse our photometric transits, describe our spot modelling approach, and compare our results to the published values. In Section~\ref{sec:obliquity_analysis} we analyse our spectroscopic transits, describe our approach for retrieving the local surface line profiles along the transit chord, and report on the derived obliquity and its inter-night variation due to stellar activity. In Section~\ref{sec:outlier_identification} we detail our analysis to identify active regions based on spot occultations in photometry and spectroscopy. In Section~\ref{sec:subplanet_profile_analysis} we discuss the local line profiles from occulted active regions, describe our detailed analysis of the line profiles to make an estimate of the associated magnetic field, and compare our results to the Sun and to main-sequence Sun-like stars. In Section~\ref{sec:centre-to-limb} we explore the centre-to-limb (CLV) effects of the local line profiles. In Section~\ref{sec:discussion} we offer some additional remarks on the rotation period and 3D obliquity of WASP-85\,A, before finally concluding in Section~\ref{sec:conclusion}.
\sisetup{
range-phrase=--, 
range-units=single, 
separate-uncertainty,
multi-part-units=single
}

\section{Target, observations, and data reduction}
\label{sec:target_observations}

\begin{table}
    \sisetup{separate-uncertainty=false,group-digits=false}
    % \sisetup{separate-uncertainty, group-digits=integer}
    \renewcommand{\arraystretch}{1.2}
    \small
    % \sisetup{round-mode=places}
    % \centering
    \caption{Target information. $^1$Simbad $^2$\citet{brown2014} $^3$\citet{andrae2018} $^4$\citet{torres2010}}
    % \begin{tabular*}{\linewidth}{@{\extracolsep{\fill}}
    \begin{tabular}{
    % \begin{tabular}{@{\extracolsep{\fill}}
    lcclc
    }
    % S[table-format=4.0(3)] % teff
    % S[table-format=1.1(1)] % logg
    % S[table-format=1.2(2)] % feh
    % S[table-format=1.1(1)] % vsini
    % c}
        \toprule
        \toprule
        Parameter & \multicolumn{2}{c}{Value} & Unit & Source \\
        % \midrule
        \cmidrule(lr){2-3}
        & WASP-85\,A & WASP-85\,B \\
        % gaia dr3 ID for A: 3909745223886018432
        % gaia dr3 ID for B: 3909745223886018560
        \midrule
        $\alpha$ & \multicolumn{2}{c}{\ra[angle-symbol-over-decimal]{11;43;38.01}} & &  \href{https://simbad.cds.unistra.fr/simbad/sim-basic?Ident=WASP-85&submit=SIMBAD+search}{[1]} \\
        $\delta$ & \multicolumn{2}{c}{\ang[angle-symbol-over-decimal,minimum-integer-digits=2,retain-explicit-plus]
        {+06;33;49.4}} & & \href{https://simbad.cds.unistra.fr/simbad/sim-basic?Ident=WASP-85&submit=SIMBAD+search}{[1]} \\
        Distance & \multicolumn{2}{c}{\SI{141 \pm 1}{}} & \si{\parsec} & \manualcite{brown2018}{2} \\
        Sky separation & \multicolumn{2}{c}{\SI{1.5 \pm 0.1}{}} & \si{\arcsec} & \manualcite{brown2018}{2} \\
        Separation & \multicolumn{2}{c}{\SI{210 \pm 22}{}} & \si{\au} & \manualcite{brown2018}{2} \\
        $V_\mathrm{mag}$ & 11.2 & 11.9 &  & \manualcite{brown2014}{2} \\
        $M_\star$ & \SI{1.04 \pm 0.07}{} & \SI{0.88 \pm 0.07}{} & \si{\Msun} & \manualcite{andrae2018}{3}, \manualcite{torres2010}{4} \\
        % $R_\star$ & \SI{0.96 \pm 0.13}{} & \SI{0.77 \pm 0.13}{} & \si{\Rsun} \\ % original from discovery paper
        $R_\star$ & \SI{0.99 \pm 0.02}{} & \SI{0.77 \pm 0.13}{} & \si{\Rsun} & \manualcite{andrae2018}{3}, \manualcite{torres2010}{4} \\ % radius_flame (gspphot) (B component has no reported values in Gaia)
        % $R_\star$ & \SI{1.0115 \pm 0.0028}{} & -- & \si{\Rsun} \\ % radius_gspphot
        {$T_\mathrm{eff}$} & \SI{5685 \pm 65}{} & \SI{5250 \pm 90}{} & {K} & §\ref{data:spectroscopy}, \manualcite{brown2014}{2} \\
        {$\log{g}$} & \SI{4.48 \pm 0.11}{} & \SI{4.61 \pm 0.14}{} & {cgs} & §\ref{data:spectroscopy}, \manualcite{brown2014}{2} \\
        {[Fe/H]} & \SI{0.08 \pm 0.10}{} & \SI{0.00 \pm 0.15}{} & log$\,\odot$ & §\ref{data:spectroscopy}, \manualcite{brown2014}{2} \\
        {$v\sin{i_\star}$} & \SI{3.13 \pm 0.02}{} & \SI{3.32 \pm 0.82}{} & {\si{\kilo\metre\per\second}} & §\ref{sec:obliquity_analysis}, \manualcite{brown2018}{2}\\
        % {$v\sin{i_\star}$} & \SI{1.89 \pm 0.98}{} & \SI{3.32 \pm 0.82}{} & {\si{\kilo\metre\per\second}} & §\ref{sec:obliquity_analysis}\\
        $P_\mathrm{rot}$ & 15.5 & 11.5 & \si{\day} & §\ref{sec:rotation_period}, §\ref{sec:discussion} \\
        $i_\star$ & \qty{79 \pm 7}{} & -- & \unit{\degree} & §\ref{sec:discussion} \\
        % \midrule 
        & \multicolumn{2}{c}{\text{WASP-85\,Ab}} & & \\
        \midrule

        $P$ & \multicolumn{2}{c}{\periodbnounit{}} & \si{\day} & §\ref{sec:transit_analysis} \\
        $T_0$  & \multicolumn{2}{c}{\tpribnounit{}} & BJD$_\mathrm{TDB}$ & §\ref{sec:transit_analysis} \\
        % $r/R_\star$ & \multicolumn{2}{c}{\SI{0.1319 \pm 0.0016}{}} & \si{\Rstar} & §\ref{sec:transit_analysis} \\
        $r/R_\star$ & \multicolumn{2}{c}{\SI{0.136750 \pm 0.000075}{}} & \si{\Rstar} & §\ref{sec:transit_analysis} \\
        $T_{14}$ & \multicolumn{2}{c}{\SI{0.10942 \pm 0.00042}{}} & \si{\day} & §\ref{sec:transit_analysis} \\
        $b$ & \multicolumn{2}{c}{\SI{0.135 \pm 0.079}{}} & \si{\Rstar} & §\ref{sec:transit_analysis} \\
        $a/R_\star$ & \multicolumn{2}{c}{\SI{8.724 \pm 0.091}{}} & $R_\star$ & §\ref{sec:transit_analysis}\\
        $i$  & \multicolumn{2}{c}{\SI{89.11 \pm 0.53}{}} & \si{\degree}  & §\ref{sec:transit_analysis}\\
        $e$ & \multicolumn{2}{c}{0} & & \manualcite{brown2014}{2} \\
        $a$ & \multicolumn{2}{c}{\SI{0.04013 \pm 0.00083}{}} & au  & §\ref{sec:transit_analysis}\\
        $M_\mathrm{p}$ & \multicolumn{2}{c}{\SI{1.25 \pm 0.03}{}} & \si{\Mjup} & \manualcite{brown2018}{2} \\
        $R_\mathrm{p}$ & \multicolumn{2}{c}{\SI{1.270 \pm 0.030}{}} & \si{\Rjup} & §\ref{sec:transit_analysis} \\
        $\lambda$ & \multicolumn{2}{c}{\SI{-1 \pm 1}{}} & \si{\degree} & §\ref{sec:obliquity_analysis} \\
        $\psi$ & \multicolumn{2}{c}{\SI{11 \pm 7}{}} & \si{\degree} & §\ref{sec:obliquity_analysis} \\

        \bottomrule
        \label{table:spectroscopic_parameters}
        \end{tabular}
\end{table}

\begin{table*}
    \sisetup{separate-uncertainty=false}
    % \sisetup{separate-uncertainty, group-digits=integer}
    \renewcommand{\arraystretch}{1.2}
    \small
    % \sisetup{round-mode=places}
    % \centering
    \caption{Observing log. $^aS/N$ at \SI{550}{\nano\metre}. $^b$Spectrograph fibre size are in units of \unit{\arcsec}, while photometry pixel scale is \unit{\arcsec\perpix}. $^c$Brackets denote camera ID.
    $^d$CH\_PR220008\_TG000101.
    $^e$CH\_PR220008\_TG000501.
    $^f$CH\_PR220008\_TG000801.
    $^g$CH\_PR220008\_TG000701.}
    % \begin{tabular*}{\linewidth}{@{\extracolsep{\fill}}
    % \begin{tabular}{
    \begin{tabular}{@{\extracolsep{\fill}}
    ll
    % c
    S[table-format=1.2]% sky res
    S[table-format=3.0]% exptime
    c
    ccc
    % l
    }
    % Transit date, Instrument, Nobs, texp, [wl range, R, SNR@500], [band, sigma (ppt)]
        \toprule
        \toprule
        % night:  2021-02-12
% euler:  0.000894
% ngts:  0.00202
% spec:  0.004216
% night:  2021-02-20
% euler:  0.000841
% ngts:  0.00206
% spec:  0.003969
% night:  2021-03-16
% euler:  0.000788
% ngts:  0.00206
% spec:  0.004708
            % 2 & 95 & \multirow{3}{*}{$\left.\begin{array}{l}
            %     \SI{30}{\second}\\
            %     \SI{30}{\second}\\
            %     \SI{45}{\second}
            %     \end{array}\right\rbrace\times\SI{35}{cycles}$} \\
        % & & & & & \multicolumn{2}{c}{Spectroscopic information} & \multicolumn{1}{c}{Photometric information} \\
        % \cmidrule(lr){6-7} \cmidrule(lr){8-9}
        Transit date & Instrument & {Sky resolution} & {$t_\mathrm{exp}$} & {Wavelength range / band} & $R$ & $\langle S/N \rangle^a$ \\%& Band \\%& $\langle\sigma\rangle$ \\
        & & {(\unit{\arcsec} or \unit{\arcsec\perpix})$^b$} & {(s)} & {(\si{\nano\metre})} & & \\%& \\%& (ppt) \\
        % \midrule
        %\mathrm{pixel}^{-1}$
        % \cmidrule(lr){2-3}
        % & WASP-85A & WASP-85B \\
        \midrule
        \multicolumn{8}{c}{\textbf{Campaign A}} \\
        \midrule
        % \cmidrule(lr){1-8}
        % \multicolumn{5}{c}{\textit{\textbf{12 February 2021}}} \\
        % \multicolumn{5}{c}{\textit{12 February 2021}} \\

        % NGTS aperture size: 3 pix (15'')
        % Eulercam aperture:
        % SPECULOOS aperture: multitudes of 4 pixels, or 1.4''. Options are: 1/2, 1/sqrt(2), 1, sqrt(2), 2, 2sqrt(2), 4, 5, 6, 7, 8, 10, 12

        % Spec aperture
        % 2021-02-12: 5'' (index 28, 15.22 pixels) is "best" according to default metric, stick with that
        % 2021-02-20: 5.4'' (index 30, 16.27 pixels) (i35 is "best",  but i30 looks more consistent with other light curves and espresso data.
        % 2021-03-16: i30? 5.4'' not sure, maybe best to stay consistent

        2021 February 12 & ESPRESSO & 1 & 400 & \SIrange{380}{788}{} & \num{138000} & 52 & \\ %&  \hspace{-1em}\rdelim\}{23}{*}[${}$ \textbf{Campaign A}] \\
        & EulerCam & 0.23 & 65 & $V$  & \\%& $V$ \\%& \num{0.89} \\
        & NGTS (801)$^c$ & 5 & 10 & ${\sim}VRI$  & \\%& $V+R+I$ \\%& \num{0.068} \\
        & NGTS (804) & 5 & 10 & ${\sim}VRI$ & \\%& $V+R+I$ \\%& \num{0.068} \\
        & NGTS (809) & 5 & 10 & ${\sim}VRI$ & \\%& $V+R+I$ \\%& \num{0.068} \\
        & NGTS (810) & 5 & 10 & ${\sim}VRI$  & \\%& $V+R+I$ \\%& \num{0.068} \\
        & NGTS (811) & 5 & 10 & ${\sim}VRI$  & \\%& $V+R+I$ \\%& \num{0.068} \\
        & SPECULOOS & 0.35 & 10 & $g'$ & \\%& $g'$ \\ %& \num{4.2} \\
        % \cmidrule(lr){1-8}
        \midrule

        % \multicolumn{5}{c}{\textit{20 February 2021}} \\  
        2021 February 20 & ESPRESSO  & 1 & 400 & \SIrange{380}{788}{} & \num{138000} & 49 \\
        & EulerCam & 0.23 & 75 & $V$  \\%& & & $V$ \\%& \num{0.84} \\
        & NGTS (801) & 5 & 10 & ${\sim}VRI$ \\%& & & $V+R+I$ \\%& \num{2.0} \\
        & NGTS (804) & 5 & 10 & ${\sim}VRI$ \\%& & & $V+R+I$ \\%& \num{2.1} \\
        & NGTS (809) & 5 & 10 & ${\sim}VRI$ \\%& & & $V+R+I$ \\%& \num{2.0} \\
        & NGTS (810) & 5 & 10 & ${\sim}VRI$ \\%& & & $V+R+I$ \\%& \num{2.1} \\
        & SPECULOOS & 0.35 & 10 & $g'$ \\%  & & & $g'$ \\% & \num{4.0} \\        
        % \cmidrule(lr){1-8}
        \midrule

        % \multicolumn{5}{c}{\textit{16 March 2021}} \\
        2021 March 16 & ESPRESSO & 1 & 400 & \SIrange{380}{788}{} & \num{138000} & 49 \\
        & EulerCam & 0.23 & 80 & $V$ \\% & & & $V$ \\%& \num{0.79} \\
        & NGTS (801) & 5 & 10 & ${\sim}VRI$ \\%& & & $V+R+I$\\% &  \\
        & NGTS (804) & 5 & 10 & ${\sim}VRI$ \\%& & & $V+R+I$ &  \\
        & NGTS (809) & 5 & 10 & ${\sim}VRI$ \\%& & & $V+R+I$& \\
        & NGTS (810) & 5 & 10 & ${\sim}VRI$ \\%& & & $V+R+I$ &  \\
        & SPECULOOS & 0.35 & 10 & $g'$  \\%& & & $g'$\\% & \num{4.7} \\
        \midrule
        \multicolumn{8}{c}{\textbf{Campaign B}} \\
        \midrule
        % \cmidrule(lr){1-8}

        % \multicolumn{5}{c}{\textit{17 April 2021}} \\
        2021 April 17 & HARPS-N & 1 & 600 & \SIrange{378}{691}{} & \num{115000} & 27 \\%\hspace{-1em}\rdelim\}{8}{*}[${}$ \textbf{Campaign B}]\\
        & CHEOPS$^d$ & 1 & 60 & ${\sim}BVRI$\\% & & &  $B+V+R+I$\\%  \\
        % \cmidrule(lr){1-8}
        \midrule

        % \multicolumn{5}{c}{\textit{7 March 2022}} \\
        2022 March 7 & HARPS-N & 1 & 600 & \SIrange{378}{691}{} & \num{115000} & 40 \\
        & CHEOPS$^e$ & 1 & 60 & ${\sim}BVRI$\\%  & & & $B+V+R+I$  \\
        % \cmidrule(lr){1-8}
        \midrule

        % \multicolumn{5}{c}{\textit{10 March 2022}} \\
        2022 March 31 & CHEOPS$^f$ & 1 & 60 & ${\sim}BVRI$\\% & & &  $B+V+R+I$ \\
        % \cmidrule(lr){1-8}
        \midrule

        % \multicolumn{5}{c}{\textit{8 April 2022}} \\
        2022 April 8 & HARPS-N & 1 & 600 & \SIrange{378}{691}{} & \num{115000} & 27 \\
        & CHEOPS$^g$ &1 & 60 & ${\sim}BVRI$\\%  & & & $B+V+R+I$ \\
        
        % \midrule 
    
        \bottomrule
        \label{table:observing log}
        \end{tabular}
\end{table*}

WASP-85A is a G5 main-sequence star ($T_\mathrm{eff} = \teffA{}$, $M_\star = \massA{}$) in a binary star system, with a projected separation of 210 au from its K0 dwarf companion WASP-85\,B ($T_\mathrm{eff} = \teffB{}$, $M_\star = \massB{}$). The hot Jupiter WASP-85Ab ($M_\mathrm{p} = \massb{}$, $R_\mathrm{p} = \radiusb{}$) orbits the primary star WASP-85A every $P = \periodb{}$ in a circular orbit \citep{brown2014}.
% and aligned prograde orbit \citep{brown2014,mocnik2016}.
%
% The planet was discovered by the WASP survey \citep{pollacco2006} and characterised using transit light curves from $K2$ Campaign 1, EulerCam on the Swiss 1-m and the 60-cm TRAPPIST-South telescope at ESO's La Silla Observatory -- and with radial velocities from SOPHIE on the 1.93-m at Observatoire de Haute-Provence near Marseille in France, CORALIE on the Swiss 1-m, and HARPS on ESO's 3.6-m telescope also located at La Silla Observatory \citep{brown2014}. 

The planet was discovered by the WASP survey \citep{pollacco2006} and characterised using transit photometry from $K2$ Campaign~1, EulerCam, and TRAPPIST-South -- and radial velocities from SOPHIE, CORALIE, and HARPS \citep{brown2014}. \citet{mocnik2016} later analysed the 80-day duration $K2$ photometry to study the rotational variability of WASP-85A induced by star spots covering its surface. Using the autocorrelation method on the rotationally modulated light curve (e.g. \citealt{mcquillan2013}) they found a rotation period of \ProtA{}. A second analysis was carried out on the spot occultations by the transits of WASP-85Ab, which suggested a rotation period of \ProtAspot{} and that the planet's orbit is aligned with the star's rotation axis with a projected spin-orbit angle of $|\lambda| < \SI{14}{\deg}$ \citep{mocnik2016}. This latter rotation period is also consistent with the dominant period of \qty{14.6 \pm 1.5}{\day} found in WASP photometry \citep{brown2014}. The host star, binary, and planet parameters are summarised in Table~\ref{table:spectroscopic_parameters}.

In this work we present new data from multiple observing campaigns of WASP-85A for the purpose of observing spot occultations during transits of WASP-85A simultaneously in photometry and spectroscopy. The data gathered during these campaigns include transit photometry from NGTS, EulerCam, SPECULOOS, CHEOPS, and spectroscopy from HARPS-N and ESPRESSO -- all of which are detailed below.

The ESPRESSO data presented in this work were analysed for the planet's atmosphere in transmission spectroscopy in \citet{jiang2023}, where they report tentative detections of H$\alpha$, \ion{Ca}{ii}, and \ion{Li}{i}, and confirm a low stellar obliquity ($\lambda = \SI{-16 \pm 3}{\deg}$) for the system by measuring the Rossiter-McLaughlin effect. As part of our analysis, we also provide a measurement of the projected and unprojected obliquity of the system using a similar approach, but additionally accounting for the effect of spot-occultations. We detail our observations below, and refer to the summary in Table~\ref{table:observing log}.

\subsection{Photometry}

\subsubsection{NGTS}
We observed WASP-85A with multiple apertures of the NGTS observatory located at ESO's Paranal Observatory, Chile. NGTS consists of 12 individual 20-cm diameter robotic telescopes that can be operated independently. Images in the NGTS band (a custom filter between \qtyrange{520}{890}{\nano\metre}, approximately $V+R+I$, see Figure~\ref{fig:bandpass}) were collected for several hours per night with an exposure time of \SI{10}{\second} for 155 nights between UT nights 2021 February 2 and 2022 May 14. The majority of the observations were scheduled for stellar activity monitoring of WASP-85A using a single camera. However, on UT nights 2021 February 12, 20, and March 16 NGTS observed transits of WASP-85Ab using five, four, and four cameras, respectively, each night to complement the spectroscopic transit observations with ESPRESSO that are described in §\ref{data:spectroscopy}. All transits were observed in full with sufficient out-of-transit baseline, except for the transit on 2021 February 12 with camera ID 811 where only pre-transit baseline and the first half of the transit was captured. Data were reduced using standard aperture photometry routines and detrended for systematics as described in \citet{wheatley2018}. The NGTS photometry from our long-term monitoring is shown in Figure~\ref{fig:ngts_monitoring}.

\subsubsection{SPECULOOS}
We observed 3 transits of WASP-85Ab on 2021 February 12, 2021 February 20, 2021 March 16 using Ganymede, one of four identical 1-m telescope units of the SPECULOOS-Southern observatory, located at Cerro Paranal, Chile. Images were taken with the Sloan-$g'$ filter (Figure~\ref{fig:bandpass}) in sequence with a \SI{10}{\second} exposure time. Given the brightness of the target, we used a defocused configuration to avoid saturating the detector. As a result, the mean FWHM for each of the three nights were \qty{5.24}{\arcsec}, \qty{4.85}{\arcsec}, and \qty{4.90}{\arcsec}, respectively. The data reduction and photometric extraction were done using a custom pipeline built with the \software{prose} package \citep{prose_soft,Prose_2022MNRAS}. For each of the datasets, we extracted the photometry of all the stars in the field for a range of 40 circular apertures. We performed differential photometry, and selected the optimal apertures to reduce the white and red noise in the light curves, corresponding to radii of \qty{5.02}{\arcsec}, \qty{6.23}{\arcsec}, and \qty{5.71}{\arcsec} respectively. 

% \textred{[more detail here for reduction etc, keep it to a short paragraph if possible]}.

\subsubsection{EulerCam}
We observed three transits of WASP-85Ab using the EulerCam imager on the Swiss 1.2-m Euler telescope located at La Silla Observatory, Chile, on UT nights 2021 February 12, 2021 February 20, 2021 March 16. These images were taken in the Geneva-$V$ filter (Figure~\ref{fig:bandpass}) with an exposure time of \SI{65}{\second}, \SI{75}{\second}, and \SI{80}{\second} on their respective nights. The raw full-frame images are corrected for bias, over-scan and flats using the standard EulerCam reduction pipeline as described in \citet{lendl2012}. The differential aperture photometry is performed by placing circular apertures with radii ranging from 14 to 50 pixels on the target star and reference stars. We chose the optimal aperture by minimising the photometric scatter of the consecutive images.  

% \textred{[Hritam: add EulerCam reduction details here, keep it to a short paragraph if possible]}

\subsubsection{CHEOPS}
We observed four transits of WASP-85\,Ab on UT nights 2021 April 17, 2022 March 7, 2022 March 31, and 2022 April 8 using ESA's CHEOPS satellite \citep{benz2021}, observed as part of the Guest Observers programmes CH\_PR220008 (PI Cegla). CHEOPS is in low Earth orbit and has a \SI{30}{\centi\metre} aperture with a broadband photometer covering the \qtyrange{380}{1100}{\nano\metre} wavelength range. All visits were automatically processed with the latest version of the CHEOPS Data Reduction Pipeline (DRP) \citep{Hoyer2020}. The DRP conducts standard image calibration and instrumental and environment artifact correction (e.g. cosmic-ray impacts, smearing trails of nearby stars, and background variations). We used the {\sc pycheops} package \citep{Maxted2022} to retrieve the CHEOPS light curves produced via aperture photometry with radii from 15-40\,pixels, and selected the data within the minimum RMS for each CHEOPS visit.

\subsubsection{Estimating flux contamination from WASP-85\,B}
\label{sec:dilution}
While the pixel scale of the EulerCam and SPECULOOS detectors in principle is small enough (\SI{0.23}{\arcsec\perpix} and \SI{0.35}{\arcsec\perpix}) to resolve the two stars separated by \SI{1.5}{\arcsec},
The apertures used in all our photometry have a radius larger than the sky separation of the WASP-85 binary. The two stars are therefore fully blended in our transit light curves, and thus, we need to account for flux dilution or third light effect \citep{southworth2010,southworth2012}. We developed an open-source tool \software{third-light}\footnote{https://github.com/chakrah/third-light} to estimate the flux ratio $f_\mathrm{B}/f_\mathrm{A}$ as seen by each of our instruments to properly scale the transit depth of our observations. We assume no contamination from additional background sources.

\texttt{third-light} uses synthetic spectra from the PHOENIX spectral library \citep{husser2013} to estimate the flux ratio between the secondary and the primary star. The synthetic spectra are parameterised by the effective temperature, surface gravity, and metallicity of the stars. We convolved the spectra with specific bandpasses to calculate the required flux ratio for different instruments. We find that the flux ratio $f_\mathrm{B}/f_\mathrm{A}$ between the two stars is \num{0.455 \pm 0.025} for EulerCam $V$, \num{0.376 \pm 0.067} for SPECULOOS $g'$, \num{0.439 \pm 0.057} for NGTS, and \num{0.433 \pm 0.089} for CHEOPS. We use these values as priors in our analysis in Section~\ref{sec:transit_analysis}.

\subsection{High-resolution spectroscopy}
\label{data:spectroscopy}
\subsubsection{ESPRESSO}
We observed WASP-85A using the ESPRESSO spectrograph on the VLT, located at Cerro Paranal, Chile, on UT nights of 2021 February 12, 2021 February, 20, and 2021 March 16 during predicted transits of WASP-85Ab \citep{pepe2021}. We used the HR2x1 mode ($R=\num{138000}$) and obtained respectively 47, 41, and 36 spectra in sequence with an exposure time of \SI{400}{\second}. The seeing during each sequence ranged from \qtyrange{0.3}{1.5}{\arcsec}, \qtyrange{0.4}{0.8}{\arcsec}, and \qtyrange{0.4}{0.7}{\arcsec}. We discarded exposures with seeing $\qty{>1}{\arcsec}$, which ensures that the contamination from WASP-85\,B is $\qty{<0.5}{\percent}$ and therefore negligible.
% which combined with the \SI{1}{\arcsec} fibre size of the ESPRESSO spectrograph means that contamination from WASP-85\,B (\SI{1.5}{\arcsec} away) was negligible.

% seeing > 1 for the 6-7 exposures after transit on first night, could contribute to the rapid change in slope

The spectra were reduced with version 3.0.0 of the ESPRESSO Data Reduction Systems (DRS), and cross-correlation functions (CCFs) were computed using a G8 line mask. The median signal-to-noise ratio (S/N) at \SI{550}{\nano\metre} for each night is 52, 49, and 49, respectively -- which translates to a disc-integrated radial velocity precision of about \SI{1}{\metre\per\second}. 

We used the \software{PAWS}\footnote{\url{https://github.com/alixviolet/PAWS}} code to perform an analysis of the ESPRESSO data to derive stellar spectroscopic parameters \citep{freckelton2024}. All spectra outside of transits were co-added to form a high-S/N spectrum in the \qtyrange{420}{680}{\nano\metre} wavelength range which were compared to synthetic spectra from the ATLAS9 grid \citep{kurucz2005}. We determine the effective temperature, surface gravity and metallicity from an equivalent width analysis of \ion{Fe}{i} and \ion{Fe}{ii} lines, giving $T_\mathrm{eff} = \SI{5780 \pm 110}{\kelvin}$, $\log{g} = \SI{4.57 \pm 0.1}{}$ and $\ion{[Fe/H]}{} = \SI{0.03 \pm 0.07}{}$. The Doppler broadening of the lines was derived from a spectral synthesis where the best fit is $v\sin{i_\star} = \SI{1.89 \pm 0.98}{\kilo\metre\per\second}$. This value is somewhat lower than what we derive from the spectroscopic transits in Section~\ref{sec:obliquity_analysis} but still compatible ($1.3\sigma$) given the large uncertainty on the Doppler broadening value.

%% ESPRESSO data is in BJD_UTC, have converted
\subsubsection{HARPS-N}
We also observed WASP-85A using the HARPS-N spectrograph, mounted on the 3.6-m TNG telescope on Roque de los Muchachos Observatory in La Palma, Spain \citep{cosentino2014}. During transits of WASP-85Ab on UT nights of 2021 April 17, 2022 March 7, and 2022 April 8, we obtained respectively 27, 40, and 27 spectra with an exposure time of \SI{600}{\second}. The S/N at \SI{550}{\nano\metre} ranged from \qtyrange{20}{32}{}, \qtyrange{34}{46}{}, and \qtyrange{19}{35} each night, translating to a disc-integrated radial velocity precision of \qtyrange{3}{7}{\metre\per\second}. We reduced the spectra with the HARPS-N DRS  version 2.3.5 \citep{dumusque2021} which is based on the ESPRESSO DRS, and CCFs were similarly computed using a G9 line mask. 

Due to somewhat variable observing conditions, some HARPS-N exposures were discarded because the auto-guider that keeps the fibre on the target star would temporarily jump to WASP-85\,B instead of tracking WASP-85\,A. These exposures are easily identifiable as their disc-integrated radial velocities jump by up to \SI{60}{\metre\per\second} compared to the slope of the orbit. We removed five pre-transit exposures from 2021 April 17 and two exposures from 2022 March 7 close to mid-transit. 

\section{The rotation period of WASP-85A}
\label{sec:rotation_period}

\begin{figure*}
    \centering
    \includegraphics{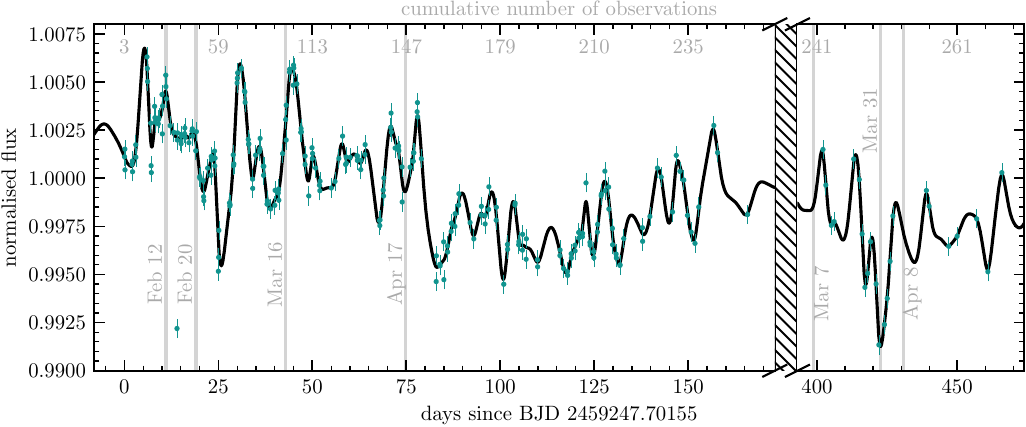}
    \caption{NGTS long-term monitoring of the WASP-85 binary system showing rotational variability in the light curve due to the evolution of spots on one or both stars. Each data point (\textit{green}) represents a \SI{2}{\hour} average of \SI{13}{\second} exposures. The individual uncertainties of the average points are fixed to \SI{0.5}{\ppt}, which we have chosen as a lower limit to the precision due to atmospheric scintillation. The vertical lines (\textit{grey}) indicate transits observed with various facilities in this work. The solid line (\textit{black}) is a Gaussian process regression fit to the data to help visualise the variability, composed of the sum of a non-periodic trend and two \software{celerite} \texttt{RotationTerm} kernels with primary periods of \qty{16.3}{\day} and \qty{11.5}{\day} found from the BGLS periodogram.}
    \label{fig:ngts_monitoring}
\end{figure*}

\begin{figure}
    \centering
    \includegraphics{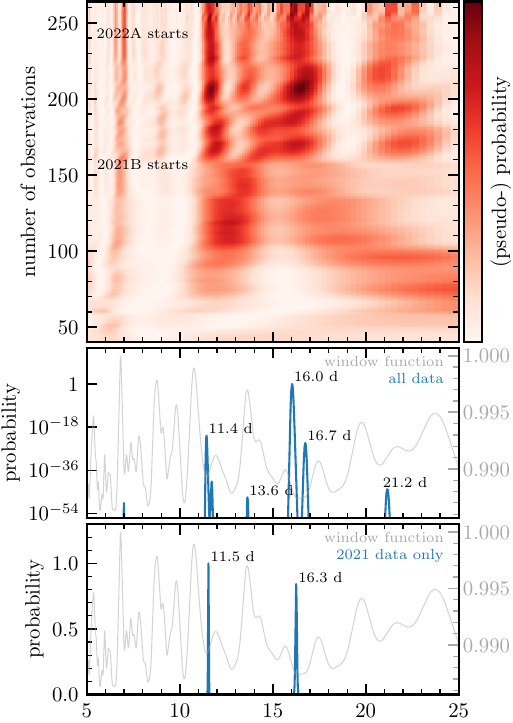}
    \caption{\textit{Top panel:} Stacked BGLS. For every \SI{2}{\hour} data point added to the light curve the BGLS periodogram is recomputed to visualise the increasing evidence for a periodic signal over time with more data. \textit{Bottom panel:} Bayesian generalised Lomb-Scargle periodogram (BGLS) of the light curve in Figure~\ref{fig:ngts_monitoring}. The height of the peaks give relative probabilities between each period. Grey lines are periodograms of the window function.}
    \label{fig:bgls}
\end{figure}

\begin{table}
    \sisetup{separate-uncertainty=true}
    \renewcommand{\arraystretch}{1.2}
    \small
    % \sisetup{round-mode=places}
    \centering
    \label{table:rotation_analysis}
    \caption{Rotation periods obtained from NGTS monitoring data. $^a$With probability $p=0.84$ $^b$With $p=1$.
    % approach with $v\sin{i_\star}=\qty{3.13\pm0.02}{\kilo\metre\per\second}$ and $R_\star = \qty{0.99\pm0.02}{\Rsun}$.
    Bold values indicate the adopted rotation periods for WASP-85\,A (\qty{15.8}{\day}) and WASP-85\,B (\qty{11.5}{\day}).}
    \begin{tabular}{@{\extracolsep{\fill}}
    llll}
        \toprule
        \toprule
        Method & Dataset & $P_\mathrm{rot}$  & Source \\
        & & (d)& \\
        % \cmidrule(lr){1-3}
        \midrule
        \multicolumn{4}{c}{\textbf{This work}}\\
        \midrule
        % BGLS all & \SI{16.0}{} & -- \\
        BGLS & NGTS all & 16.0 & This work \\
        BGLS & NGTS 2021 & 16.3$^a$ & This work \\
        BGLS & NGTS 2021 & \textbf{11.5}$^b$ & This work \\
        % BGLS & NGTS 2021B & \SI{16.9}{} & -- & 87 \pm 3 \\
        % BGLS & NGTS 2022A & \SI{12.9}{} & -- & 54 \pm 2 \\
        % ACF 2021A & \SI{13.8}{} & -- \\
        % ACF 2021B & \SI{13.8}{} & -- \\
        ACF & NGTS 2021A & $13.8 \pm 0.3$ & This work \\
        Transit correlation & EulerCam & \textbf{15.5} & This work \\
        % \midrule
        % \multicolumn{4}{c}{\textbf{Other work}}\\
        % \midrule
        ACF & $K2$ & 13.6 & \citeauthor{mocnik2016} \\
        Recurring spots & $K2$ & $15.1 \pm 0.6$ & \citeauthor{mocnik2016} \\
        Recurring spots & $K2$ & $15.2 \pm 0.3$ & \citeauthor{dai2018} \\
        % $^a$R-M $v_\mathrm{eq}$ & \SI{16.0}{} & \SI{0.3}{} \\
        \bottomrule
    \end{tabular}
\end{table}

% \begin{figure}
%     \centering
%     \includegraphics{figures/acf_periodogram_smooth1.5day_1_2.pdf}
%     \caption{Autocorrelation function (ACF) of 2021A and 2021B from the light curve in \ref{fig:ngts_monitoring}. The primary and subsequent integer multiple peaks are annotated, which are used to derive a measurement of the rotation period.}
%     \label{fig:acf}
% \end{figure}

The monitoring light curve of WASP-85AB with NGTS camera 9 is shown in Figure~\ref{fig:ngts_monitoring}. We observed the system over a period of 466 days (two seasons) on 154 individual nights at a cadence of \SI{10}{\second} ranging from a few hours per night to the whole night. We make cuts on airmass ($<\!1.6$) and perform an iterative $3\sigma$ clip on the flux measurements to exclude significant outliers, which leaves us with \num{147318} data points. These are averaged to \SI{2}{\hour} bins to reduce computational complexity in our further analysis and more clearly visualise the spot-induced rotational modulation in Figure~\ref{fig:ngts_monitoring}. We fix the uncertainty per \SI{2}{\hour} bin to \SI{500}{\ppm}, as we consider this to be the noise floor due to atmospheric scintillation given our observation setup. The root-mean-square (rms) for camera 9 has been demonstrated to follow a white noise law down to \SI{400}{\ppm} per \SI{45}{\minute} bin, therefore we consider \SI{500}{\ppm} a conservative choice \citep{bryant2020}. 

We follow multiple approaches to determine the rotation period of WASP-85A. First, we compute the Bayesian generalised Lomb-Scargle periodogram (BGLS) using the full dataset shown in the middle panel of Figure~\ref{fig:bgls} \citep{mortier2015}. We also compute the stacked BGLS by recomputing the BGLS by iteratively adding more data \citep{mortier2017}, these are shown in the top panel of Figure~\ref{fig:bgls}. Spot lifetimes on stars have lifetimes of several rotations \citep{basri2022,giles2017}, so because there is a gap of \qty{\sim 235}{\day} between the 2022 data and 2021 data, we also compute the BGLS periodogram using 2021 data only which we show in the bottom panel of Figure~\ref{fig:bgls}. Following the work of \citet{mcquillan2013,mcquillan2014} we also compute the autocorrelation function (ACF) of the light curve. Since the ACF is dependent on continuous coverage and can be sensitive to long term variability, we only compute the ACF using the first \qty{80}{\day} of 2021 data, i.e. 2021\,A.
% We first interpolate the light curve onto a regular grid to fill in large gaps, but set the flux to zero where we have no data. The ACF is then smoothed using a Gaussian kernel of width \SI{2.5}{\day}. The primary peak of the ACF is provisionally selected as the rotation period,
% then we convert the time lag of the remaining peaks to epochs. Each subsequent peak that is within \SI{20}{\percent} of an integer number epoch is kept, which in this case amounts to four additional peaks, annotated in Figure~\ref{fig:acf}. The period located at the ACF peaks as a function of epoch number is then used to fit a linear model $\mathbf{y} = m\mathbf{x} + b$ with a white noise term $\sigma_y$. We adopt the best-fit slope $m$ as the rotation period, and its uncertainty $\sigma_m$ as the uncertainty on the rotation period.
% and the best-fit $\sigma_y$ as its uncertainty.

Due to blending of the two stars in the light curve the rotation period we determine for WASP-85\,A differs between methods, and to some extent between seasons due to spot evolution, as evidenced in the light curve in Figure~\ref{fig:ngts_monitoring}.
% There are two main peaks are observed in the top panel at roughly $t = \SI{31}{\day}$ and \SI{45}{\day} that are of similar amplitude and shape, and both are also followed by a smaller peak about \SI{\sim5}{\day} later. The absence of a third clear peak at \SI{\sim 58}{\day} suggests that new spots are forming, flattening out the light curve and making inferences about the rotation period more difficult. Because spot lifetimes on stars have lifetimes of several rotations \citep{basri2022,giles2017}, we perform periodogram analyses on three individual sections of the light curve, labelled 2021A, 2021B, and 2022A in Figure~\ref{fig:ngts_monitoring}. 

The BGLS result using all available data is \SI{16.0}{\day} with no other  candidate periods. When using the 2021 data in isolation we find two strong candidate periods at \qty{16.3}{\day} with probability $p=0.84$ and \qty{11.5}{\day} with $p=1$. In contrast, using the first \qty{80}{\day} of data from 2021 the ACF method returns $P_\mathrm{rot}=\qty{13.8 \pm 0.3}{\day}$, and one can see from the stacked periodogram plot that this is the dominant period with BGLS as well in the first half of 2021. As the second half of data from 2021 is included the BGLS periogram splits into two solutions, which we attribute to the two signals from the WASP-85 binary, though from this we cannot be sure which one belongs to which star. 
% It therefore appears that the rotation period of WASP-85\,A is close to \qty{16}{\day}, while the rotation period of WASP-85\,B is close to \qty{11.5}{\day}.
The results are summarised in Table~\ref{table:rotation_analysis}, and in Section~\ref{sec:discussion} we return to the question of the rotation periods.

% For 2021A we find $P_\mathrm{rot} = \SI{13.8 \pm 0.3}{\day}$ with the ACF analysis, and \SI{13.4}{\day} using the BGLS method. In 2021B and 2022A there were too many gaps in the data for a reliable ACF analysis, but we obtained \SI{16.9}{\day} and \SI{12.9}{\day} with BGLS, respectively. The combined BGLS estimate using all available data is \SI{16.0}{\day}. The results are summarised in Table~\ref{table:rotation_analysis}.

\section{Transit light curve modelling}
\label{sec:transit_analysis}

\begin{figure*}
    \centering
    \includegraphics{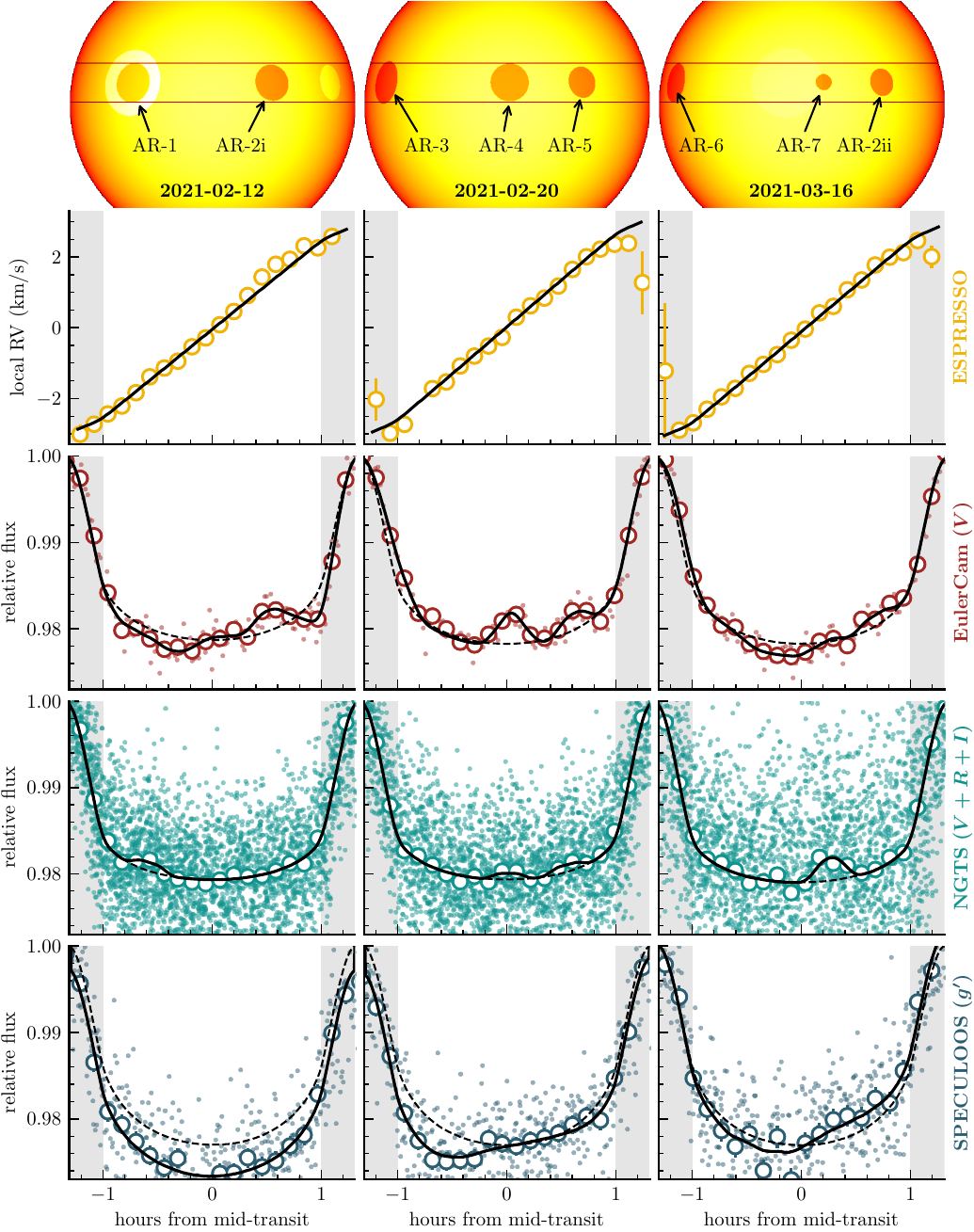}
    \caption{Transit observations from Campaign A (\textit{left to right column:} 2021 February 12, 20, and March 16) using (\textit{top to bottom}) ESPRESSO, EulerCam, NGTS, and SPECULOOS.  The ESPRESSO data show the local (subplanet) velocities based on the reloaded Rossiter-McLaughlin analysis (Section~\ref{sec:reloaded_rm}). The images in the top row are best-fit spot models of WASP-85\,A for each night based on EulerCam data. For the photometric data the coloured points are individual exposures, while the larger circles with white centres are binned to the same exposure time as individual ESPRESSO exposures (\SI{400}{\second}).
    The data has been corrected for the dilution and background trend. 
    % Data points with a coloured dot in the centre means they are classified as an outlier point according to Section~\ref{sec:outlier} and Figure~\ref{fig:prob_campaignA}.
    The median transit models from the MCMC posterior distribution are shown without (\textit{solid black}) and with (\textit{dashed}) the spot model.
    % using an Gaussian likelihood function convolved with the exponential distribution for the in-transit data.
    Transit ingress/egress regions are shaded grey, while the highlighted central regions are transit phases between the second and third contact points.}
    \label{fig:observations_espr}
\end{figure*}

\begin{figure*}
    \centering
    \includegraphics{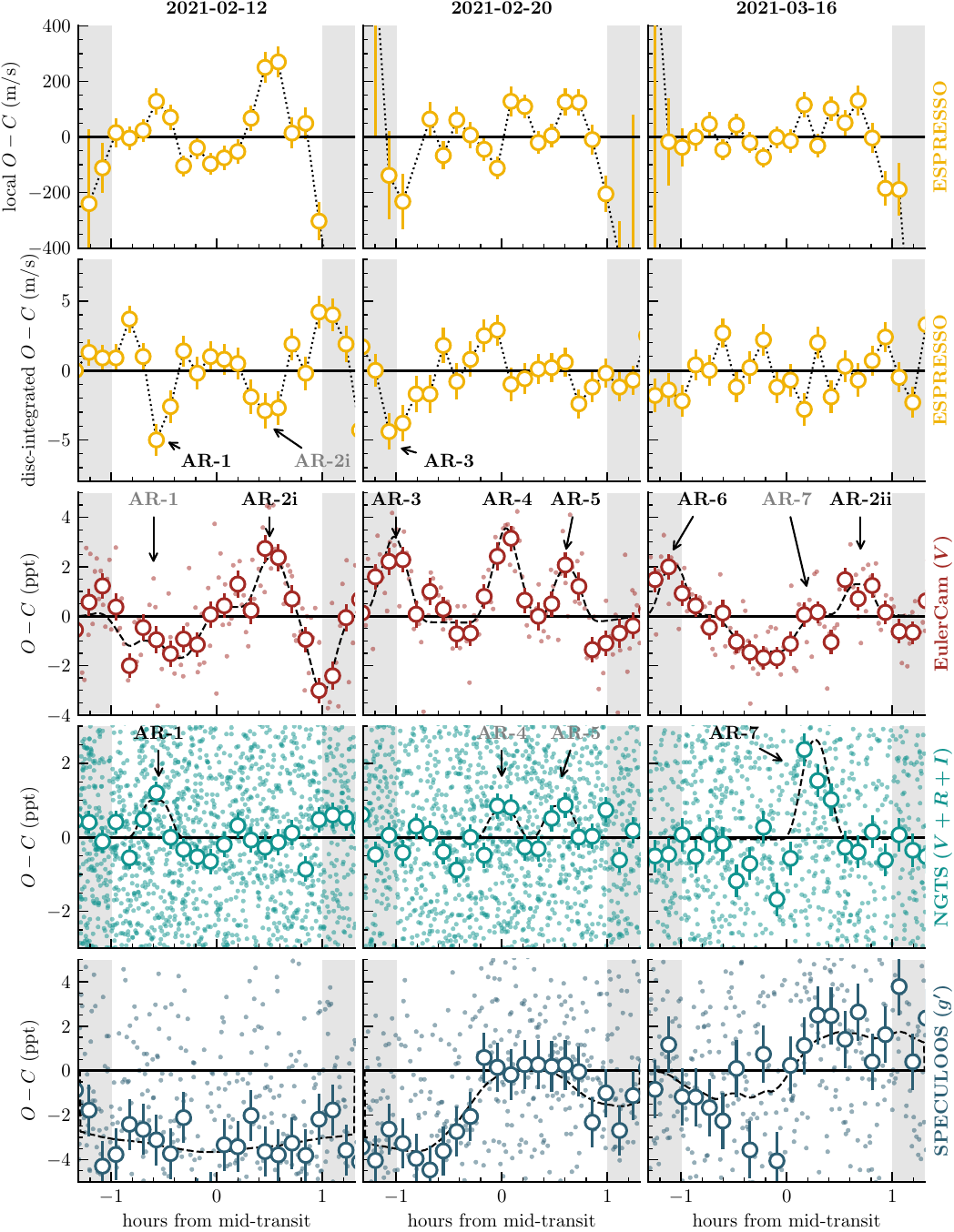}
    \caption{Transit residuals from subtracting the median transit model (without spots) from the data. The spot model for the in-transit data (\textit{dashed}) is also shown, with annotations indicating spot occultations of each active region, based on an outlier analysis of the residuals from EulerCam and NGTS. An annotation in black is classified as an outlier in that dataset, while grey annotations are outliers in a different dataset.}
    \label{fig:observations_espr_res}
\end{figure*}

The photometric transit modelling was performed in several steps: \textit{i}) We modelled the transit light curves assuming no spots are present. \textit{ii)} We analysed the in-transit residuals for signatures of spot occultations (Section~\ref{sec:outlier_identification}) \textit{iii)} Using  the spot occultations identified from \textit{ii)} we fit for the spot sizes and contrast to match the observed light curves.

\subsection{Spot-free model}
\label{subsec:spot_free_model}
Our spot-free global model was built using version 5.10.4 of \software{pymc} \citep{salvatier2016}, a Bayesian probabilistic programming library for \software{python}. We used \software{exoplanet} \citep{foreman-mackey2019} to generate limb-darkened transit light curves, fitting for the limb darkening coefficients $c_1$, $c_2$ assuming a quadratic model, mid-transit time $T_0$, impact parameter $b$, and the natural logarithms of the orbital period $P$, transit duration $T_{1,4}$, and planet-to-star radius ratio $r/R_\star$ to constrain them to positive values. We incorporated a prior on the stellar radius $R_\star = \SI{0.99 \pm 0.02}{\Rsun}$ based on Gaia DR3 stellar parameters catalogue \citep{fouesneau2023}. 

% A primary challenge of modelling transits without including spots is that the spot occultation signatures will bias the fit towards shallower transits and dilute the signal when searching for its signature in the residuals. Variable spot activity across different transit nights will also lead to variations in the transit light curve which may also bias the fit. This is further complicated by the fact that the transit light curve is diluted by the companion WASP-85\,B, which can also vary between nights and change the transit depth. 

% We account for these effects in several ways. 
% We place a constraining Gaussian prior on the planet-to-star radius ratio $r/R_\star = \SI[group-digits=false]{0.136749 \pm 0.000073}{}$ and transit duration ($T_{14} = \SI[group-digits=false]{0.10816 \pm 0.00004}{\day}$) published in \citet{mocnik2016} which is based on fitting 30 consecutive transits (with spot occultations removed) of WASP-85Ab observed with \textit{K2} Campaign 1.

We include a dilution\footnote{The flux ratio between the B and A stellar components, sometimes also referred to as ``third light``.} term to each photometric bandpass to account for the shallower transit due to light from the secondary star. Their values are controlled by a Gaussian prior with mean and standard deviations determined in Section~\ref{sec:dilution}.
% Variable spot activity across different transit nights can also lead to variations in the apparent transit depth which may also bias the fit. We account for this effect by re-scaling the transit depth by a factor $f_\mathrm{spot}$ which we fit for, and which is applied to each separate night and photometric bandpass. 

Each ground-based light curve was also modelled with a flux offset $a_0$ and a quadratic polynomial in time, $p^2_t \equiv a_1t + a_2t^2$, to correct for airmass effects over the course of the observations. In addition, the model may include one or several first or second-degree polynomials using sky background flux, FWHM of the photometric aperture, or aperture position in the $y$-direction as independent variables to correct for brightness variability induced by changes in these auxiliary measurements. These additional baselines share the intercept $a_0$ and do therefore not have additional intercepts. 
The CHEOPS light curves were mainly detrended against the roll angle of the spacecraft using first-order sine and cosine terms \citep{Maxted2022}, as well as first degree polynomials of the background flux. 
Our final light curve model then takes the form
\begin{align}
% f = \frac{(1 + f_\mathrm{spot} * f_0) + D}{1 + D} bkg
% \mathbf{y} = \left(\frac{f_\mathrm{spot}\,\mathbf{y_0}}{1 + D} + 1\right) \times \left(\bar{y} + \sum_{x} \mathbf{p}^{(n)}_x \right),
\mathbf{y} = \left(\frac{\mathbf{y_0}}{1 + D} + 1\right) \times \left(\bar{y} + \sum_{x} \mathbf{p}^{(n)}_x \right),
\end{align}
where bold-faced notation indicates the parameter is an array, $\mathbf{y_0}$ is the zero-normalised transit light curve,
% $f_\mathrm{spot}$ is the transit re-scaling factor due to spots,
$D = f_\mathrm{B}/f_\mathrm{A}$ is the dilution from WASP-85B, $\bar{y}$ is the mean flux offset, and $\mathbf{p}^{(n)}_x$ is a polynomial of degree $n$ with coefficients $c_{n,x}$ and $x$ as the independent variable, where the summation is over the chosen regressors, i.e. a combination of $x \in \{t, \mathrm{sky}, \mathrm{FWHM}, \delta y\}$.

% (m + a_0\,t + a_1\,t^2 + b_0\,\mathrm{sky} + b_1\,\mathrm{sky^2})
Finally, to avoid biasing the fit if spot occultations are present, we include a Gaussian process model to describe the in-transit data only. We use a Matern-3/2 covariance function provided by the \software{celerite2} package \citep{foreman-mackey2018}. This kernel is parametrized by the amplitude $\sigma_\mathrm{GP}$ and timescale $\rho_\mathrm{GP}$ which we fit for.

We sample the parameters of the model using the NUTS Markov chain Monte Carlo sampler incorporated in \software{pymc}. We initialise two separate chains whose starting positions are found by iteratively finding the maximum likelihood solution for different combinations of parameters in turn, and run them for \num{1000} steps after discarding an initial \num{1500} tuning steps. We inspected the $\hat{R}$ metric and the time series of the chains which both confirmed the two chains were fully converged and therefore approximated the posterior distribution of the model. We report the mean values and \SI{68}{\percent} ($1\sigma$) credible interval of the physical and orbital parameters in Table~\ref{table:spectroscopic_parameters}, while the full list of parameters and their priors are reported in Table~\ref{table:priors_table}. % and the best-fitting models to the light curves in Figure~\ref{fig:observations_espr}.

\sisetup{group-digits=false}
The median transit model from the posterior of the MCMC is overplotted on the data in Figure~\ref{fig:observations_espr} for Campaign A and in Figure~\ref{fig:observations_harpsn} for Campaign B. We obtain $P = \SI{2.6556746 \pm 0.0000012}{\day}$ which agrees to within a fraction of a second of the $K2$ analysis in \citet{mocnik2016}. We update the ephemeris of the system with $T_0 = \SI{2459258.825010 \pm 0.000090}{}$ BJD$_\mathrm{TDB}$. The value we get for the projected area on the star $r/R_\star = 0.1337 \pm 0.0016$ is $1.9\sigma$ different (our uncertainty) from the published value in \citet{mocnik2016}, namely $r/R_\star = \SI{0.1367485 \pm 0.000073}{}$. Our uncertainties are larger by a factor $\sim\!22$ which is due to several reasons. \textit{i)} we fit 23 transit light curves, compared to 30 transits from $K2$ which have higher S/N than our ground-based light curves. \textit{ii)} \citet{mocnik2016} adopt a fixed value for the light contamination from WASP-85\,B, whereas we include the uncertainty on the dilution factor in our fit.
%and \textit{iii)} the transit depth re-scaling factor $f_\mathrm{spot}$ accounting for variations in transit depth due to variable spot coverage on the stellar disc between transits, is largely degenerate with the radius ratio $r/R_\star$. 
Nevertheless, because we adopt a somewhat higher value for the stellar radius from \satellite{Gaia} DR3 compared to \citet{mocnik2016} ($R_\star = \SI{0.99 \pm 0.02}{\Rsun}$ vs. $R_\star = \SI{0.935 \pm 0.023}{\Rsun}$), we obtain a $1\sigma$ agreement on the planet radius $R_\mathrm{p} = \SI{1.288 \pm 0.030}{\Rjup}$. The transit duration we obtain $T_{14} = \SI{0.10915 \pm 0.00037}{\day}$ is $\sim\!\!2.7\sigma$ longer compared to the published value of $T_{14} = \SI{0.10816 \pm 0.00004}{\day}$. Despite this, our orbital solution leads to a physical separation $a = \SI{0.04005 \pm 0.00091}{\au}$ which is fully consistent with \citet{mocnik2016}. In conclusion, the physical parameters of the planet and its orbit are consistent with literature values. The full results and the priors are found in Table~\ref{table:priors_table}.

\subsection{Spot model}
\label{section:spot_model}
Our transit model, including active region occultations, was built using \texttt{PyTranSpot} \citep{pytranspot, sage}. The code uses a pixelation approach to project a stellar sphere, transit chord, and active regions onto a 2-dimensional Cartesian grid of pixels. Each active region is parametrised by its location (latitude and longitude), size, and contrast relative to the clear photosphere. A contrast greater than one corresponds to faculae, while a contrast less than one corresponds to star spots. The transit parameters are identical to those adopted for the spot-free model in Section~\ref{subsec:spot_free_model}, and we sampled the posterior distribution using  \software{emcee} \citep{goodman2010,emcee}. 

For Campaign A, the SNR of simultaneous transits observed with EulerCam, SPECULOOS, and NGTS varies significantly between observations. Furthermore, instrumental systematics differ across telescopes, limiting multicolour identification of active regions. Potential active region occultations are first identified by subtracting the median spot-free model and analysing the residuals (see Section~ \ref{sec:outlier_identification}). To constrain the properties of active regions, we fitted the \texttt{PyTranSpot} model to each light curve, adopting wide uniform priors on their size and contrast. To limit the degeneracies between the parameters, we fixed the spot latitude to 9.4$^{\circ}$, corresponding to the centre of the transit chord, and imposed Gaussian priors on the longitude of each active region, centred on the estimates derived from the spot-free residual analysis. For planetary orbital parameters, we adopted wide, uniform priors on mid-transit time and planet-to-star radius ratio, while fixing all remaining orbital parameters to the values obtained from the spot-free model. The resulting best-fit stellar surface maps are shown Figure~\ref{fig:observations_espr}, while the spot occultation light curves are shown in Figure~\ref{fig:observations_espr_res}. For Campaign B, orbital gaps in the CHEOPS light curves and the low SNR of the HARPS-N observations limit reliable identification of active regions occultations. Therefore, we exclude these light curves from the spot-modelling analysis.

\section{Determining the stellar obliquity}
\label{sec:obliquity_analysis}

We analyse the spectroscopic data in two ways to determine the projected obliquity of the star, $\lambda$. First we perform the velocimetric, i.e., ``classical'' Rossiter-McLaughlin analysis which models the disc-integrated radial velocities of the star during transit. As a second approach, we use the reloaded Rossiter-McLaughlin method, which recovers and models the local surface radial velocity of the star along the transit chord.

\subsection{``Classical'' Rossiter-McLaughlin}
\label{section:classical_rm_analysis}

\begin{figure*}
    \centering
    \includegraphics{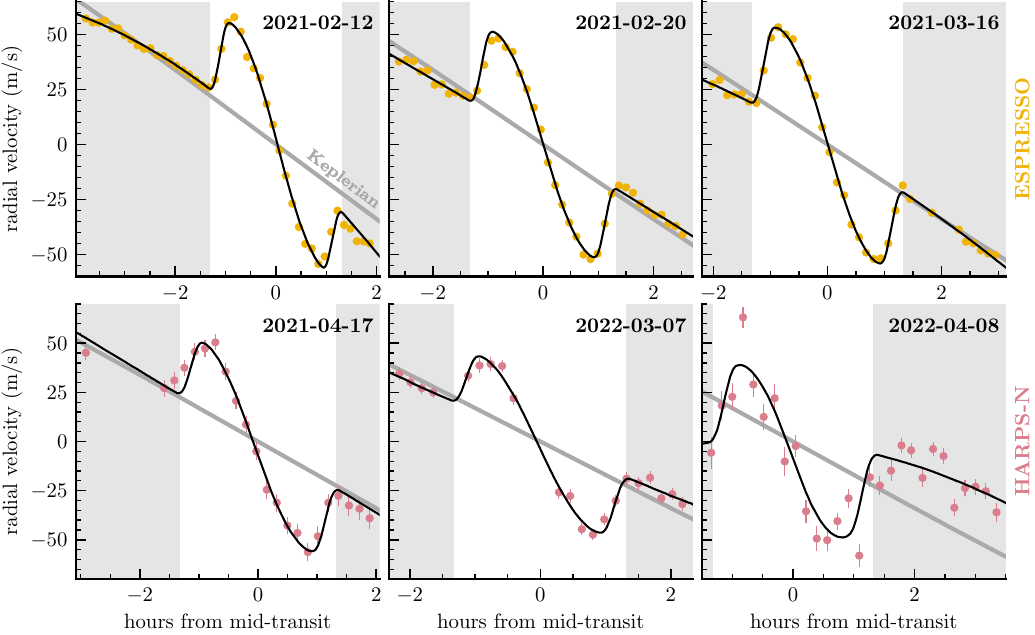}
    \caption{Rossiter-McLaughlin fit to disc-integrated radial velocities (i.e. ``classical'' Rossiter-McLaughlin analysis) from ESPRESSO and HARPS-N. Coloured points are the radial velocities and black lines are the best-fitting models to the data. The highlighted central regions indicate the transit window, grey areas are out of transit, while the grey sloping line is the Keplerian orbit expected from the orbital solution. Transit dates are indicated in each panel.}
    \label{fig:classical_rm}
\end{figure*}

We use the \citet{boue2013} Rossiter-McLaughlin model to model the in-transit disc-integrated radial velocities from ESPRESSO and HARPS-N. The model is provided by the publicly available \software{ARoME} library, however we use a Python wrapper provided by the \software{PyARoME}\footnote{\url{https://github.com/andres-jordan/PyARoME}} software.

The orbital parameters $a/R_\star$ and $i$ are controlled by Gaussian priors from our transit light curve analysis, while $P$, $T_0$ and $r/R_\star$ are fixed to the median values from the same analysis in order to make the best comparison to the reloaded Rossiter-McLaughlin analysis. The obliquity parameters $v\sin{i_\star}$ and $\lambda$ are free parameters allowed to range between \qtyrange{0}{5}{\kilo\metre\per\second} and \qtyrange[range-phrase=\ \text{to}\ ]{-180}{180}{\degree}, respectively. The \citet{boue2013} model further depends on the intrinsic width of the stellar absorption lines, $\beta_0$, and the width of the best-fitting Gaussian to the observed CCF, $\sigma_0$. We set the intrinsic width to the instrumental resolution $\beta_0 = c/R = \SI{2.19}{\kilo\metre\per\second}$ and \SI{2.61}{\kilo\metre\per\second} for the two instruments, where $c$ is the light speed and $R$ is the resolving power of the instrument. The average out-of-transit CCF has a night-dependent broadening $\sigma_0 = \qtyrange[range-phrase=\text{--}]{3.15}{3.17}{\kilo\metre\per\second}$ for ESPRESSO and \qtyrange{3.09}{3.11}{\kilo\metre\per\second} for HARPS-N. There is also an option to include line broadening due to macro-turbulence, which we set to $\zeta = \SI{2.97}{\kilo\metre\per\second}$ using the \citet{doyle2014} relation for the stellar parameters of WASP-85\,A. 

Because of the presence of spots on the stellar surface, the radial velocity slope during transit deviates from the expected slope due to the Keplerian orbit alone, shown in Figure~\ref{fig:classical_rm}. We fit a polynomial to the out-of-transit disc-integrated radial velocity data to get an accurate slope. For the ESPRESSO data we found that a second-order polynomial fit best, while the HARPS-N data was best-fit by a first-order polynomial. The coefficients of the polynomials are included as free parameters in the fit.

We run 300 ``walkers`` for 3000 steps each using the \software{emcee} \citep{foreman-mackey2019a} MCMC ensemble sampler, thin the chains by the autocorrelation length, and report the 68th percentile credible interval in Table~\ref{table:obliquity_analysis}. 
% removing post-transit on 2021-02-12:
%vsini           = 3.075 +/- 0.021
%ell             = -1.25 +/- 2.53
For the three different nights observed with ESPRESSO the measured value of $\lambda$ ranges from \qty{-17 \pm 4}{\degree} on 12 February 2020 to \qty{-4 \pm 5}{\degree} on the remaining two nights, a difference of \qty{13}{\degree} or a $2\sigma$ difference. While spots could be responsible for such a large difference between nights \citep{oshagh2018}, we note that the local seeing after egress rises above \qty{1}{\arcsec}, reaching \qty{1.5}{\arcsec} by the end of observations. This could mean that some light from the neighbouring WASP-85\,B might be entering the fibre and contributing to the slope deviating from the expected Keplerian. We repeated the fit on 12 February 2020 by excluding the last 7 exposures whose seeing was \qty{>1}{\arcsec} and obtained $\lambda = \qty{-1\pm4}{\degree}$, bringing its value more in line with those from the other nights. The $v\sin{i_\star}$ measured on each night also differs, but here it is the night of 2021 March 16 that measures \qty{3.24 \pm 0.04}{\kilo\metre\per\second}, compared to a lower value on the first two nights of \qty{2.99 \pm 0.04}{\kilo\metre\per\second} and \SI{3.01 \pm 0.04}{\kilo\metre\per\second} -- a difference of \qty{0.27}{\kilo\metre\per\second} and a $3.7\sigma$ separation of the two distributions which we attribute to the difference in spot coverage.

The results from HARPS-N are significantly different from ESPRESSO. From the former instrument we obtain $\lambda = \qty{37 \pm 10}{\degree}$ (2021 April 17), \qty{35\pm11}{\degree} (2022 March 7) and \qty{39 \pm 14}{\degree} (2022 April 8), all of which are ${>}3\sigma$ at odds from any of the ESPRESSO results. The projected velocities are also different, with values of $v\sin{i_\star} = \qty{4.3 \pm 0.5}{\kilo\metre\per\second}$, \qty{3.6 \pm 0.4}{\kilo\metre\per\second}, and \qty{4.8 \pm 0.7}{\kilo\metre\per\second} for the same order of nights as above. The likely reason for this discrepancy can be found in the observing log from all three HARPS-N nights. They all point to the fact that observing conditions were such that the WASP-85 binary could not be resolved most of the time, with a switch to manual guiding being necessary as the auto-guider could not track the correct component. It is therefore likely that all HARPS-N data suffer from a non-significant contamination from WASP-85\,B, therefore we interpret these results cautiously. 

% 2021 04 17 log: WASP25A is very close to the companion. It is not possile to guide on the star without including companion. 
% Variable seeing between 1.2 arcsec and 2.7 arcsec 

% 2022 03 07 log: The WASP85A transit observation was challenging because of the close proximity of the companion. Originally the seeing was excellent and the auto-guider worked well but as it worsened slightly part-way through the transit we had to switch to manual guiding to keep the correct target in the fibre. This resulted in a jump in the DRS RVs, implying the companion may have been contaminating the exposures immediately before the switch to manual guiding.

% 2022 04 08 log: WASP-85Ab transit observations between UT20:25-01:25 in variable 1.1-1.6" seeing conditions. The binary not resolved, manually guiding for 5 hours via telescope offsets trying to keep the western part of the combined source in the fiber.

With the combined fit using all ESPRESSO + HARPS-N transits we get $v\sin{i_\star} = \qty{3.07 \pm 0.02}{\kilo\metre\per\second}$ and $\lambda = \qty{-1\pm3}{\degree}$, owing to the higher S/N dataset from ESPRESSO compared to HARPS-N. This value of $v\sin{i_\star}$ is just about in agreement ($1.2\sigma$) with the result from Doppler broadening of \qty{1.89 \pm 0.98}{\kilo\metre\per\second} from Section~\ref{data:spectroscopy}.

% With the combined fit using all ESPRESSO + HARPS-N transits we get $v\sin{i_\star} = \qty{3.06 \pm 0.03}{\kilo\metre\per\second}$ and $\lambda = \qty{-9\pm3}{\degree}$ % this is without removing the last 7 points from 2021-02-12

% For the three different nights the measured value of $\lambda$ ranges from \SI{-18 \pm 4}{\degree} on 12 February 2020 to \SI{-4 \pm 5}{\degree} and \SI{-3 \pm 4} on the remaining two nights, a difference of up to \SI{15}{\degree}. The  $v\sin{i_\star}$ measured on each night also differs, but here it is the night of 2021 March 16 that measures \SI{3.34 \pm 0.04}{\kilo\metre\per\second}, compared to a lower value on the first two nights of \SI{3.07 \pm 0.06}{\kilo\metre\per\second} and \SI{3.09 \pm 0.04}{\kilo\metre\per\second} -- a difference of \SI{0.26}{\kilo\metre\per\second} or $>\!6$ standard deviations.

\subsection{Reloaded Rossiter-McLaughlin}
\label{sec:reloaded_rm}
We also use the reloaded Rossiter-McLaughlin method to determine the projected obliquity of the star. The method is well-established and described in detail in various literature \citep[e.g.][]{cegla2016,bourrier2017,kunovachodzic2021}. In summary, the disc-integrated CCFs are normalised by the transit light curve, resampled onto a common velocity grid by removing the radial velocity slope, and finally brought into the star's rest frame by removing the systemic velocity. The subplanet line profiles, CCF$_\mathrm{p}$ -- referring to the part of the disc-integrated CCF that is hidden behind the planet during transit -- is then obtained from taking the difference between the disc-integrated in-transit CCFs and the average out-of-transit CCFs. For the normalisation step we use the transit model with spot occultations to normalise the disc-integrated CCFs. A choice also has to be made about what radial velocity slope to remove: the expected Keplerian or the observed slope fit with a second order polynomial. The latter is a result of line shape deformations of the disc-integrated CCFs and is therefore not a pure Doppler shift, which may impact the results. We found that both methods lead to a similar inter-night spread in the values for $\lambda$, but not enough of a difference to affect any of our conclusions. The $v\sin{i_\star}$ remained largely consistent within uncertainties. To make a fair comparison to the classical analysis in the previous subsection, we remove the observed radial velocity slope.
% keplerian
% 3.12 +- 0.04, -2 -+ 2
% 3.02 +- 0.05, -7 +- 2
% 3.12 +- 0.05, -13 +- 2

% slope
% 2.97 +- 0.04, -8 +- 2
% 3.07 +- 0.04, 3 +- 2
% 3.10 +- 0.04, 2 +- 2
We fit Gaussian functions to the subplanet profiles and determine the local radial velocity across the transit chord (top panels in Figures~\ref{fig:observations_espr} and \ref{fig:observations_harpsn} for the ESPRESSO and HARPS-N data, respectively), and fit a solid-body rotation model to the data using \software{olio}\footnote{\url{https://github.com/vedad/olio}}.
% We also experimented with including a centre-to-limb convective blueshift term, though it did not improve the fit.
We sample the posterior distribution of $v\sin{i_\star}$, $\lambda$ with Gaussian priors on $R_\star/a$ and $\cos{i}$ from the photometric analysis using \software{emcee}. We run 200 ``walkers'' for 3000 steps each and thin the chains by the autocorrelation length and report the 68th percentile credible interval in Table~\ref{table:obliquity_analysis}.

% harpsn first night low snr, same with last night, both show misalignment because out of transit data baseline skews lambda. Middle night highest snr and shows most aligned solution, with consistent vsini with espresso

% slope
% 2.97 +- 0.04, -8 +- 2
% 3.07 +- 0.04, 3 +- 2
% 3.10 +- 0.04, 2 +- 2

% combined (slope)
% 3.00 \pm 0.04 km/s, 0 +- 1 deg

For the ESPRESSO data, the reloaded analysis gives $\lambda = \qty[multi-part-units=repeat]{-8 \pm 2}{\degree}$ (2021 February 12), \qty[multi-part-units=repeat]{3 \pm 2}{\degree} (2021 February 20), and \qty[multi-part-units=repeat]{2 \pm 2}{\degree} (2021 March 16), and $\vsini{} = \qty{2.97 \pm 0.04}{\kilo\metre\per\second}$, \qty{3.07 \pm 0.04}{\kilo\metre\per\second}, \qty{3.10 \pm 0.04}{\kilo\metre\per\second} for the same nights. While the \vsini{} is largely consistent between nights, $\lambda$ is consistent with an aligned orbit for the latter two nights, with a somewhat larger measurement of the misalignment for 2021 February 12 which we attribute to a poorly constrained slope because we removed the post-transit data due to possible contamination from WASP-85\,B. Indeed if we remove the Keplerian slope instead of the fitted slope we get is $\lambda = \qty[multi-part-units=repeat]{-2 \pm 2}{\degree}$ and $\vsini{} = \qty{3.12 \pm 0.04}{\kilo\metre\per\second}$, which is more consistent with the latter two nights.

The HARPS-N results are $\lambda = \qty[multi-part-units=repeat]{-4 \pm 9}{\degree}$ (2021 April 17), \qty[multi-part-units=repeat]{10 \pm 9}{\degree} (2022 March 7), and \qty[multi-part-units=repeat]{75 \pm 2}{\degree} (2022 April 8), and $\vsini{} = \qty{3.00 \pm 0.13}{\kilo\metre\per\second}$, \qty{2.85 \pm 0.12}{\kilo\metre\per\second}, and \qty{10.59 \pm 0.78}{\kilo\metre\per\second} for the same order of nights. While the latter two nights are in agreement with ESPRESSO in both \vsini{} and $\lambda$, the night of 2022 April 8 is a significant outlier in both parameters which we attribute to a combination of lack of pre-transit baseline, poor S/N, and poor seeing which likely introduced light from WASP-85\,B into the fibre. The average S/N was ${<}25$ during the transit, compared to the other two nights where the average S/N was roughly 28 and 40. The two first nights have a projected obliquity consistent with ESPRESSO, i.e. an aligned orbit, while the projected velocities too are largely consistent between nights and instruments.

Using all available ESPRESSO and HARPS-N data (excluding 2022 April 8), we obtain $v\sin{i_\star} = \SI{3.00 \pm 0.04}{\kilo\metre\per\second}$ and $\lambda = \ang{0} \pm \ang{1}$, which is within a $1\sigma$ agreement with the classical analysis.
% old analysis below
% Using all available ESPRESSO and HARPS-N data, we obtain $v\sin{i_\star} = \SI{3.13 \pm 0.02}{\kilo\metre\per\second}$ and $\lambda = \ang{-1} \pm \ang{1}$ from the Rossiter-McLaughlin analysis. The results for individual transit nights are summarised in Table~\ref{table:obliquity_analysis}, and illustrated in Figure~\ref{fig:corner_red_blue}. Overall we find excellent agreement between nights in $v\sin{i_\star}$ ($<\!\!1\sigma$) both for ESPRESSO and HARPS-N, with the exception of 2022 April 8 where $v\sin{i_\star}$ is $3\sigma$ from the combined result. While the value of $\lambda$ is low in all cases, it differs by up to 6 standard deviations between nights. The most discrepant night is 2021 February 12 for which we find a small, but non-zero misalignment of $\lambda = \ang{-10} \pm \ang{2}$, which is \SI{16}{\degree} different from the measurement on 2021 March 16, which is the least discrepant night that most closely agrees with the combined analysis.

% this needs updating too, later...
When separating the data into ``blue'' and ``red'' (relating to the two separate ESPRESSO CCDs covering different wavelength ranges, see Figure~\ref{fig:bandpass}) we find for 2021 February 12 $\lambda = \ang{-21} \pm \ang{4}$ for the red arm, which is ${\sim}3\sigma$ discrepant from the blue arm ($\lambda = \ang{-2} \pm \ang{3}$) and the white light result. Spot contrasts tend to be higher towards blue wavelengths, thus \textit{a priori} we would have expected the blue arm to be more discrepant than the red if the difference in the measured obliquity is due to spots. Instead, it may be a result of fewer lines in the CCF mask in the red arm, which reduces its S/N and skews the measured obliquity. A similar difference is seen for 2021 March 16, where the blue and red are up to \ang{27} discrepant from the white light, with the $v\sin{i_\star}$ also up to \qty{0.3}{\kilo\metre\per\second} different. For 2021 February 20 both $v\sin{i_\star}$ and $\lambda$ are in agreement for the blue, red and white light. 
% Instead, it may just be a result of the lower S/N of the red arm data that skews the measured obliquity.
% We see a similar effect when analysing the HARPS-N data in isolation. The lowest S/N data was taken on 2022 April 8 with HARPS-N, where we obtain $v\sin{i_\star} = \SI{4.37 \pm 0.44}{\kilo\metre\per\second}$ and $\lambda = \ang{37} \pm \ang{8}$ which is $>\!\!3\sigma$ discrepant from the combined result (which is mainly driven by the higher S/N ESPRESSO data). For the other two nights where HARPS-N observed we find an agreement to about $1\sigma$ with ESPRESSO, though with larger uncertainties. 

\subsection{Comparison of the two methods}

\begin{figure*}
    \centering
    \includegraphics{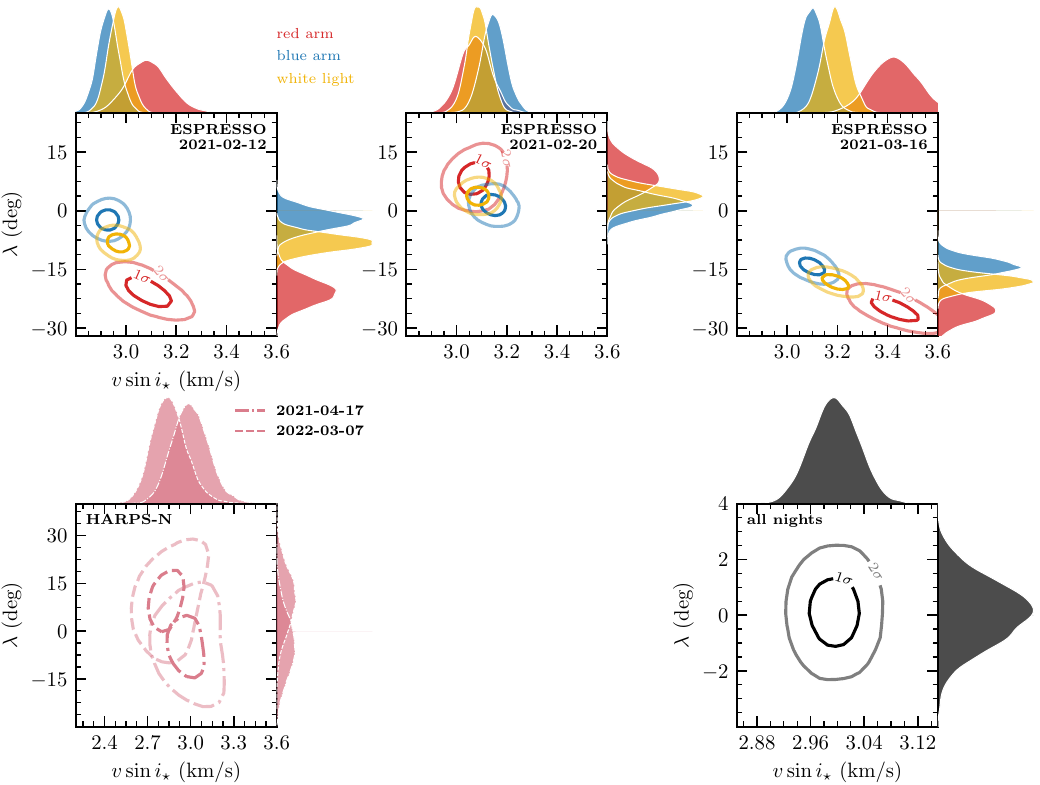}
    \caption{Rotation and obliquity results from the reloaded R-M analysis on ESPRESSO data without a spot model. \textit{Top row:} Red and blue curves are from the ESPRESSO red and blue arms (Figure~\ref{fig:bandpass}), and yellow is the combined "white" light results. Each panel represents individual nights, going left to right they are 2021 Feb 12, 2021 Feb 20, 2022 Mar 16. \textit{Bottom row:} HARPS-N results from the reloaded R-M analysis for each night (\textit{left}) -- the result from 2022 April 8 is outside the axes; and combined result from all ESPRESSO + HARPS-N data (\textit{right}).}
    \label{fig:corner_red_blue}
\end{figure*}

\begin{table*}
    \sisetup{separate-uncertainty=true}
    \renewcommand{\arraystretch}{1.2}
    \small
    % \sisetup{round-mode=places}
    \centering
    \caption{Summary of obliquity parameters obtained from the reloaded Rossiter-McLaughlin analysis for different configurations of instruments and models.}
    \begin{tabular}{
    l
    S[table-format=1.2(2)]%, round-mode=uncertainty, round-precision = 2, scientific-notation = false, drop-zero-decimal=false]
    S[table-format=-1.0(1), round-mode=uncertainty, round-precision = 1, scientific-notation = false, drop-zero-decimal=false]
    S[table-format=1.2(2)]
    S[table-format=-2.0(2)]
    S[table-format=1.2(2)]
    S[table-format=-1.0(1), round-mode=uncertainty, round-precision = 1, scientific-notation = false, drop-zero-decimal=false]
    }
        \toprule
        \toprule
         & \multicolumn{2}{c}{\textit{Spot model}} & \multicolumn{4}{c}{\textit{Spot-free model}}  \\ 
        \cmidrule(lr){2-3} \cmidrule(lr){4-7} 
        & \multicolumn{2}{c}{\textit{Reloaded R-M}} & \multicolumn{2}{c}{\textit{Reloaded R-M}} & \multicolumn{2}{c}{\textit{Classical R-M}} \\ 
        \cmidrule(lr){2-3} \cmidrule(lr){4-5} \cmidrule(lr){6-7}
        Configuration & $v\sin{i_\star}$ & $\lambda$ & $v\sin{i_\star}$ & $\lambda$ & $v\sin{i_\star}$ & $\lambda$ \\
        & {(\unit{\kilo\metre\per\second})} & {(\unit{\degree})} & {(\unit{\kilo\metre\per\second})} & {(\unit{\degree})} & {(\unit{\kilo\metre\per\second})} & {(\unit{\degree})} \\
        \midrule
        \textit{All nights} \\
        ESPRESSO + HARPS & 3.00 \pm 0.04 & 0 \pm 1 & 3.07 \pm 0.04 & -8 \pm 1 & 3.07 \pm 0.02 & -1 \pm 3 \\[3pt]

        \textit{2021 Feb 12} \\
        ESPRESSO & 2.97 \pm 0.04 & -8 \pm 2 & 3.12 \pm 0.04 & -2 \pm 2 & 2.99 \pm 0.04 & -17 \pm 4  \\
        ESPRESSO red arm & 3.09 \pm 0.09 & -21 \pm 4 & {--} & {--} & {--} & {--} \\
        ESPRESSO blue arm & 2.93 \pm 0.04 &  -2 \pm 3 & {--} & {--} & {--} & {--}\\[3pt]
        
        \textit{2021 Feb 20} \\
        ESPRESSO & 3.07 \pm 0.04 & 3 \pm 2 & 3.07 \pm 0.04 & 3 \pm 2 & 3.01 \pm 0.04 & -4 \pm 5 \\
        ESPRESSO red arm & 3.08 \pm 0.06 & 8 \pm 4 & {--} & {--} & {--} & {--}\\
        ESPRESSO blue arm &  3.15 \pm 0.05 & 1 \pm 3 & {--} & {--} & {--} & {--}\\[3pt]
        
        \textit{2021 Mar 16} \\
        ESPRESSO & 3.10 \pm 0.04 & 2 \pm 2 & 3.10 \pm 0.04 & 2 \pm 2 & 3.24 \pm 0.04 & -4 \pm 4\\
        ESPRESSO red arm & 3.43 \pm 0.09 & -25 \pm 3 & {--} & {--} & {--} & {--}\\
        ESPRESSO blue arm &  3.10 \pm 0.05 & -14 \pm 2 & {--} & {--} & {--} & {--}\\[3pt]

        \textit{2021 Apr 17} \\
        HARPS-N & {--}& {--}& 3.0 \pm 0.13 & -4 \pm 9 & 4.3 \pm 0.5 & 37 \pm 10 \\[3pt]
        
        \textit{2022 Mar 7} \\
        HARPS-N & {--}&{--} & 2.85 \pm 0.12 & 10 \pm 9 & 3.6 \pm 0.4 & 35 \pm 11 \\[3pt]

        \textit{2022 Apr 8} \\
        HARPS-N & {--}& {--}& 10.59 \pm 0.78 & 75 \pm 2 & 4.8 \pm 0.7 & 39 \pm 14 \\[3pt]
        \bottomrule
    \end{tabular}
    \label{table:obliquity_analysis}
\end{table*}

When fitting the data from ESPRESSO and HARPS jointly the results from both classical and reloaded analyses are in perfect agreement with one another. To summarise, we obtained  $\vsini{} = \qty{3.07 \pm 0.02}{\kilo\metre\per\second}$ and $\lambda = \qty[multi-part-units=repeat]{-1\pm3}{\degree}$ from the classical analysis, and $v\sin{i_\star} = \SI{3.00 \pm 0.04}{\kilo\metre\per\second}$ and $\lambda = \ang{0} \pm \ang{1}$ from the reloaded analysis -- both of which indicate that the orbit of WASP-85\,Ab is aligned with the projected spin axis of WASP-85\,A. The agreement between the two methods is largely due to the consistency of the results from ESPRESSO, owing to the higher S/N of the ESPRESSO data (a factor ${\sim}2$ improvement in the S/N per pixel at \qty{550}{\nano\metre}).

The HARPS-N results in isolation differ significantly between the two methods by ${\sim}2\ \text{to}\ 5\sigma$. While we saw that the classical result suffered significantly from contamination from WASP-85\,B due to poor conditions, its effect on the derived \vsini{} and $\lambda$ is less obvious with the reloaded method; in fact the HARPS-N results with the reloaded method are in $1\sigma$ agreement with ESPRESSO. This is somewhat expected; the reloaded method does not assume any specific shape for the disc-integrated line profile. If the observed profile is in fact a composite from two stars, subtracting it from the in-transit data should largely remove the contaminating contribution (as well as that of the target star), provided that the contamination remains approximately constant during the observing sequence. The last point in particular is perhaps the reason for why the reloaded result for HARPS-N on 2022 April 8 is significantly different from the combined result ($\vsini{} = \qty[multi-part-units=repeat]{10.59 \pm 0.78}{\kilo\metre\per\second}$, $\lambda = \qty[multi-part-units=repeat]{75 \pm 2}{\degree}$). From inspecting the shape of the disc-integrated line profiles throughout the night there appears to be a linear trend in the contrast, owing to the S/N gradually improving over the course of the night. Since there is only one pre-transit exposure compared to twelve post-transit exposures, the averaged out-of-transit profile does not match the true in-transit baseline. This mismatch artificially deepens the contrast of the local line profiles and can bias the inferred obliquity parameters.

In conclusion, the two methods are overall in agreement under two conditions: \textit{i)} the fibre is free of contamination from neighbouring stars, and  \textit{ii)} the line profile shape does not change too much over the observing sequence. If \textit{i)} is not true -- such as for WASP-85 when the components are unresolved -- the reloaded method appears to have an advantage as it can remove the contaminating light and give an accurate measurement of \vsini{} and $\lambda$. However, note that if the transiting companion causing the R-M signal itself is the contaminant (e.g. with FGK-M dwarf eclipsing binaries, \citealt{swayne2021,swayne2024,davis2024}) the contamination is time-variable and can still significantly bias the spin-orbit angle measurement \citep{kunovachodzic2020}. If \textit{ii)} is not true and there is a change in the line profile contrast or FWHM, e.g. due to S/N changes, then the classical analysis may have an advantage as it is less sensitive to the shape changes.

\section{Identifying active regions}
\label{sec:outlier_identification}

Figure~\ref{fig:observations_espr_res} (Campaign A) and Figure~\ref{fig:observations_harpsn} (Campaign B) show the residuals of our data after subtracting the median models from our posterior (Section~\ref{subsec:spot_free_model} and \ref{sec:obliquity_analysis}). In the following we will first consider Campaign A. The top two rows of panels show the spectroscopic ESPRESSO residuals during transit for both local and disc-integrated radial velocities, corresponding to the output from the reloaded and classical Rossiter-McLaughlin analyses, respectively. The dotted lines simply connect the data points for visual aid and is not a model. The bottom three rows of panels show the photometric data in turn for EulerCam, NGTS and SPECULOOS. The EulerCam light curves have the highest S/N and thus shows the most in-transit variability. On each night during transit there is a correlated signal clearly deviating from the expected ``white'' noise  which may be due to an inhomogeneous stellar disc. If this is the case the positive residuals in photometry would indicate the planet is occulting a dark region (spot) and the negative residuals could be due to an occultation of a brighter region (plage or faculae).

In radial velocity assigning a dark or bright feature based on the direction of the peaks is not as straightforward because it depends on the phase of the transit and the net convective blueshift. For systems with low obliquity, the net velocity shift of the disc-integrated Rossiter–McLaughlin effect decreases toward the projected stellar spin axis because the projected rotation speed (\vsini{}) is small. As a result, occultations of active regions produce stronger signatures near the stellar limb but are largely hidden at disc centre.

% The net velocity of the visible stellar disc (excluding rotation) is the sum of convective upflows in granules and downflows in intergranular lanes. Summed over the full disc the net effect is a blueshift commonly referred to as the net convective blueshift. 
In active regions magnetic fields suppress local convective motions.
Normally, surface granulation consists of bright upflows and dark downdrafts, which results in a net blueshift in the stellar spectrum. 
When these convective velocities are reduced, the affected region appears relatively redshifted. The effect is stronger in dark spots compared to faculae due to stronger magnetic fields being associated with the former. As a planet occults such a region, the removal of this redshifted contribution produces a relative blueshift in the disc-integrated spectrum. Therefore both bright and dark regions should produce a net blueshift in disc-integrated radial velocities, which means we have to resort to photometry to break this degeneracy.
% If active regions on the visible disc strongly suppress the overall convective blueshift, then occultation of a comparatively ``quiet'' region can produce an apparent redshift. -> true but the redshifted points are definitely in facular region so can't be this

% As a result, we consider any change in the residuals of the disc-integrated radial velocity during transit as an indication that the planet is occulting an active region, but without speculating on whether it is a dark or a bright region.

To quantify locations of likely spot occultations (instead of relying on visual confirmation) we perform regression on the residuals from the transit fit using a mixture model, whereby any data point belongs to either an inlier or outlier population (see e.g. \citealt{hogg2010}). We apply this approach to both campaigns for the classical R-M residuals and all photometry residuals. The reloaded R-M data is excluded because the analysis does not provide an out-of-transit baseline to compare the in-transit residuals to. In practice we use \software{pymc} and set up two likelihood functions, one for each population, which allows us to marginalise over the outlier class and get a measure of probability that any given point is an outlier\footnote{We follow the approach described in \citet{pymc_ex_outlier} and \url{https://gist.github.com/dfm/5250dd2f17daf60cbe582ceeeb2fd12f}}. 
% with slope $m$ and intercept $b$ fixed to zero.}.
For each residual light curve the inlier population is assumed to be a normal distribution with a zero mean and standard deviation given by the formal uncertainties. The outliers are also modelled as a normal distribution with zero mean, but with a different standard deviation $\sigma_\mathrm{out}$. We also sample the parameter $f_\mathrm{out}$, which is the fraction of points that are outliers. Using these 2 parameters ($\sigma_\mathrm{out}$, $f_\mathrm{out}$) we can generate models for both ``good`` and ``bad`` data points and perform inference on the observed data. At each step of the MCMC we track the probability that each point is an outlier and use its median posterior value as the probability that any given point is an occultation of an active region.
We sample the parameters of the model using four independent chains that are run for \num{10000} tuning (burn-in) steps and a final \num{1000} steps from which we obtain our posterior distribution.
% Those points that have a probability $\geq\!\SI{90}{\percent}$ of being an outlier (spot occultation) are highlighted in Figure~\ref{fig:observations_espr_res}. 

In Figure~\ref{fig:prob_campaignA} and \ref{fig:prob_campaignB} we show the outlier probabilities of all points for Campaign A and B respectively, and below we discuss each night in detail.
% In Figure~\ref{fig:prob_campaignA} and \ref{fig:prob_campaignB} we show the outlier probabilities of all points, and below we discuss each night in detail. 
In the following we will refer to the outlier probabilities as $p$. We use ``brightening'' to refer to residuals that tend upwards i.e. spot occultations, and ``darkening'' for residuals tending downwards, i.e. occultations of bright regions (plage or faculae). All active regions that were identified in the next three subsections are drawn on a simulated disc in the top row of Figure~\ref{fig:observations_espr}. The occultations of these regions are annotated in Figure~\ref{fig:observations_espr_res} in black if they are detected with $p>0.9$, and grey if $p<0.9$ but the region is $p>0.9$ in another photometric band.

\subsection{2021 February 12}

In the first column in Figure~\ref{fig:prob_campaignA} there is strong evidence of in-transit variability in ESPRESSO, EulerCam and NGTS, with multiple peaks showing $p > 90\%$ chance of being outliers indicated by yellow points. EulerCam shows a steady darkening starting at the second exposure during ingress which may be due to a plage or facular region at the limb of the star. The facular region itself has an outlier probability $p>0.9$. This bright region is quickly followed by a rapid brightening around phase $-0.1$ which could be due to a spot occultation. However given that the brightening is relative to the local photosphere (which is brighter than the average), the occultation of this potential spot brings the residual closer to the average (zero) and is therefore not flagged as an outlier in EulerCam. Despite this limitation of the method, the NGTS data does detect a significant brightening with  $p>0.9$ indicating a spot occultation, but no evidence of a bright region. Bright facular regions are hotter patches of gas relative to the quiet photosphere whose emergent spectrum is shifted slightly towards shorter wavelengths. Since EulerCam data uses the $V$-band filter it is more sensitive to faculae than the broadband NGTS filter (Figure~\ref{fig:bandpass}). Conversely the emergent spectra of dark spots are shifted towards redder wavelengths and are therefore still detectable by NGTS. The ESPRESSO data too shows two outliers in the same region with $p>0.9$, changing rapidly from a \qty{4}{\metre\per\second} redshift to a \qty{5}{\metre\per\second} blueshift. As discussed earlier the occultation of a spot or faculae both lead to a blueshift in the disc-integrated radial velocities, so the blueshifted outlier is consistent with a (dark) spot occultation when interpreted in tandem with the EulerCam and NGTS photometry. We will refer to the combined facular and spot region as active region 1, or AR-1 for short and use the suffixes ``b'' and ``d'' when referring to the bright and dark component, respectively. 

The redshifted outlier can be explained by reduced convective blueshift and viewing geometry. Faculae are concentrations of evacuated magnetic flux tubes that suppress convection and lower energy transport. At disc centre they appear slightly darker, but toward the limb the evacuated depressions expose the hot granule walls, making faculae appear brighter and giving them disproportionate weight in the flux average. Because they are both brighter and locally more redshifted, limb faculae shift the disc-integrated velocity toward the red. The systemic velocity fitted in the classical R–M model is offset: it combines the true stellar velocity with the net convective blueshift. Limb faculae reduce this blueshift, biasing the observed systemic velocity redward. When such a facula is occulted, the bright, redshifted contribution is removed, shifting the disc-integrated value back to the blue. Because the net convective blueshift is baked into the systemic velocity, it leads to an apparent redshift in the residual velocities.

Turning our attention to EulerCam again, the transit progresses towards a darker region with another brightening ($p>0.9$) that peaks around phase $0.008$ with amplitude $3\,\mathrm{ppt}$, followed by a darkening near egress at the limb ($p>0.9$). The dark region is not detected in NGTS, whereas the bright region may be with $p > 0.8$. The bright region is a clear detection in ESPRESSO ($p>0.9$), again producing a redshift due to similar arguments as explained in the previous paragraph. The dark region only has an outlier probability of 0.7--0.8, likely affected by reduced sensitivity near disc-centre. We will refer to the dark spot as active region 2i, or AR-2i, for reasons that will become clear later.

The SPECULOOS $g'$ data has comparatively lower S/N than EulerCam and the multi-camera NGTS data, therefore we are unable to identify individual active region occultations. The $g'$-band data also suffers from more correlated noise as seen from the out-of-transit residuals, making physical interpretations difficult. However, we see from Figures~\ref{fig:observations_espr}, \ref{fig:observations_espr_res}, and \ref{fig:prob_campaignA} that the transit depth in the $g'$ filter is about 3\,ppt deeper on 2021 February 12 than the other bands and any other night. This could indicate that the stellar disc as a whole is dimmer on this particular night by \qty{\sim12}{\percent} in the $g'$-band. Alternatively -- given the $g'$ filter is the most sensitive to faculae -- the two faculae seen in EulerCam may extend across most of the visible surface so that the planet is constantly occulting regions that are \qty{\sim14}{\percent} brighter than average photosphere.

% During ingress there is a steady darkening in photometry which may be due to plage or faculae at the limb of the star, where the NGTS and EulerCam data shows that this is quickly followed by a rapid brightening which could be due to a spot occultation. This rapid change is also seen in ESPRESSO data. We will label this region from phase \qtyrange{-0.9}{-0.5}{\hour} as active region 1, or AR-1.  The transit progresses through a somewhat darker region and peaks another time at around \SI{0.5}{\hour} after mid-transit with an amplitude of $\sim\!\!\SI{2.5}{\ppt}$, also with $p_\mathrm{out} > 0.9$. After the peak it drops to a minimum of about \SI{-2}{\ppt} at \SI{1}{\hour} after mid-transit, potentially due to a plage ($p_\mathrm{out} > 0.9$). These rapid changes are also seen in ESPRESSO. We will refer to the region between \qtyrange{0.3}{1.2}{\hour} as AR-2. The SPECULOOS light curve from this night do not quite reproduce the same variability due to lower S/N or potentially correlated noise of instrumental and/or environmental origin.

\subsection{2021 February 20}
 
On the second night (second column in Figure~\ref{fig:prob_campaignA}) we see three significant brightenings indicative of spot occultations with $p>0.9$: one during ingress with an amplitude of $\sim\!\!\SI{2}{\ppt}$, another at mid-transit with an amplitude of $\sim\!\!\SI{3}{\ppt}$, followed by a $\sim\!\!\SI{2}{\ppt}$ peak at phase 0.012. In the NGTS data none of the peaks stand out strongly, with only the second and third reaching moderate outlier probabilities ($p=\qtyrange[range-phrase=\text{--}]{0.7}{0.8}{}$). However, the clear detection of these events in the EulerCam light curve updates our prior expectation, and within a Bayesian framework this strengthens the case that the same peaks are also present in NGTS. In ESPRESSO only the first peak is detected clearly ($p>0.9$) with a \qty{5}{\metre\per\second} blueshift, as one would expect for a spot occultation. Because the second peak is at mid-transit our sensitivity to outliers is greatly diminished. The third peak only has a moderate outlier probability ($p=\qtyrange[range-phrase=\text{--}]{0.7}{0.8}{}$) with a \qtyrange{2}{3}{\metre\per\second} blueshift, however again from Bayesian arguments our prior expectation strengthens the case that this peak is also detected in ESPRESSO. We will labels these active regions AR-3, AR-4 and AR-5.

The SPECULOOS $g'$ data again suffers from correlated noise as seen from the out-of-transit residuals making spot occultation inferences challenging. In the raw light curve there is a significant trend before transit which appears correlated with changes in the sky background flux. Our attempt to model this trend regressing against the background flux, FWHM and centroid changes, and an airmass trend did not produce a perfect fit, causing some residual signal to remain. 

% We will label the peaks AR-3 to AR-5 going forward. From the outlier analysis in Section~\ref{sec:outlier}, all three peaks due to a dark spot and the one peak due to a bright spot (plage) are assigned $>\!\!90\%$ probability of being an outlier, i.e. $p_\mathrm{out} > 0.9$. AR-3 is seen in ESPRESSO as well with $p_\mathrm{out} > 0.9$, while AR-5 is detected at $p_\mathrm{out} > 0.8$. AR-4 is mid-transit so we do not expect the disc-integrated radial velocities to be sensitive to it. The active regions AR-2 and AR-3 are also visible in NGTS (third row in Figure~\ref{fig:observations_espr_res}), where they are also assigned a $>\!\!90\%$ outlier probability. The GP model is negligible in NGTS data because the data from a single camera does not have high enough S/N to fit any other signal than white noise. However when stacking the data from multiple cameras the spot occultations become apparent. In SPECULOOS data (fourth row) AR-2 and 3 are not individually resolved due to low S/N, but the data shows evidence for a brightening at the expected location, though the outlier probability is only about $0.6$. AR-5 is also seen in SPECULOOS data with $p_\mathrm{out} > 0.9$, as would be expected for two dark spots followed by a bright region such as a plage. 

\subsection{2021 March 16}

On the final night of Campaign A (third column in Figure~\ref{fig:prob_campaignA}) the EulerCam light curve show a brightening with $p>0.9$ at ingress indicating a dark spot at the limb. This spot is not detected in NGTS, but ESPRESSO does show a \qtyrange[range-phrase=\text{--}]{2}{3}{\metre\per\second} blueshift with a moderate outlier probability of \qtyrange[range-phrase=\text{--}]{0.7}{0.8}{}. Again from Bayesian arguments we consider this dark spot detected in ESPRESSO. We will label this dark spot AR-6. The EulerCam light curve continues with a gradual darkening, peaking right before mid-transit around phase $-0.05$ with an amplitude just shy of $\qty{-2}{\ppt}$. This bright region appear detected even in broadband NGTS with $p>0.9$. Immediately after mid-transit there is a clear brightening in EulerCam at phase 0.05, however just as for AR-1 this brightening occurs within a facular region so the occultation only brings the flux back to the average. Despite this, NGTS clearly detects this spot occultation with a peak amplitude of \qty{2.5}{\ppt}. Since the dark spot is close to mid-transit the sensitivity to outliers in ESPRESSO is reduced, so this region only has an outlier probability of \qtyrange[range-phrase=\text{--}]{0.6}{0.7}{}. Because the dark spot is embedded within the bright region we will refer to the combined region as AR-7, with the suffixes `b' and `d' used for describing the bright and dark regions just as for AR-1. After AR-7 EulerCam detects another brightening with with outlier probability $p>0.9$ around phase 0.01 with amplitude \qty{1.5}{\ppt}. This peak is not detected in NGTS nor ESPRESSO. In Sections~\ref{sec:rotation_period} and \ref{sec:discussion} we derive the rotation period of WASP-85\,A and find that on this night the same hemisphere is visible as on 2021 February 12 to within a few degrees. As a result we hypothesise that this dark spot is the second iteration of AR-2 detected on the first night, so we label it AR-2ii.

We also make three more hypotheses. In EulerCam at phase around -0.01 there is one exposure that is temporarily brightened compared to the surrounding area. Because it is in a somewhat brighter area the outlier probability is low and NGTS data does not detect it, so it could be consistent with a small or low contrast spot (with a temperature similar to the surrounding photosphere). Its location coincides with AR-1d detected on 2021 February 12, so this particular brightening could be the evolved version of AR-1d. Similarly on the same night there is a temporary brightening at phase 0.05 with $p = \qtyrange[range-phrase=\text{--}]{0.7}{0.8}{}$ which again is not detected in NGTS so we cannot claim a bona-fide spot, but this too could also be consistent with a small or low contrast spot because its location is the same as AR-7d on 2021 March 16 which is detected. If so, AR-7d could be the evolved version of this the small putative spot on 2021 February 12. Similar arguments can be used for AR-6 on 2021 March 16, which has a potential counterpart at ingress on 2021 February 12 with $p = \qtyrange[range-phrase=\text{--}]{0.7}{0.8}{}$. 

While we are unable to identify active region occultations from the SPECULOOS $g'$ light curve, the residuals look qualitatively similar to the EulerCam data: a darkening due to half the hemisphere dominated by faculae, followed by a brightening due to the other hemisphere being dominated by dark spots. 

\subsection{HARPS-N and CHEOPS}

In Campaign B we do not find any evidence of outliers in the HARPS-N disc-integrated residuals for 2021 April 17 and 2022 March 7 (Figure~\ref{fig:prob_campaignB}). The uncertainties on the individual exposures are \qtyrange{3}{5}{\metre\per\second}, which is close to the maximum velocity shift observed for active region occultations in the ESPRESSO data, suggesting our sensitivity to detecting active regions with HARPS-N is reduced. The final night of 2022 April 8 shows three significant outliers, but as we discussed in Section~\ref{sec:obliquity_analysis} the data has low S/N and is significantly affected by light from WASP-85\,B entering the fibre. As a result we are unable to draw any conclusions from the HARPS-N data about the activity of WASP-85\,A on 2021 April 17, 2022 March 7 and 2022 April 8. 

The precision on the CHEOPS photometry is better than any other photometric data studied in this work. However the data contains gaps due to CHEOPS' orbit which last up to \qty{30}{\minute}. These gaps are often accompanied with a flux trend as can for example be seen in the CHEOPS residuals on 2022 March 7 in Figure~\ref{fig:prob_campaignB}. Moreover, given the complications with the HARPS-N data discussed above, and the lack of simultaneous photometry from another facility means it is challenging to draw conclusions on the presence of active regions based on CHEOPS data only.

On 2021 April 17 there is no significant outlier in the residuals, which suggests the photosphere is quiet. On 2022 March 7 there is one outlier event around phase -0.007 with $p>0.9$, however there is a gap in the data at the limb where faculae would normally be brighter, which makes it somewhat difficult to draw a conclusion on the nature of the outlier.
On 2022 March 31 there are two significant outliers with $p>0.9$ at the limb near egress. The first point is at the start of a new CHEOPS orbit, making it unlikely that it is due to an active region.
% The general trend appears to be a darkening of the residuals after mid-transit which is consistent with this hypothesis. The two darkened points are interrupted by a brightening event at phase 0.0125 which may be due to a central dark spot, though its moderate outlier probability $p = \qtyrange[range-phrase=\text{--}]{0.6}{0.7}{}$ prevents us from firmly concluding on it. These outliers are immediately preceded by a small gap in the data which also affects our ability to claim a solid detection of an active region.
On the final night, 2022 April 8, there are two significant outliers ($p>0.9$) around phase -0.01. However both points are associated with the end and start of a CHEOPS orbit so are likely caused by spacecraft systematics. Later, after mid-transit, there is a brightening ($p>0.9$)  at phase 0.005 with an amplitude of around \qty{0.6}{\ppt} which is not obviously associated with the CHEOPS orbit. This is the only candidate dark spot from CHEOPS, but due to lack of simultaneous photometry we cannot confirm its nature. At egress there is another outlier with $p = \qtyrange[range-phrase=\text{--}]{0.8}{0.9}{}$, but this too is likely caused by spacecraft systematics.

\section{Line profiles from active regions}
\label{sec:subplanet_profile_analysis}

In the previous section we identified eight unique active regions spread across campaign A: seven dark spots -- two of which are located in bright regions with faculae -- and one facular area without a dark spot. In this Section we study the shape of the line profiles in these active regions.

\subsection{Line shapes}
\label{sec:subplanet_line_shapes}

\begin{figure*}
    \centering
    \includegraphics{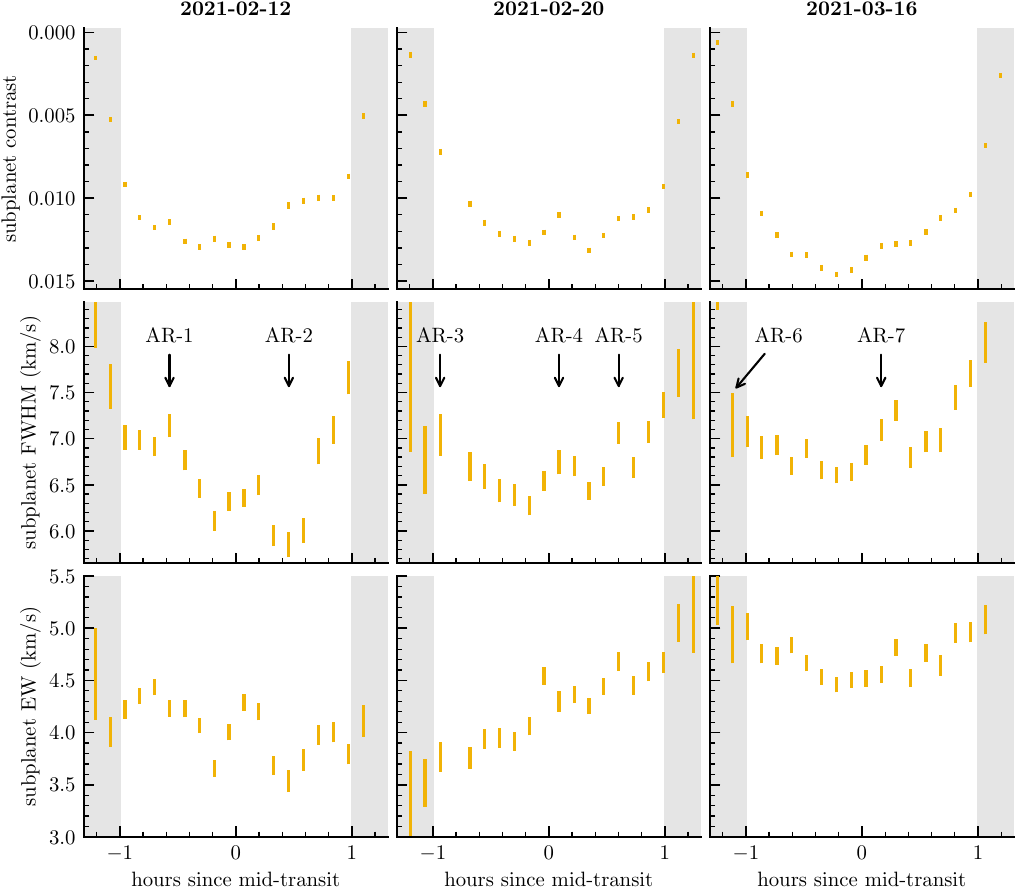}
    \caption{Subplanet profile shape parameters for each ESPRESSO transit as function of transit phase. In all panels yellow vertical lines are the uncertainties on the measurement. The line profile depth $a_\mathrm{p}$ (\textit{top row}) and line profile width FWHM$_\mathrm{p}$ (\textit{middle row}) are derived from Gaussian fits to the profiles. \textit{Bottom:} Equivalent width (EW) of the profiles, i.e. a rectangle of unit height with a width such that its area is equivalent to the area of the profile.}
    \label{fig:local_profiles}
\end{figure*}

In the first two rows of Figure~\ref{fig:local_profiles} we show the contrast and FWHM obtained from Gaussian fits to the local CCF profiles on each night. These line profiles are obtained using the reloaded Rossiter-McLaughlin approach in Section~\ref{sec:reloaded_rm}, and as a reminder are normalised by the spot + transit model on each respective night to accurately reflect the continuum levels of the in-transit profiles. The overall shape of the contrast curve will therefore resemble our full transit model in Figure~\ref{fig:observations_espr}. In addition to the overall transit light curve shape, variations in line strength are observed at locations corresponding to active regions. These signatures persist even when the subplanet profiles are normalised using a spot-free transit model, though with reduced amplitude. 

As for the FWHM, 3D magnetohydrodynamic simulations predict that line profiles broaden toward the limb, where horizontal flows become increasingly aligned with our line of sight and therefore contribute more strongly to the observed broadening \citep{beeck2013}. We \textit{broadly} observe this effect for all three nights. Besides this general trend we also observe clear changes in the line profile during spot occultations, as shown by the annotations of the active region occultations in  Figure~\ref{fig:local_profiles}. Subplanet profiles from all but three active regions are broader compared to the general trend with limb angle. The exceptions and AR-6 which is at the very limb with low S/N; and AR-2 which is narrower. The broadening in the remaining active regions is significant: the signal-to-noise of the feature amplitudes range from 3.7--6.2, calculated as $|A|\sqrt{N}/\delta y$ where $A$, $N$ and $\delta y$ is the change in FWHM, number of points in the active region, and uncertainty on the FWHM, respectively. For AR-2 the narrower width of the subplanet line has a signal-to-noise of 11.1.
% 3.7, 11.1, 1.3, 5.9, 3.6, 0, and 6.2 for active regions 1--7, respectively.

% To explain both the narrowing and the broadening observed in active regions, we consider the role of the magnetic field in forming mature dark spots.
% The composite profile from a subplanet area on the stellar disc contains components from both upflows in granules and downflows in intergranular lanes. Each component produces a characteristic absorption line shape: the upflows produce a blueshifted line that is nearly symmetric, though it is often characterised by a steeper slope in the red wing than the blue due to the vertical velocity gradient. The line from intergranular lanes is wider with an extended red wing due to turbulent flows and a steep velocity gradient \citep{cegla2013,beeck2015a,cegla2018}. The convection in granular cells brings with it magnetic flux tubes from the interior which concentrate in the intergranular lanes. As magnetic pressure builds in the lanes it pushes gas out thereby lowering the local gas density and reducing absorption in the lanes. As a consequence the composite profile becomes more granule-like, i.e. narrower with a steeper red wing.

To explain the observed narrowing and broadening in active regions, we consider the role of local magnetic field concentrations in the photosphere. Our observed subplanet profiles combine contributions from granules and intergranular lanes. At disc centre, granules produce nearly symmetric, blueshifted lines with slightly steeper red wings, while lanes yield broader, redshifted lines with extended red wings \citep{cegla2013}. Magnetic flux, advected by convection, accumulates in the lanes where the magnetic pressure grows and evacuates gas, ultimately lowering its density and weakening the absorption in the lanes. The convection in the local area is also suppressed, making the region appear darker. At this stage the magnetic feature may be a pore or a ``proto''-spot, and the resulting composite profile may appear more granule-like, i.e. narrower, with a somewhat steeper red wing. As more magnetic flux accumulates (\qtyrange{>2}{3}{\kilo\gauss})  convection in the region becomes almost fully suppressed so the pore develops into a star spot. At this stage the magnetic field strength is such that spectral lines split into multiple components (Zeeman effect). At very high spectral resolutions ($R>500\,000$) the splitting can be resolved \citep{lohner-bottcher2018a}, but at more moderate resolving powers it instead manifests as a broadening of the line \citep{lienhard2023}. 

The line width within an active region is then a net effect of narrowing of the lines due to reduced thermal broadening and convective suppression, and broadening due to the Zeeman effect. We interpret AR-2 as a proto-spot where the local temperature is lower than the surrounding photosphere, with a weak enough magnetic field to have insignificant Zeemann broadening. Our observations indicate that active regions 1, 3--5, and 7 are broadened compared to the local photospheric lines -- and appear significantly darker -- which we interpret as fully formed star spots broadened by a magnetic field of a few \unit{\kilo\gauss}. In Section~\ref{sec:zeeman_broadening} we model the changes in line broadening to put constraints on the star spot magnetic field strengths.

% We interpret this as a dark pore where the intergranular lanes are significantly evacuated of gas, causing the local line profile to narrow.

% As the magnetic flux further increases it almost fully inhibits energy transport to the surface. The local area becomes cooler, and a dark pore is formed.
% For strong local magnetic fields (\qtyrange{>2}{3}{\kilo\gauss}) convection in the region becomes almost fully suppressed so the pore develops into a star spot. At this stage the magnetic field strength is such that spectral lines split into multiple components (Zeeman effect). At very high spectral resolutions ($R>500\,000$) the splitting can be resolved \citep{lohner-bottcher2018a}, but at more moderate resolving powers it instead manifests as a broadening of the line \citep{lienhard2023}. 

In the bottom row of Figure~\ref{fig:local_profiles} we compute the equivalent width (EW) of the local line profiles. 
% In 3D magnetohydrodynamic simulations of the \ion{Fe}{i} line at \qty{617.3}{\nano\metre} the centre-to-limb behaviour of equivalent width depends on stellar temperature.
% For cooler stars, the EW decreases toward the limb because the continuum intensity of the line decreases faster than the line core intensity owing to a shallow temperature gradient along the line of sight, i.e., the abundance of neutral iron along sight lines does not change significantly as a function of $\mu$.
In the Sun the centre-to-limb variation of the equivalent width depends on the line studied, and the underlying physics is not always well-understood with some lines growing stronger and others becoming weaker at the limb, or showing no limb dependence at all \citep{lind2017}. In general the effect is thought to be a combination of the temperature gradient, velocity fields, formation height and line saturation.

% In stars like the Sun and WASP-85\,A,  increases toward the limb based on 3D MHD simulations of the \ion{Fe}{i} line at \qty{617.3}{\nano\metre} \citep{beeck2013}.
% This is because of the temperature gradient observed towards the limb. At disc centre we probe deeper into the photosphere where the temperature is highest. The \ion{Fe}{i} line is depopulated by excitation and ionisation above \qty{7000}{\kelvin}, leaving less neutral iron for absorption thus weakening the line (lower EW). 
% At the limb the higher, cooler layers of the photosphere contain comparatively more neutral iron than the disc-centre and therefore produce stronger absorption, leading to a stronger line (higher EW) \citep{beeck2013}.

% Because our CCF line mask mainly consists of \ion{Fe}{i} and \ion{Fe}{ii} lines
We see a clear line strengthening at both limbs on 2021 March 16. On February 12 the subplanet CCF profiles are stronger at the approaching limb, but weaker at the receding limb.  On 2021 February 20 the opposite is observed. The linear trends may potentially arise from line depth variations in the disc-integrated profiles caused by changes in signal-to-noise. For the first two nights the S/N varied by \qtyrange{\sim20}{30}{\percent} from start to end of transit, while on 2021 March 16 the S/N was relatively constant in comparison with up to \qty{12}{\percent} variation. Alternatively, the linear trends may, at least partly, be explained by the surface temperature gradient due to active regions on the respective nights. Darker regions indicate a cooler atmosphere on average which could lead to deeper lines when normalised to its local continuum \citep{gray2005}. On 2021 February 20 our spot model in Figure~\ref{fig:observations_espr} indicates that the receding (right) hemisphere is comparatively darker  than the approaching (left) hemisphere due to the distribution of spots, which could explain a line strengthening in the former. 
% Following similar arguments, a facular region represents a hotter atmosphere which further weakens the line due to depopulation of \ion{Fe}{i}. 
Following similar arguments, a facular region represents a hotter atmosphere which could weaken the subplanet CCF profile.
On 2021 February 12 the line from the proto-spot in AR-2 is also expected to be weakened owing to the evacuated gas in intergranular lanes, thus the combined effect of AR-2 and the large facular region after it could lead to comparatively weak line formation on the receding hemisphere compared to the approaching one. 

% On 2021 February 12 the S/N exhibits a linear trend with a negative slope, ranging from 58 to 38 from start to end of transit. On February 20 the slope is also linear but positive, starting at 43 and ending at 53. The S/N on 2021 March 16 is relatively constant in comparison ranging from 47--53. 

\subsection{Bisectors}

\begin{figure*}
    \centering
    \includegraphics{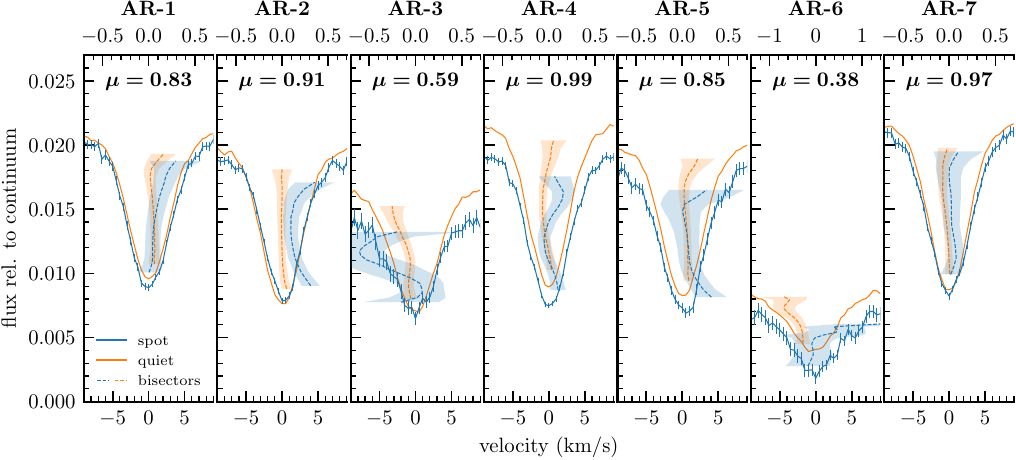}
    \caption{Subplanet CCF profiles from active regions (AR) 1--7 with their disc position $\mu$ annotated. In each panel the solid line with errorbars (\textit{blue}) are data from one exposure within each active region, encapsulating a circular area of radius equal to $R_\mathrm{p}/R_\star$, or equivalently \qty{90}{\mega\metre}. A ``quiet'' subplanet CCF is shown for comparison (\textit{orange solid line}), which is an average of the 8 closest profiles (using all ESPRESSO data) whose average value of $\mu$ is equal to the active region. The \textit{dashed} lines are the bisectors of the respective CCFs, with their uncertainty visualised as the shaded area. The velocity scale of the bisectors is shown on the secondary $x$-axis at the top of each panel in units of \unit{\kilo\metre\per\second}.}
    \label{fig:ar_ccf}
\end{figure*}

We now turn our attention to the bisectors from active regions. In Figure~\ref{fig:ar_ccf} we show the CCF subplanet profiles from each active region that contains a dark spot or pore. The blue, solid lines are the observed subplanet profiles in active regions and is a sum of contributions from the local, surrounding photosphere and the dark region itself.  The blue profiles are an average of $n=1\text{--}3$ subplanet profiles if the active region was occulted in multiple exposures. We compare these to the orange, solid lines which are an average of the 8 closest (in $\mu$) profiles across all three nights observed with ESPRESSO, after excluding active regions with dark spots. As such the orange lines can be considered a close match to what the quiet profile at the same limb angle would look like without the presence of a spot. The continua of the quiet profiles are adjusted to reflect the local continuum in the photosphere surrounding the dark spot/pore. This is because AR-1 and AR-7 are embedded in locally bright regions (faculae) so their continua should be brighter than the quiet profile.

The dashed lines in their respective colours show the bisectors of the two profiles, with the uncertainty of the bisectors visualised in the shaded region. For quiet profiles close to disc centre ($\mu \gtrsim 0.9$) we see the characteristic ``C''-shaped bisectors originating from the combination of line components from granulation and intergranular lanes. We will discuss the centre-to-limb variation of the bisectors in Section~\ref{sec:centre-to-limb}.
% At higher limb angles the bisector shape starts to deviate from it characteristic behaviour, which we discuss further in Section~\ref{sec:centre-to-limb}.
In all active regions except AR-3 -- which is observed at a high limb angle with low S/N -- the mean of the bisector is redshifted compared to the quiet profiles. Since magnetic fields suppress convection we expect a redshift compared to the quiet profiles, which we also discuss in Section~\ref{sec:centre-to-limb}.
Besides the relative velocity shift, bisectors from several active regions deviate from the quiet profiles as a function of line depth. AR-1 and AR-7 are indistinguishable from their quiet counterparts within uncertainties, while AR-6 has very low S/N so we exclude these from further discussions. Others, like AR-2 and AR-5 retain a ``C''-shape and may even show a stronger feature than the quiet profiles. Perhaps the most curious effects are seen in AR-3 and AR-4; both of which show inversions in the bisector as function of line depth.
The bisector shape as a function of line depth traces the vertical velocity structure, since the line wings largely originate from deeper layers of the photosphere, while the line core largely forms higher in the atmosphere. In both cases the largest deviations from the quiet profile occur near the wings, where the magnetic field should be strongest, which may indicate that the magnetic field is affecting the vertical velocity structure. Alternatively, at least for AR-3 which is viewed at the limb, the velocity flows from the Evershed effect may start to influence the line profile \citep{evershed1909,solanki2003a}.

\subsection{Magnetic fields from Zeeman broadening}
\label{sec:zeeman_broadening}

\begin{figure*}
    \centering
    \includegraphics{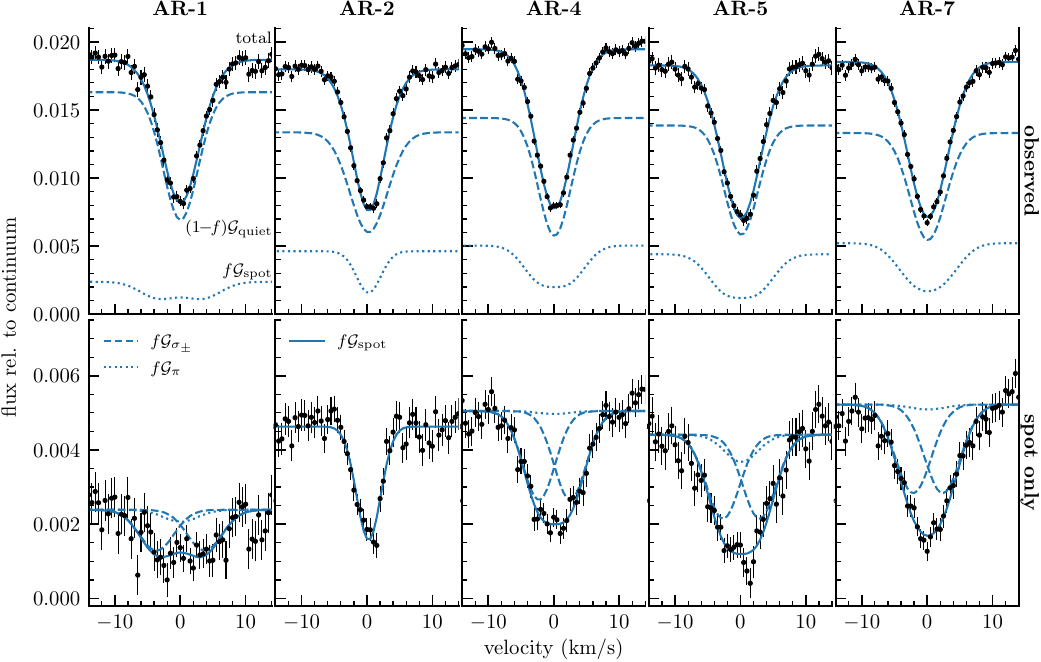}
    \caption{\textit{Top:} Fits of the combined spot + quiet model to the observed subplanet CCFs in active regions using the approach in Section~\ref{sec:zeeman_broadening}. AR-3 and AR-6 are excluded due to low S/N. \textit{Bottom:} The spot component of the fit, representing our closest estimate of the line profile from a magnetic spot broadened by a Zeeman triplet with two symmetric components $\sigma_\pm$ and a central component $\pi$. }
    \label{fig:active_regions}
\end{figure*}

% \begin{figure*}
%     \centering
%     \includegraphics{figures/spot_residuals_final2.pdf}
%     \caption{Observed line profiles of selected active regions (\textit{yellow}), and a Gaussian or double Gaussian fit to the data (\textit{black}). These are obtained from stacking \qtyrange{2}{3}{} subplanet profiles during occultations of regions and subtracting a Gaussian model with shape parameters expected from a ``quiet'' surface (Figure~\ref{fig:local_profiles}). AR-1 and AR-2 are fit with a double Gaussian model (\textit{black}) to model the Zeeman splitting of the  occulted on 12 February 2021, while AR-3 was occulted on 20 February 2021.}
%     \label{fig:spot_residuals}
% \end{figure*}

In this section, we describe how we model the local CCF profiles of active regions to infer the magnetic field strengths required to reproduce the observed broadening (Section~\ref{sec:subplanet_line_shapes}). We assume the line profile from the active region can be modelled as a weighted sum of Gaussians
\begin{align}
    \mathcal{G} = f\gspot{} + (1-f)\gquiet{},
    \label{eq:full_model}
\end{align}
where \gspot{} and \gquiet{} are the models describing the spot and quiet photosphere respectively, and $f$ is the fraction of the subplanet area covered by a spot. In the following will use the subscripts ``q'' (quiet) and ``s'' (spot) to refer to the two components. The components are each described by its own continuum level $c$, line depth $a$, and line width $\sigma$. 
% To first order, we will assume that the line width of the two components are the same, such that $\sigma_\mathrm{s} = \sigma_\mathrm{q} = \sigma$. 
For the quiet model we have
\begin{align}
    \mathcal{G}_\mathrm{quiet} = c_\mathrm{q} - a_\mathrm{q}\exp{\left( -\frac{(v-v_\mathrm{q})^2}{2\sigma_\mathrm{q}^2} \right )},
\end{align}
where $v_\mathrm{q}$ is the central velocity of the quiet photosphere. For AR-2 the combined line profile is narrower than the surrounding photosphere. We then simply have that
\begin{align}
    \mathcal{G}_\mathrm{spot} = c_\mathrm{s} - a_\mathrm{s}\exp{\left( -\frac{(v-v_\mathrm{s})^2}{2\sigma_\mathrm{s}^2} \right )} \qquad\qquad\quad \mathrm{(narrowing)}
\end{align}
with $\sigma_\mathrm{s} < \sigma_\mathrm{q}$. For broadened line profiles we will assume that the observed broadening within an active region is
only caused by the Zeeman effect due to a strong magnetic field. The observed broadening is due to the splitting of the line into multiple components. We adopt the classical case of the Zeeman triplet consisting of a central, unperturbed $\pi$ component modelled as a Gaussian $\mathcal{G}_\pi$ and two symmetric $\sigma$ components $\mathcal{G}_{\sigma_\pm}$ displaced by ${\pm}\Delta v$ from $\mathcal{G}_\pi$. The spot model is then a sum of three components, i.e.

% For AR-4 the magnetic field vector is observed along the line of sight since $\mu = 1$, we therefore assume only the $\sigma_+$ and $\sigma_-$ components are non-zero \citep{kochukhov2021}. The spot model is then a sum of two components $\mathcal{G}_+$ and $\mathcal{G}_-$, each with a velocity shift $+\Delta v$ and $-\Delta v$ respectively compared to the central line $v_\mathrm{s}$, i.e. 
\begin{align}
\mathcal{G}_\mathrm{spot} &= c_\mathrm{s} - a_\mathrm{s}\left[ \frac{1}{2}(\mathcal{G}_{\sigma_+} + \mathcal{G}_{\sigma_-}) + \eta \mathcal{G}_\pi \right] \qquad \mathrm{(broadening)} \\
&= c_\mathrm{s} - \frac{a_\mathrm{s}}{2} \left[
   \exp\left(-\frac{(v - (v_\mathrm{s} + \Delta v))^2}{2\sigma^2}\right) \right. \notag \\
&\hspace{5em} + \left. \exp\left(-\frac{(v - (v_\mathrm{s} - \Delta v))^2}{2\sigma^2}\right) \right. \notag \\
&\hspace{10em} + \left. 2\eta \exp\left(-\frac{(v - v_\mathrm{s})^2}{2\sigma^2}\right) \right].
\end{align}
The relative strength (depth) between the $\pi$ and $\sigma$ components depends on the viewing angle $\theta$ of the magnetic field vector along the line of sight. This dependency is controlled by the dimensionless $\eta$ parameter, which controls the ratio of the intensities ($I$) from the $\pi$ and $\sigma$ components \citep{stenflo1994,bruls1997}
\begin{align}
    I_\pi &\propto \sin^2{\theta},\\
    I_{\sigma_+} + I_{\sigma_-} &\propto 1 + \cos^2{\theta},
\end{align}
so that their relative strength is
\begin{align}
    \eta = \frac{\sin^2{\theta}}{1+\cos^2{\theta}}.
\end{align}
We will also assume that the magnetic field vector is primarily perpendicular to the stellar surface such that the viewing angle of a magnetic field in an active region is simply $\theta = \cos^{-1}{\mu}$.

The total number of parameters in the combined model is 9, however a number of these can be constrained from our observed data. The continua are relative to unity after we normalised the CCFs using the transit model in Section~\ref{sec:reloaded_rm}. The continuum $c_\mathrm{q}$ of the quiet photosphere is then just the subplanet flux $f_\mathrm{p}$ at the disc position $\mu$. The line depth $a_\mathrm{q}$ of the quiet model can be found by linearly interpolating the subplanet contrast curve in Figure~\ref{fig:local_profiles} after removing the exposures in active region. A similar approach can be used to compute the width of the quiet lines, $\sigma_\mathrm{q}$. The continuum of the spot model $c_\mathrm{s}$ is simply the continuum of the quiet photosphere multiplied by the contrast of the spot. We follow the same definition of the spot contrast as \citet{morris2017}:
\begin{align*}
c = 1 - I_\mathrm{spot}/I_\mathrm{quiet},
\end{align*}
where $I$ denotes the mean intensity inside the spot and in the quiet photosphere, such that a ``low'' spot contrast corresponds to a spot that has a similar intensity to the surrounding photosphere. Then the continuum of the spot model is
\begin{align*}
c_\mathrm{s} = c_\mathrm{q}(1-c) = f_\mathrm{p}(1-c).
\end{align*}
To determine the spot area fraction $f$ we need to know the size of the spot. We assume that the spots are circular and have homogeneous intensity. As noted in  \citet{morris2017}, assuming the size of the spot is smaller than the planet, the area fraction is
\begin{align}
    f = \left(\frac{R_\mathrm{spot}}{R_\mathrm{pl}}\right)^2 = c A / f_\mathrm{p},
    \label{eq:spot_area_fraction}
\end{align}
where $A$ is the amplitude of the brightening during the spot occultation, which we calculate from the difference between our unspotted model $f_\mathrm{p}$ and the spotted model at the position of each active region. The only unknown parameter is the spot contrast, $c$. In our spot modelling in Section~\ref{section:spot_model} the spot contrast and size are strongly correlated and relatively unconstrained even when fixing the latitude of the spots to the transit chord. To constrain $c$ we therefore follow the approach of \citet{morris2017} by calculating the area-weighted average of the spot contrast in the umbra and penumbra using measurements of the Sun, with a penumbral area four times that of the umbra \citep{solanki2003a}. We assume a uniform distribution of spot contrast in the umbra between \qtyrange{0.5}{0.8}{} and penumbra between \qtyrange{0.15}{0.25}{}. The resulting distribution of the average spot contrast is approximately Gaussian with a mean of 0.29 and standard deviation of 0.03 which we adopt as a prior in our analysis, and using (\ref{eq:spot_area_fraction}) yields a spot area fraction $f = \qtyrange[range-phrase=\text{--},range-units = single]{\sim20}{50}{\percent}$ in the active regions we have identified.

The average separation $\Delta v$ of the split components is directly proportional to the total strength of the magnetic field $B$ according to \cite[e.g.][]{reiners2012a,kochukhov2021}:
\begin{equation}
    \Delta v = 1.4\times 10^{-3} g_\mathrm{eff} \lambda_0 B,
    \label{eq:magnetic_field}
\end{equation}
where $g_\mathrm{eff}$ is the effective Land\'{e} factor of a given line (characterising its magnetic sensitivity), and $\lambda_0$ is its rest wavelength in \qty{}{\nano\metre}. The velocity is in units of \qty{}{\kilo\metre\per\second} and $B$ is in units of \qty{}{\kilo\gauss}.
% In Figure~\ref{fig:spot_residuals} we show the observed residual profiles (\textit{yellow}) in active regions, i.e. areas with increased magnetic activity. As a reminder, these are obtained from removing the expected profile for a ``quiet'' surface within the area occulted by the planet, and as such can be thought of as isolated profiles from active regions, which we will refer to as a spot profile. The spot profiles from AR-1 and AR-2 shows that the line profile has been split into two components which we attribute to the Zeeman effect. 
% We make an attempt to estimate the local magnetic field strength using a simple toy model. We fit the sum of two Gaussian components with different means to the observed spot profile (\textit{black}, Figure~\ref{fig:spot_residuals}) and find that their average separation from the midpoint is \SI{3.4 \pm 0.6}{\kilo\metre\per\second} and \SI{2.2 \pm 0.4}{\kilo\metre\per\second} for AR-1 and AR-2 respectively.
From the VALD3 database we retrieve the mean Land\'{e} factors of all lines within the wavelength range of ESPRESSO for a star with parameters similar to WASP-85\,A. For each line in the CCF mask we find all lines within \qty{0.05}{\angstrom} and calculate the average Land\'{e} factor weighted by the individual line depths. We exclude weak and strong lines, choosing lines between with relative depths between 0.01 and 0.9. Finally, we then calculate the average of those Land\'{e} factor weighted by the CCF line depths, ESPRESSO transmission curve and the SED of the target star to find the average effective Land\'{e} factor $\langle g_\mathrm{eff} \rangle = 1.24$. We use a similar weighting scheme to calculate the average central wavelength and find $\langle \lambda_0 \rangle = \SI{489.8}{\nano\metre}$. Similar approaches have been employed in the literature in analysing average profiles derived from least-squares deconvolution \citep{morin2008}.

Our final model has then been reduced to 4 parameters for AR-2 ($c$, $a_\mathrm{s}$, $v_\mathrm{q} = v_\mathrm{s}$, and $\sigma_\mathrm{s}$). The spot contrast $c$ is constrained by a Gaussian prior as discussed earlier, while the remaining parameters are restricted to the ranges $a_\mathrm{s} \in [0,0.08]$, $v_\mathrm{q}=v_\mathrm{s} \in [-1, 1]\,\unit{\kilo\metre\per\second}$, $\sigma_\mathrm{s} \in [0,10]\,\unit{\kilo\metre\per\second}$. The median solution from our posterior is shown in the top row, second column in Figure~\ref{fig:active_regions}. We find that profile width of the active region AR-2 is $\sigma_\mathrm{s} = \qty{1.86 \pm 0.15}{\kilo\metre\per\second}$, or \qty{35}{\percent} narrower compared to the surrounding quiet photosphere (\qty{2.86}{\kilo\metre\per\second}).

For our broadened model we fit for an additional parameter $B \in [0,10]\,\unit{\kilo\gauss}$ where the magnetic field strength is related to $\Delta v$ through (\ref{eq:magnetic_field}). In our model $\sigma_\mathrm{s}$ and $B$ are clearly highly correlated, so we restrict $\sigma_\mathrm{s}$ the range $[1.86, \sigma_\mathrm{q}]$. The lower bound is to account for the fact that the individual components of the Zeeman split lines may be narrower than the quiet photosphere due to suppressed convection so its lowest value is set to the result from our fit of AR-2. The upper bound is set to the line width in the quiet photosphere, $\sigma_\mathrm{q}$, which is roughly \qtyrange{2.7}{2.9}{\kilo\metre\per\second} depending on $\mu$. We find that the magnetic fields required to broaden the observed local line profiles range from \qtyrange{2.7}{4.4}{\kilo\gauss}, with an average value \qty{3.1 \pm 0.2}{\kilo\gauss}. The results for each active region is summarised in Table~\ref{table:active_regions}, and the individual fits are shown in Figure~\ref{fig:active_regions}.

\begin{table}
    \sisetup{
    % table-alignment-mode = format,
% table-number-alignment = center
    % separate-uncertainty=true,
  table-align-uncertainty = true,
  % retain-unity-mantissa = false,
  % table-number-alignment = center,
  table-space-text-pre = +
  }
    \renewcommand{\arraystretch}{1.2}
    \small
    % \sisetup{round-mode=places}
    \centering
    \label{table:active_regions}
    \caption{Characteristics of active regions. AR-6 was excluded from the analysis due to low S/N. $^a$Assuming spot contrast $c=1-I_\mathrm{spot}/I_\mathrm{quiet} = 0.29 \pm 0.03$.}
    % \begin{tabular}{@{\extracolsep{\fill}}
    \begin{tabular}{
    lcc
    S[table-format=+3.0(3), retain-explicit-plus]
    }
        \toprule
        \toprule
        Label & $f$ & $B$ & {$\Delta v$} \\
        & (\unit{\percent} of visible disc) & (\unit{\kilo\gauss}) & {(\unit{\metre\per\second})} \\
        % \cmidrule(lr){1-3}
        \midrule
        AR-1 & $0.31 \pm 0.03$ & $4.4 \pm 0.6$ & +176(49) \\ 
        AR-2 & $0.57 \pm 0.09$ & -- & +333(54) \\ 
        AR-3 & $0.92 \pm 0.09$ & $2.8 \pm 0.8$ & -196(97) \\ 
        AR-4 & $0.60 \pm 0.06$ & $2.8 \pm 0.5$ & +199(53) \\ 
        AR-5 & $0.57 \pm 0.05$ & $3.3 \pm 0.5$ & +148(50) \\
        AR-6 & -- & -- & +120(152) \\
        AR-7 & $0.65 \pm 0.06$ & $2.7 \pm 0.4$ & +108(45) \\ 
        
        % AR-1 & $17 \pm 2$ & $0.31 \pm 0.03$ & $4.4 \pm 0.6$ \\ 
        % AR-2 & $32 \pm 5$ & $0.57 \pm 0.09$ & $4.4 \pm 0.6$ \\ 
        % AR-3 & $52 \pm 5$ & $0.92 \pm 0.09$ & $2.8 \pm 0.8$ \\ 
        % AR-4 & $34 \pm 3$ & $0.60 \pm 0.06$ & $2.8 \pm 0.5$ \\ 
        % AR-5 & $32 \pm 3$ & $0.57 \pm 0.05$ & $3.3 \pm 0.5$ \\ 
        % AR-7 & $36 \pm 3$ & $0.65 \pm 0.06$ & $2.7 \pm 0.4$ \\ 
        
        % AR-1 & $17 \pm 2$ & $1.4 \pm 0.1$ & $4.4 \pm 0.6$ \\ 
        % AR-2 & $32 \pm 5$ & $1.9 \pm 0.2$ & -- \\ 
        % AR-3 & $52 \pm 5$ & $2.4 \pm 0.2$ & $2.8 \pm 0.8$ \\
        % AR-4 & $34 \pm 3$ & $2.0 \pm 0.1$ & $2.8 \pm 0.5$ \\ 
        % AR-5 & $32 \pm 3$ & $1.9 \pm 0.1$ & $3.3 \pm 0.5$ \\ 
        % % AR-6 & -- & -- & -- \\ 
        % AR-7 & $36 \pm 3$ & $2.0 \pm 0.1$ & $2.7 \pm 0.4$ \\ 
        \bottomrule
    \end{tabular}
\end{table}

\subsection{Comparison to $B$-measurements in other stars}

\begin{figure*}
    \centering
    \includegraphics{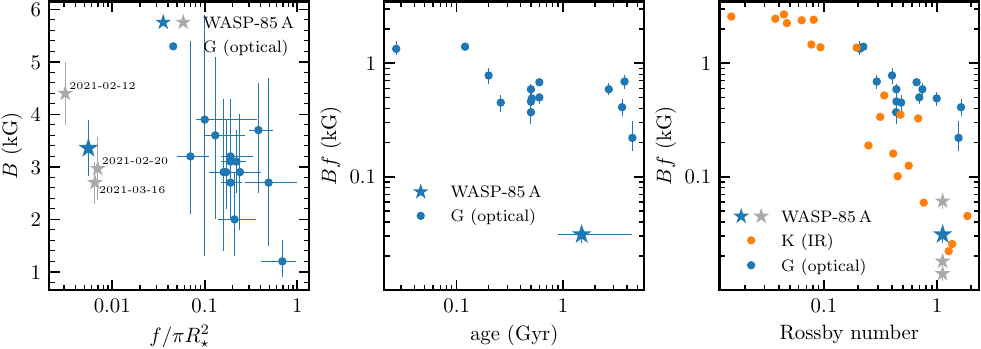}
    \caption{Average unsigned magnetic field measurements from optical Stokes $I$ spectra of solar type stars ($T_\mathrm{eff} = \qtyrange[range-phrase=\text{--}]{5570}{5998}{\kelvin}$). Data taken from \citet{kochukhov2020} and references therein. Blue points are exclusively measured from Zeeman broadening of magnetically sensitive lines, while our measurement is derived from the local line profile broadening in spot occultations detected in ESPRESSO cross-correlation functions (CCF). The age range of WASP-85\,A span \qtyrange{0.6}{3}{\giga\year}, where the lower bound is set by the \ion{Li} abundance, and the upper bound is set by the isochrone model \citep{brown2014}. No age uncertainties are shown for the \citet{kochukhov2020} sample.}
    \label{fig:magnetic_field_comparison}
\end{figure*}

As shown by \citet{komori2025}, the broader and more asymmetric line profiles observed in sunspots cannot be explained by a cooler atmosphere alone, but require magnetic broadening from strong local fields. Our measurement of the magnetic field strengths from star spots on WASP-85\,A (\qtyrange{2.7}{4.4}{\kilo\gauss}) is similar to measurements on the Sun. In sunspots the magnetic field strength peaks in the umbra (the darkest region) where it reaches typical values of \qtyrange{2}{3.7}{\kilo\gauss}, though values up to \qty{6.1}{\kilo\gauss} have been measured \citep{livingston2006}. 
% and decreases to \qtyrange{0.7}{1}{\kilo\gauss} towards the periphery \citep{solanki2003a}.
In light bridges -- bright, narrow regions between sunspot cores -- magnetic fields of up to \qty{8.2}{\kilo\gauss} have been measured \citep{castellanosduran2020}. 

Magnetic field strengths from spatially resolved regions have never been directly measured on stars other than the Sun. With the exception of interferometry, stellar surfaces are never resolved and only disc-integrated information is available. Stellar surface mapping techniques such as Zeeman-Doppler imaging (ZDI) attempt to reconstruct large-scale magnetic field vector maps of the surface by using spectropolarimetric data taken over one or several rotational cycles using a combination of Stokes $I$ and $V$ profiles. However ZDI is not sensitive to small-scale magnetic components such as spots, and the Stokes $V$ component only recovers a small fraction (a few percent) of the total magnetic field due to cancellation of opposite polarities \citep{see2019}.

% With this method one obtains the surface-averaged global magnetic field as seen in (mostly) circularly polarised light, $\langle B_V \rangle$. One should note that ZDI is sensitive to large-scale surface magnetic field structures but not small-scale magnetic components such as spots. 

Another group of methods rely on analysing the broadening of selected magnetically sensitive (unsaturated) lines using high resolution intensity spectra. Often lines in the infrared are chosen due to the larger Zeeman splitting, and consequently a lot of work has been carried out on M dwarfs \citep{kochukhov2021}, though optical studies in Sun-like stars have also been successful \citep{reiners2012a}. In these cases a distribution of the field strength is assumed, parametrised by a magnetic filling factor $f$ and local field strength $B$, which can be added together to form a multi-component model that one attempts to reconcile with the observed spectral line to obtain an average total (unsigned) field strength, $\langle B \rangle = Bf = \sum_i B_if_i$. 

The aforementioned techniques are usually applied to slow rotators, since even modest rotation ($v\sin{i_\star} \gtrsim \SI{5}{\kilo\metre\per\second}$) can be detrimental to detecting the subtle effects of Zeeman broadening in disc-integrated spectra. However, one can overcome this obstacle by analysing the broadening of saturated lines -- a method referred to as Zeeman intensification -- extending the sample of $\langle B \rangle$ to a wider range of stars including young systems  \citep{basri1992}. Using the Zeeman intensification technique, \citet{kochukhov2020} studied 15 main-sequence dwarfs of spectral type K1-G0 (\qtyrange{5000}{6000}{\kelvin}) across a range of ages and found that local magnetic field strength $B$ remained at \SI{\sim 3}{\kilo\gauss} across all ages, which is consistent with our findings on WASP-85\,A. They also found that the mean field strength $\langle B \rangle = Bf$ dropped from \SI{>1}{\kilo\gauss} in stars younger than \SI{120}{\mega\year} to a steady \SI{<1}{\kilo\gauss} in stars \qtyrange{0.3}{4}{\giga\year} old.

To compare with the wider literature, we make simplifying assumptions to estimate the average magnetic field of WASP-85\,A in Stokes $I$, given by 
$Bf = \sum_i B_if_i$.
% With a rotation period of \qtyrange{15.5}{16}{\day}, our observations on 2021 February 12 and 20 sample opposite hemispheres and thus provide near-complete coverage of the surface of WASP-85\,A. % apparently the literature normalises to visible disc, not full surface area so this is not important
We assume that all magnetic flux is confined to the identified spots on each respective night, and that no additional unocculted spots contribute. The spot filling factors $f$ used in Section \ref{sec:zeeman_broadening} are relative to the subplanet region during an exposure; here we convert them to fractions of the visible stellar disc. This gives a spot coverage of 
\qty{0.31\pm0.03}{\percent}, \qty{2.09\pm0.12}{\percent}, and \qty{0.65\pm0.06}{\percent} on 2021 February 12, 20, and March 16, respectively. The disc-averaged (unsigned) magnetic flux is $\langle B \rangle = Bf = \qty{14 \pm 2}{\gauss}$, \qty{61 \pm 9}{\gauss}, and \qty{18 \pm 3}{\gauss} on the same nights. These results are visualised in Figure~\ref{fig:magnetic_field_comparison}, where we compare with the sample of G dwarfs presented in \citet{kochukhov2020} from Zeeman line broadening analyses in optical intensity spectra. These measurements represent average values from observations at multiple epochs. Therefore we present the $Bf$ values for each night we observed (\textit{grey}), as well as the average value across all nights (\textit{orange}). The individual magnetic field strengths $B$ in our active regions (left panel) all cluster around similar values to the \citeauthor{kochukhov2020} sample, which -- together with observed magnetic fields in Sun spots discussed earlier -- gives us confidence that our analysis in Section~\ref{sec:zeeman_broadening} is reasonable. Our values of the magnetic filling factor $f$, however, are smaller by about an order of magnitude compared to the \citeauthor{kochukhov2020} sample. Our method is more sensitive to local magnetic fields in spatially resolved spots, which cannot be resolved from Zeeman broadening in disc-integrated intensity spectra. As a result, the product $Bf$ on WASP-85\,A differs significantly from the literature values by about an order of magnitude at similar ages (Figure~\ref{fig:magnetic_field_comparison}, middle panel). 
% Although we can detect small-scale fields, and for fainter stars than traditional methods (Figure~\ref{fig:magnetic_field_comparison}, right panel) 
However, we are only sensitive to a single band of latitude. If spots exist outside of the active latitude we have assumed, our value of $Bf$ is more accurately interpreted as a lower bound. Magnetic fields can also arise from bright surface features such as faculae, which are typically more extended than dark spots. These regions may contribute significantly to the filling factor $f$ and affect the disc-integrated spectra, but not necessarily the local CCF variations along the transit chord. 

The strength of stellar magnetic fields is empirically observed to correlate with rotation, or more fundamentally with the Rossby number, reflecting the efficiency of the stellar dynamo \citep[e.g.][]{saar1996,reiners2009,marsden2014,vidotto2014,see2017}. In the right panel of Figure~\ref{fig:magnetic_field_comparison} we show the relation of the disc-averaged magnetic field as a function of rotation, expressed as the Rossby number $P_\mathrm{rot}/\tau_c$ where $P_\mathrm{rot}$ is the rotation period and $\tau_c$ is the convective turnover time, calculated from the empirical relation of \citet{noyes1984}. An important caveat here is that the data shown in this panel are derived using two different methods. The blue data in this panel is the same \citet{kochukhov2020} G dwarf sample discussed earlier, whose measurements are derived using the Zeeman intensification technique on optical intensity spectra using four \ion{Fe}{I} lines near \qty{550}{\nano\metre}. The orange data are measurements compiled from \citet{reiners2012a} which are based on Zeeman broadening observed at infrared (IR) wavelengths, typically \ion{Ti}{i} lines near \qty{2.2}{\micro\metre}. We find that our disc-averaged measurements is consistent with the $Bf-\mathrm{Ro}$ relation measured from IR techniques, but not those measured from Zeeman intensification in the optical. The discrepancy between magnetic field strengths measured in the optical and infrared is well established, although its origin remains unclear \citep{reiners2012a,hahlin2023}. 

With a rotation period of \qtyrange{15.5}{16}{\day}, our observations on 2021 February 12 and March 16 sample the same hemisphere and show a consistent value of $Bf$. However the active regions that contribute to the magnetic fields are different for the two nights according to our analysis. On 2021 February 12 the magnetic field originates in AR-1, while on 2021 March 16 it is AR-7. The two nights are separated by 32 days, or two stellar rotation periods, which is in the typical range of lifetimes of spots \citep{basri2022}. This suggests that by 2021 March 16, AR-1 has mostly dissipated, and AR-7 has appeared as a mature spot. 
On 2021 February 20 we observed the opposite hemisphere, where we measure $Bf = \qty{61\pm9}{\gauss}$ -- roughly four times higher than on the other two nights which is due to the detection of three active regions (3, 4, and 5), all of which have observed line broadening. The discrepancy between the two hemispheres is not unusual in the Sun during activity maxima.
% \qty{7.7 \pm 0.2}{\percent} and an average (unsigned) surface magnetic field strength of $\langle B \rangle = Bf = \qty{248 \pm 27}{\gauss}$.
In the Sun the unsigned disc-averaged magnetic field varies significantly overs it 11-year cycle. During activity minimum in 2008--2010 $Bf$ was \qty{\sim4}{\gauss} with a spread of around \qty{1}{\gauss}. During solar maximum however the average differs depending on the type of instrument and method of observation, for example the ground-based SOLIS/VSM instrument \citep{balasubramaniam2011} whose data are available online\footnote{\footnote{\url{https://nso.edu/data/nisp-data/solis-data/mean-field-time-series/}}} shows an average value around \qty{9}{\gauss}, while the space-based SDO/HMI magnetograms show an average around \qty{25}{\gauss} \citep{stenflo2012}. Regardless of the average value, however, the spread in $Bf$ during activity maxima is significantly higher than during minima on timescales of a rotation period. The Sun therefore shows a significant asymmetry in the total averaged magnetic field depending on the visible disc, which our observations indicate is the case for WASP-85\,A as well. Using the maximum value from our observations, we find that the total magnetic flux on WASP-85\,A is at least three times higher than for the Sun at maximum using SDO/HMI measurements. This appears consistent with observations that suggest the Sun is relative quiet in photometry compared with many Kepler solar-type stars \citep{chaplin2011a,mcquillan2012,basri2013,reinhold2020}; a pattern that also appears to be reflected in chromospheric activity \citep{radick1998,hall2007,lockwood2007}.

\section{Centre-to-limb effects in local line profiles}
\label{sec:centre-to-limb}

\begin{figure*}
    \centering
    \includegraphics{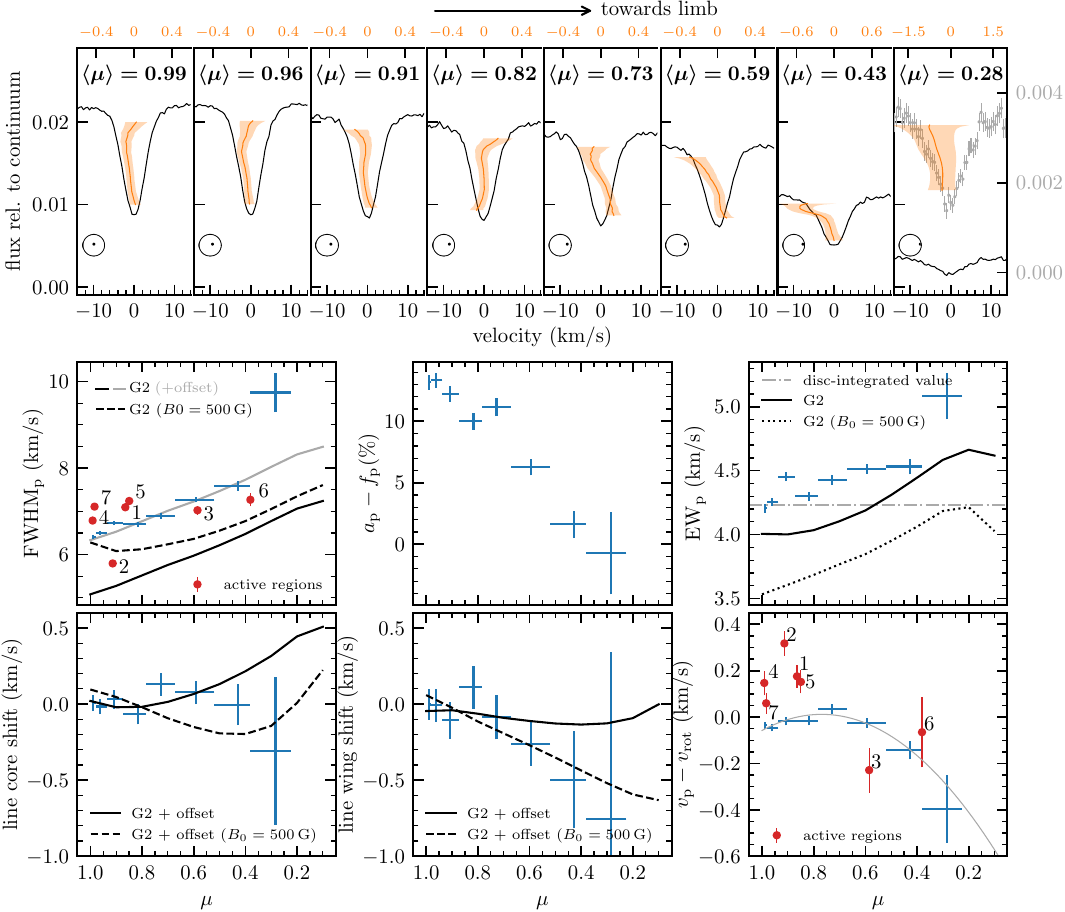} 
    \caption{\textit{Top row:} Average subplanet profiles as a function of limb angle $\mu$ using all three ESPRESSO transits. The rotational velocity has been removed, and exposures identified as spot occultations have been removed. The \textit{black line} in each panel is an average of 6 profiles, with its mean value of $\mu$ annotated. Profile towards the right are closer to the limb, with an approximate position annotated by the circle (star) and dot (planet) within each panel. The \textit{orange line} is the bisector of the line, calculated between \qty{15}{\percent} (line wing) and \qty{90}{\percent} (line core) of maximum depth, and the shaded area its uncertainty. The secondary $x$-axis on top of the panels give the values of the bisectors in units of \unit{\kilo\metre\per\second}. The remaining panels below are quantities derived from these average profiles. \textit{Middle left:} Width of a Gaussian fit to quiet profiles (\textit{blue}) and spots (\textit{red}) with vertical errors denoting the uncertainty on the fit, and horizontal errors are the range of $\mu$ values included in the bin. \textit{Middle centre:} Difference between observed line profile depth (from a Gaussian fit) and expected depth from adopted limb darkened transit model. \textit{Middle right:} Equivalent width. \textit{Bottom left:} The average velocity of the line core and its uncertainty, calculated from the bisectors at \qtyrange{85}{90}{\percent} of maximum line depth. \textit{Bottom centre:} Equivalently for the line wing, calculated between \qtyrange{15}{20}{\percent} of maximum line depth. \textit{Bottom right:} The mean of the Gaussian fit to the quiet profiles (\textit{blue}) and spot profiles (\textit{red}). In the middle and bottom rows, the black lines are theoretical expectations from 3D MHD simulations with (\textit{dashed}) and without (\textit{solid}) a unipolar magnetic field with an average field strength of \SI{500}{\gauss}, calculated for the \ion{Fe}{i} line at \qty{617.3}{\nano\metre} \citep{beeck2013,beeck2015a}. For the line core and line wing the models describe the centre-to-limb convective blueshift. If an offset has been applied to the models (see legends) its value is found by minimizing the difference to the data.}
    \label{fig:local_profiles_mu}
\end{figure*}

This section explores the impact of viewing geometry on the shape of local line profiles. Our aim is to stack the local line profiles from 2021 February 12, 20, and March 16 to improve S/N and study their variation as a function of limb angle $\mu = \cos{\theta}$. To derive the local profiles we make an amendment to the reloaded Rossiter-McLaughlin analysis in Section~\ref{sec:reloaded_rm}. Previously we used the average out-of-transit disc-integrated CCF profile to estimate the systemic velocity on a given night, after removing the radial velocity slope due to the orbit of WASP-85\,Ab. This step is required to bring the velocities to the heliocentric reference frame. However, the centre of a Gaussian fit to this master profile ranged from \qty{13.489}{\kilo\metre\per\second} on 2021 February 12, \qty{13.485}{\kilo\metre\per\second} on February 20, to \qty{13.476}{\kilo\metre\per\second} on March 16 -- a  blueshift of \qty{13}{\metre\per\second} from the first to last night of ESPRESSO observations. At the time of writing there is only one known planet orbiting WASP-85\,A, and the expected reflex motion induced by the binary companion WASP-85\,B over the same time period is expected to be far smaller than the uncertainties on the radial velocity data given its physical separation. We interpret the change in the rest velocity as a combination of variations in the net convective blueshift of WASP-85\,A, and the varying distribution of active regions on each night which deform the disc-integrated line profiles in slightly different ways.
% As discussed earlier, magnetic fields suppress convective motions in the photosphere which leads to a reduction in the net convective blueshift.
% Following this line of thought, the convective velocities are most suppressed on 2021 February 12 and least suppressed on March 16.
To get a better estimate of the true systemic velocity we therefore average the master profiles from all three nights and adopt the centre of a Gaussian fit as the true systemic velocity.

We remove the rotation from the local line profiles using the best-fit rigid body model, calculate their limb angle, and create 8 bins in $\mu$ using quantile binning to ensure each bin contains approximately the same number of observations, which in our case leads to 6 exposures per bin. We exclude exposures that occult active regions identified in Section~\ref{sec:outlier_identification}. Note that due to the rapidly decreasing S/N at higher limb angles we are unable to ensure equal S/N in each bin without significantly compromising our resolution in $\mu$ near the limb. The averaged local line profiles within each bin is shown in the top row of Figure~\ref{fig:local_profiles_mu}. The viewing angle relative to disc centre ranges from \ang{8} (leftmost panel) to \ang{74} (rightmost panel), with the average value of $\mu$ in the bin annotated in each panel. In the final panel, an enhanced version of the average profile is shown for clarity. For each average profile we calculate the bisector and its uncertainty (orange) between \qtyrange{15}{90}{\percent} of maximum line depth. In the following, we use these averaged profiles and their bisectors to study the line shapes in detail.

\subsection{Bisectors and line shapes}

In Figure~\ref{fig:local_profiles_mu}, the bisectors near disc centre exhibit the characteristic ``C''-shape, owing to the combined effect of strong upflows in granules and slower downflows in the intergranular lanes. Toward the limb, the bisector becomes increasingly inclined (`\textbackslash'-shaped) as the vertical component of convective motions projects less onto the line of sight. This effect is seen in the Sun \citep{lohner-bottcher2018} and predicted from several models \citep{gray2005,frame2025}.

The middle row of Figure~\ref{fig:local_profiles_mu} shows the FWHM (left) and contrast (centre) of the local line profiles. The FWHM shows a monotonic broadening as a function of limb angle which is due to the horizontal photospheric velocities contributing to the line of sight. As a comparison we also show the 3D MHD prediction for the \ion{Fe}{i} \qty{617}{\nano\metre} line from \citet{beeck2013} for both a non-magnetic surface and a surface with a \qty{500}{\gauss} field, shown as black solid and dashed lines respectively. The local CCF represents an average over thousands of lines with varying properties, so comparing it to a single spectral line may offer limited insight. Moreover, the width of the local CCF is correlated with the true average line profile, but does not directly equal it due to being smoothed by the properties of the CCF mask. Therefore at a glance the data does not match the absolute models. However, although the absolute offset cannot be compared, the relative variation as a function of limb angle closely matches the non-magnetic model with a fitted offset shown as the grey line. For comparison we also show the FWHM of the active regions identified in this work.

The centre panel in the middle row in Figure~\ref{fig:local_profiles_mu} shows the observed local line profile depths, expressed as the difference between the observed values and those expected if the subplanet profiles contributed in proportion to the flux they block. The latter is an assumption commonly adopted in grid-based codes such as \software{SOAP} that model the surfaces of stars \citep{boisse2012,dumusque2014}. We find that the observed line depths are \qtyrange{13}{14}{\percent} deeper at disc centre, approaching agreement at the limb with an approximately linear trend at intermediate limb angles. A similar effect was noted by \citet{dravins2017a} who studied the average line profile of 26 weak \ion{Fe}{i} lines during a transit of HD 209458\,b. They found an up to \qty{50}{\percent} difference in local line depth relative to the disc-integrated spectrum, which they attributed to the fact that local line profiles are not broadened by rotation while the disc-integrated profile is, so under the assumption of conserved equivalent width the local line profiles would have to deepen. In the right panel of the middle row we show the equivalent width of the local line profiles, and how they compare to the disc-integrated value and 3D MHD models for the \ion{Fe}{i} \qty{617}{\nano\metre} line. The equivalent width appears conserved near disc centre, 
and the net CLV EW over all lines in our CCF mask appears to be stronger at the limb. Some observations of the Sun suggest that weak and moderate lines tend to become stronger towards the limb, while CLV variations of strong and saturated lines tend to get flatter or weaker \citep{lind2017,takeda2019}. 
% but the lines grow stronger towards the limb due to a high temperature gradient which gives a higher abundance of neutral iron which leads to more absorption. 
Again, the data reproduce the relative trend of the models accurately but with a different offset due to the same reasons discussed for the FWHM above. 

\subsection{Centre-to-limb convective blueshift}
In the bottom row of Figure~\ref{fig:local_profiles_mu} we show the centre-to-limb variation of the (from left to right) line core, line wing, and mean of the Gaussian fit to the averaged local line profiles. Because we have removed the rotational contribution, these measurements represent the centre-to-limb convective blueshift. The line core and line wing velocities are computed using the average value of the bisector and its uncertainty between \qtyrange{85}{90}{\percent} and \qtyrange{15}{20}{\percent} of maximum line depth, respectively. Within uncertainties the line core appears to show little centre-to-limb variation, while the line wings show an increasing blueshift towards the limb. The models are to be interpreted as relative variations since each model has a different offset applied, so the difference between the magnetic and non-magnetic models are not representative of their absolute difference. With this caveat in mind, the velocity in the line wings appears to be better described by the magnetic model compared to the non-magnetic one. In the bottom right panel we show the centre of a Gaussian fit to the line profiles, which reproduces the centre-to-limb blueshift in the wing. We find that the centre-to-limb variation can be described by a second-order polynomial (grey line) $v_\mathrm{cb} = -1.30(1-\mu)^2 + 0.60(1-\mu) - 0.06\,\unit{\kilo\metre\per\second}$. The local velocity in active regions is shown in red for comparison, and the velocity difference is tabulated in Table~\ref{table:active_regions}. All active regions, except AR-3 and AR-6 near the limb, show a significant redshift compared to the baseline. The redshift values range from \qtyrange{108}{333}{\metre\per\second}, with the highest redshift measured in AR-2 which we classified as a dark pore, compared to the others which are thought to be fully formed spots. The average value after excluding AR-3 and AR-6 is $\Delta v = \qty{185 \pm 22}{\metre\per\second}$. Our average value and range is somewhat lower than those found in sunspot umbrae, which measure around \qtyrange{300}{400}{\metre\per\second} \citep{lohner-bottcher2018a}. However, our values agree with the Doppler velocities commonly observed in penumbrae (\qtyrange{100}{200}{\metre\per\second}, \citealt{franz2009}), implying that penumbral flows likely occupy the largest fraction of the active region. Since AR-6 is near the limb its S/N is low and its velocity displacement is non-significant. For AR-3 the Doppler velocity is blueshifted by \qty{-196 \pm 97}{\metre\per\second} compared to the baseline. At high limb angles, horizontal photospheric outward flows in the penumbra (the Evershed effect) starts to dominate the line of sight velocity, and these velocities can reach upward of \qty{1}{\kilo\metre\per\second}. Since the centre-side of the penumbra is less foreshortened and brighter than the limb-side the net velocity shift may be a blueshift of a few tens percent of the horizontal velocity, similar to what we observe.

\section{Discussion}
\label{sec:discussion}

% \subsection{What is the true rotation period of WASP-85\,A?}
% \subsection{Recurring spots}
% \label{sec:recurring}
\begin{figure}
    \centering
    \includegraphics{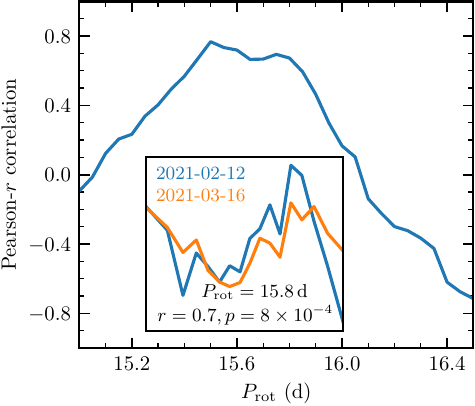}
    \caption{Pearson-$r$ sample correlation of the transit residuals from February 12 2021 and March 16 2021 as function of stellar rotation period. The inset axis shows the transit residuals at $P_\mathrm{rot} = \qty{15.8}{\day}$ with a Pearson-$r$ correlation of 0.7, implying a $p$-value of \num{0.0008}.}
    \label{fig:correlation}
\end{figure}

Our analysis of BGLS data in Section~\ref{sec:rotation_period} found two rotation periods at \qty{11.5}{\day} and \qty{16.3}{\day} but we were unable to determine which belongs to which star. Furthermore, both values are different from our ACF estimate of \qty{13.8 \pm 0.3}{\day}. The correlation between spot occultations between successive transits can be used to derive a rotation period that is not biased due to the blending of the stars, since the planet only transits WASP-85\,A. When applied to $K2$ data, \citet{mocnik2016} derived a rotation period of \SI{15.1 \pm 0.6}{\day} using this method, which is ${\sim}2\sigma$ different compared to their ACF result of \SI{13.6}{\day}, the latter of which can be influenced by blending. Using a more formalised technique, \citet{dai2018} derived a rotation period of \qty{15.2 \pm 0.3}{\day} on the same dataset using the transit chord correlation, which is incompatible with the ACF estimate in this work and of \citet{mocnik2016} at ${>}3\sigma$. Consequently we conclude that our ACF estimate is of the rotation period is not reliable due to both stars being active. The \SI{16.3}{\day} solution from our BGLS estimate is closer to the \citeauthor{dai2018}
result though still not in reasonable agreement ${<}1\sigma$, while the \qty{11.5}{\day} solution likely belongs to WASP-85\,B.

Having narrowed down the allowed rotation periods for WASP-85\,A, we investigate whether any active region in Campaign A is reoccurring from one night to another. The shortest interval between our observed transits is 8 days between 2021 February 12 to 20, and 2022 March 31 to April 8, equivalent to three orbital periods. As noted in \citet{dai2018}, the angular displacement of an active region between consecutive transits is
\begin{align}
    \Delta \theta = 2\pi \frac{nP}{P_\mathrm{rot}} \bmod 2\pi,
\end{align}
where $n$ is the number of consecutive transits and the remaining symbols have their usual meanings. The displacement between the active region from one night to another has to be $<\!\!\SI{180}{\degree}$ to be visible in both, which would be equivalent to a feature at the very start of ingress in one night being visible in the next transit at the very end of egress.
% If we use $P_\mathrm{rot} = \SI{13.8}{\day}$ as the rotation period for the star from the ACF analysis of season 2021A, we get $\Delta \theta = \SI{208}{\degree}$.
If we use $P_\mathrm{rot} = \SI{16.0}{\day}$ (BGLS analysis of all seasons) the displacement is \SI{179}{\degree}. As a result we do most likely not expect to see reoccurring active regions between these nights. What about between 2021 February 12 and 2021 March 16? Twelve orbital periods (\SI{\sim32}{\day}) separate those two nights, which leads to an angular displacement of \SI{357}{\degree}.
% \SI{111}{\degree} and 
% \SI{357}{\degree} for the two possible values of $P_\mathrm{rot}$.
% In the first case we could expect to see features in the first half the transit on 2021 February 12 re-appear in the second half of the transit on 2021 March 16, which we do not see. In the second case,
The star would have made almost exactly two full rotations between those two nights, so any feature on 12 February would approximately be at the same location on 16 March. The two EulerCam light curves on 2021 February 12 and 2021 March 16 look similar in that both have a cosine-like structure at a similar phase. To investigate this further, we translate the phase of the two transits to stellar longitudes using the approach in \citet{dai2018}, and successively calculate the Pearson-$r$ sample correlation for different values of $P_\mathrm{rot}$. The result is shown in Figure~\ref{fig:correlation}. We find the peak correlation $r = 0.78$ at $P_\mathrm{rot} = \qty{15.5}{\day}$ with a $p$-value ${<}10^{-3}$, in agreement with the \citeauthor{dai2018} (\qty{15.2 \pm 0.3}{\day}) and \citeauthor{mocnik2016} (\qty{15.1 \pm 0.6}{\day}) result from transit correlation analyses. In conclusion, we find that the rotation period of WASP-85\,A is \qty{15.5}{\day} as measured in our dataset, while the rotation period of WASP-85\,B is likely \qty{11.5}{\day}.

Based on these results we derive the stellar inclination $i_\star = \qty[separate-uncertainty-units=repeat]{79\pm6}{\degree}$ following the approach of \citet{masuda2020}. In turn this gives a 3D obliquity, i.e. the true angle (as opposed to projected) between the star's spin axis and the normal to the planetary orbital axis, $\psi = \qty[separate-uncertainty-units=repeat]{11 \pm 7}{\degree}$, in agreement with the upper limit of \qty{7}{\degree} found by \citet{dai2018} from transit chord correlation. A caveat is that this assumes no differential rotation. The stellar latitudes transited by the planet range from \qtyrange[range-phrase=\ to\ , separate-uncertainty-units=repeat]{-6}{3}{\degree}. For a Sun-like differential rotation rate ($\alpha = 0.2$) this gives an error of \qty{0.4}{\percent} on $v\sin{i_\star}$, less than the precision on our measurement. The rotation period we measured from transit correlation is dependent on the latitude of the spots if the star rotates differentially. Spot latitudes on the Sun range from \qtyrange[range-phrase=\ to\ , separate-uncertainty-units=repeat]{-30}{30}{\degree} depending on the solar cycle. Using the measured impact parameter and $r/R_\star$, it is unlikely that we would observe spot occultations if the spots are at higher latitudes than \qty{\sim12}{\degree}. At this latitude the star would rotate \qty{\sim0.9}{\percent} slower than at the equator which is also negligible.

\section{Conclusions}
\label{sec:conclusion}

We have presented results from a comprehensive observational campaign aimed at studying the active, spotted star WASP-85\,A during transits of its hot Jupiter companion. By combining photometry from EulerCam, NGTS and SPECULOOS and radial velocities from ESPRESSO (Campaign A) we identify up to six active regions occulted by the planet during three transits. We model these active regions by fitting faculae and spot parameters using a stellar grid-based code. 
Our dataset also comprises, respectively, four and three transits observed with CHEOPS and HARPS-N (Campaign B) at a later time compared to the previous observations. However due to a combination of the lower S/N of HARPS-N, gaps in the CHEOPS data, and potentially lower activity levels for WASP-85\,A, we did not detect any other spot occultations in this dataset.

We also analysed NGTS data spanning a total of roughly 250 days to derive a rotation period for WASP-85\,A and B. Because the stars are blended in all photometric apertures the results from studying the rotational signal varied between different methods. We found that the autocorrelation function (ACF) method performed poorly in a blended system where both stars are active, while the Bayesian Generalised Lomb-Scargle periodogram (BGLS) succeeded in separating two distinct signals, finding a high probability of a \qty{16.3}{\day} and \qty{11.5}{\day} period which we attributed to WASP-85\,A and B, respectively. 
The transits on 2021 February 12 and 2021 March 16 transits are separated by two stellar rotation periods exactly, allowing us to probe the same active regions twice. In the time that passed the active regions appear to have diminished in size and/or contrast, but are still barely identifiable, giving a lower limit of two rotation periods (\qty{\sim 32}{\day}) for their lifetime. A correlation analysis between the two transit residuals allowed us to pin down the rotation period of WASP-85\,A to \qty{15.5}{\day} -- a value that is unaffected by the blended binary -- from which we were able to derive a stellar inclination $i_\star = \qty{79 \pm 6}{\degree}$ and true obliquity $\psi = \qty{11 \pm 7}{\degree}$. The transit on 2021 February 20 probes the opposite hemisphere of the other two nights, giving us complete coverage of WASP-85\,A within a single rotation. 

We found that the projected obliquity $\lambda$ and $v\sin{i_\star}$ derived from a Rossiter-McLaughlin analysis differed for each night due a changing stellar surface. The largest difference in $\lambda$ was found with a classical Rossiter-McLaughlin analysis, where $\lambda$ differed by up to \qty{13}{\degree} and $v\sin{i_\star}$ by up to \qty{0.27}{\kilo\metre\per\second}, at a significance of $2$ and $3.7\sigma$ from their respective uncertainties. Adopting a ``reloaded`` Rossiter-McLaughlin approach also resulted in $\lambda$ changing up to \qty{11}{\degree}, and $v\sin{i_\star}$ \qty{0.12}{\kilo\metre\per\second} ($2.3\sigma$) between nights. Our results suggest that the reloaded Rossiter-McLaughlin analysis -- or similar variations that analyse line profiles directly -- may be better suited for active stars. When fitting all three transits, both methods were in agreement, and we find $\lambda = \ang{0} \pm \ang{1}$ and $v\sin{i_\star} = \qty{3.00 \pm 0.04}{\kilo\metre\per\second}$. The HARPS-N data were significantly affected by blending from WASP-85\,B, however we found that the reloaded method gave consistent results to the ``un-blended'' ESPRESSO data, while the classical method gave highly discrepant results, suggesting the reloaded method is a robust choice even for blended systems.  

Using the reloaded Rossiter-McLaughlin method we studied the shape of the local line profiles during spot occultations. We found that several line profiles from active regions were significantly broadened compared to the surrounding photosphere, which we attributed to Zeeman broadening due to a strong magnetic field. We modelled the CCF profiles using a two-component model that included both ``quiet'' and spot components, and found that the magnetic field required to reproduce the broadening ranged from \qtyrange{2.7}{4.4}{\kilo\gauss}, which is in agreement with solar values. Averaging these values over the visible disc we find a lower limit on the average magnetic field strength $Bf = \qty{16 \pm 3}{\gauss}$ and \qty{61 \pm 9}{\gauss} for each hemisphere. These values are up to three times the value of the Sun during activity maxima. The line-of-sight velocities from all but one active region are redshifted by \qtyrange{108}{333}{\metre\per\second} relative to the centre-to-limb convective blueshift, which is due to suppression of convective motions. These values are similar to net velocity shifts in solar penumbrae and some umbrae. The exception is an active region located at the limb whose velocity is blueshifted by \qty{-196 \pm 97}{\metre\per\second} which we speculate may be due to the Evershed effect. Owing to the fact that stars are unresolved, this is to our knowledge the first time a magnetic field strength has been measured in a spatially resolved region on a star other than the Sun.

By stacking the local profiles from each transit we were able to study centre-to-limb effects. We reproduce several centre-to-limb effects observed in the Sun and predicted from 3D magnetohydrodynamic models, including bisector reversal from a `C'-shape to a `\textbackslash'-shape, broadening of the lines towards the limb, line strengthening towards the limb, and centre-to-limb convective blueshift. While the net convective blueshift is unavailable as a consequence of our analysis, we find that the net convective blueshift is blueshifted by \qty{400}{\metre\per\second} at the limb compared to disc centre and is best described by a second-order polynomial. 

Our results show that current optical high-resolution spectrographs installed on \qty{>8}{\metre} class telescopes, such as ESPRESSO and KPF \citep{gibson2016}, make it possible to study the stellar surface in detail using the transiting planet as a probe. A promising prospect is starting to emerge where one can use the subtle changes in the line profile parameters across the stellar disc to constrain the average magnetic field of the star, opening up a whole new way to characterise transiting exoplanet systems and their magnetic environments. Looking ahead, the high-resolution ANDES spectrograph \citep{marconi2024,palle2025} -- planned as an instrument on the upcoming \qty{39}{\metre} ELT \citep{padovani2023} -- will provide unprecedented detail as it will offer near $5\times$ improvement on the S/N for equivalent exposure times.

However, while the prospects are tantalizing, such work requires detailed description of the line profile shapes. In future work we may improve our analysis by applying the least-squares deconvolution (LSD) technique with a custom line list, which has some benefits compared to the CCF \citep{lienhard2022}. While 3D radiative MHD simulations exist \citep[e.g. \software{muram},][]{vogler2005}, detailed comparisons across the full range of stellar parameters relevant to transiting systems remain limited, constraining our ability to model the stellar surface along the transit chord. Such models are already costly to compute, but further computational power is required to synthesize the full spectrum that is available with optical spectrographs \citep{witzke2021,frame2025}. 

\section*{Acknowledgements}
VK acknowledges funding from the Royal Society through a Newton International Fellowship with grant number NIF/R1/232229.
% eso
Based on observations collected at the European Organisation for Astronomical Research in the Southern Hemisphere under ESO programme 106.21EM.001, 106.21EM.002, 106.21EM.005.
% ngts
This work is based on data collected under the NGTS project at the ESO La Silla Paranal Observatory. The NGTS facility is operated by the consortium institutes with support from the UK Science and Technology Facilities Council (STFC) under projects ST/M001962/1, ST/S002642/1 and ST/W003163/1. 
% speculoos
Based on data collected by the SPECULOOS consortium. The ULiege's contribution to SPECULOOS has received funding from the European Research Council under the European Union's Seventh Framework Programme (FP/2007-2013) (grant Agreement n$^\circ$ 336480/SPECULOOS), from the Balzan Prize and Francqui Foundations, from the Belgian Scientific Research Foundation (F.R.S.-FNRS; grant n$^\circ$ T.0109.20), from the University of Liege, and from the ARC grant for Concerted Research Actions financed by the Wallonia-Brussels Federation. This work is supported by a grant from the Simons Foundation (PI Queloz, grant number 327127). This research is in part funded by the European Union's Horizon 2020 research and innovation program (grants agreements n$^{\circ}$ 803193/BEBOP), and from the Science and Technology Facilities Council (STFC; grant n$^\circ$ ST/S00193X/1, and ST/W000385/1).
CHEOPS is an ESA mission in partnership with Switzerland with important contributions to the payload and the ground segment from Austria, Belgium, France, Germany, Hungary, Italy, Portugal, Spain, Sweden, and the United Kingdom. The CHEOPS Consortium would like to gratefully acknowledge the support received by all the agencies, offices, universities, and industries involved. Their flexibility and willingness to explore new approaches were essential to the success of this mission. CHEOPS data analysed in this article will be made available in the CHEOPS mission archive (\url{https://cheops.unige.ch/archive_browser/}).
% vatsal, cis, heather
HMC, CL and VP acknowledge support from the UKRI Future Leaders Fellowship grant (MR/S035214/1, MR/Y011759/1), and VP further acknowledges support from the UKRI Science and Technology Facilities Council (STFC) through the consolidated grant ST/X001121/1.
% hritam and monika
HC and ML acknowledge support of the Swiss National Science Foundation under grant number PCEFP2\_194576. This work has been carried out within the framework of the NCCR PlanetS supported by the Swiss National Science Foundation under grants 51NF40\_182901 and 51NF40\_205606. 
% mccormac, wilson, brown
DJAB, JMCC, TGW and LD acknowledge support from the UKSA and University of Warwick.
% freckelton
AF acknowledges the support of the Institute of Physics through the Bell Burnell Graduate Scholarship Fund
% gill
SG, DB, and PJW have been supported by STFC through consolidated grants ST/P000495/1, ST/T000406/1 and ST/X001121/1.
% annelies
AM acknowledges funding from a UKRI Future Leader Fellowship, grant number MR/X033244/1 and a UK Science and Technology Facilities Council (STFC) small grant ST/Y002334/1.
% romain
RA acknowledges the Swiss National Science Foundation (SNSF) support under the Post-Doc Mobility grant P500PT\_222212 and the support of the Institut Trottier de Recherche sur les Exoplanètes (IREx).
% gillen
EG gratefully acknowledges support from UK Research and Innovation (UKRI) under the UK government’s Horizon Europe funding guarantee [grant number EP/Z000890/1].
% jenkins
JSJ greatfully acknowledges support by FONDECYT grant 1240738 and from the ANID BASAL project FB210003.
% marina
This research was in part funded by the UKRI (Grant EP/X027562/1).

%%%%%%%%%%%%%%%%%%%%%%%%%%%%%%%%%%%%%%%%%%%%%%%%%%
\section*{Data Availability}

The inclusion of a Data Availability Statement is a requirement for articles published in MNRAS. Data Availability Statements provide a standardised format for readers to understand the availability of data underlying the research results described in the article. The statement may refer to original data generated in the course of the study or to third-party data analysed in the article. The statement should describe and provide means of access, where possible, by linking to the data or providing the required accession numbers for the relevant databases or DOIs.



\bibliographystyle{mnras}
\bibliography{example} % if your bibtex file is called example.bib
\phantomsection
\label{sec:bibliography}

%\begin{thebibliography}{99}
%\end{thebibliography}

%%%%%%%%%%%%%%%%%%%%%%%%%%%%%%%%%%%%%%%%%%%%%%%%%%

%%%%%%%%%%%%%%%%% APPENDICES %%%%%%%%%%%%%%%%%%%%%

\appendix

\section{Supplementary figures}

\begin{figure*}
    \centering
    \includegraphics{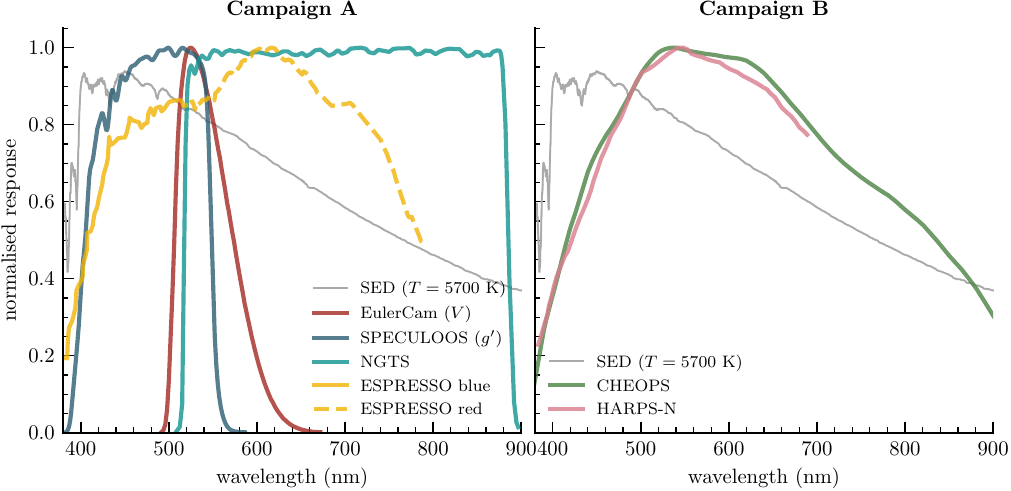}
    \caption{Wavelength response to each instrument involved in the transit observations (\textit{coloured lines}) compared to a smoothed PHOENIX theoretical spectrum \citep{husser2013} of a star with the same stellar parameters as WASP-85\,A (SED, \textit{grey line}).}
    \label{fig:bandpass}
\end{figure*}

\begin{figure*}
    \centering
    \includegraphics{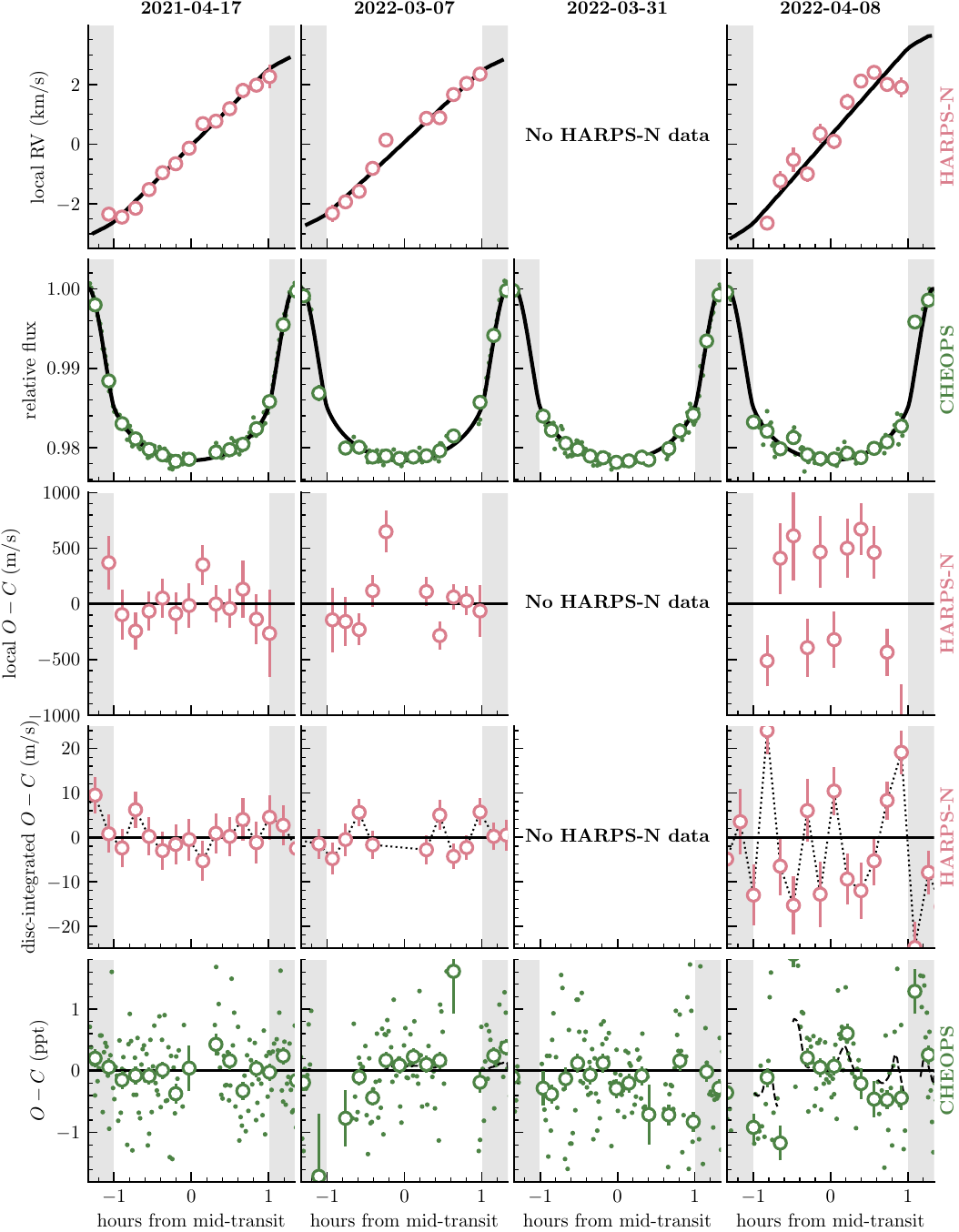}
    \caption{Transit observations from Campaign B (2021 April 17, 2022 March 7 and 31, and 2022 April 8) using CHEOPS.}
    \label{fig:observations_harpsn}
\end{figure*}

\begin{figure*}
    \centering
    \includegraphics{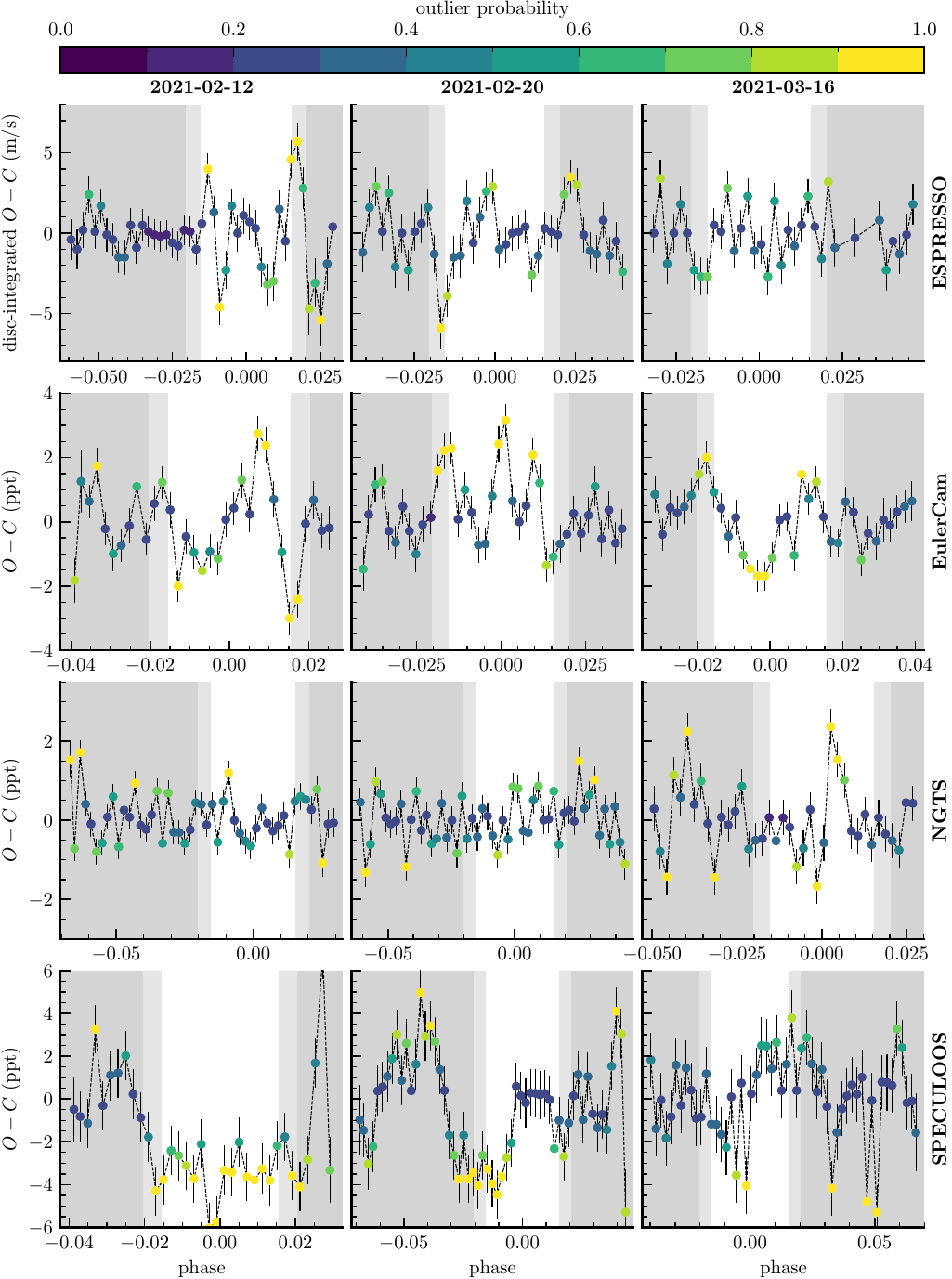}
    \caption{Transit observations from Campaign A (2021 February 12, 20, and March 16) using (\textit{top to bottom}) EulerCam, NGTS, and SPECULOOS. The data points are binned to the same exposure time as individual ESPRESSO exposures, and their colour refers to their probability of being an outlier point from the regression analysis in Section~\ref{sec:outlier_identification}. Out-of-transit regions are shaded dark grey, transit ingress/egress regions are shaded light grey, while the highlighted central regions are transit phases between the second and third contact points.}
    \label{fig:prob_campaignA}
\end{figure*}

\begin{figure*}
    \centering
    \includegraphics{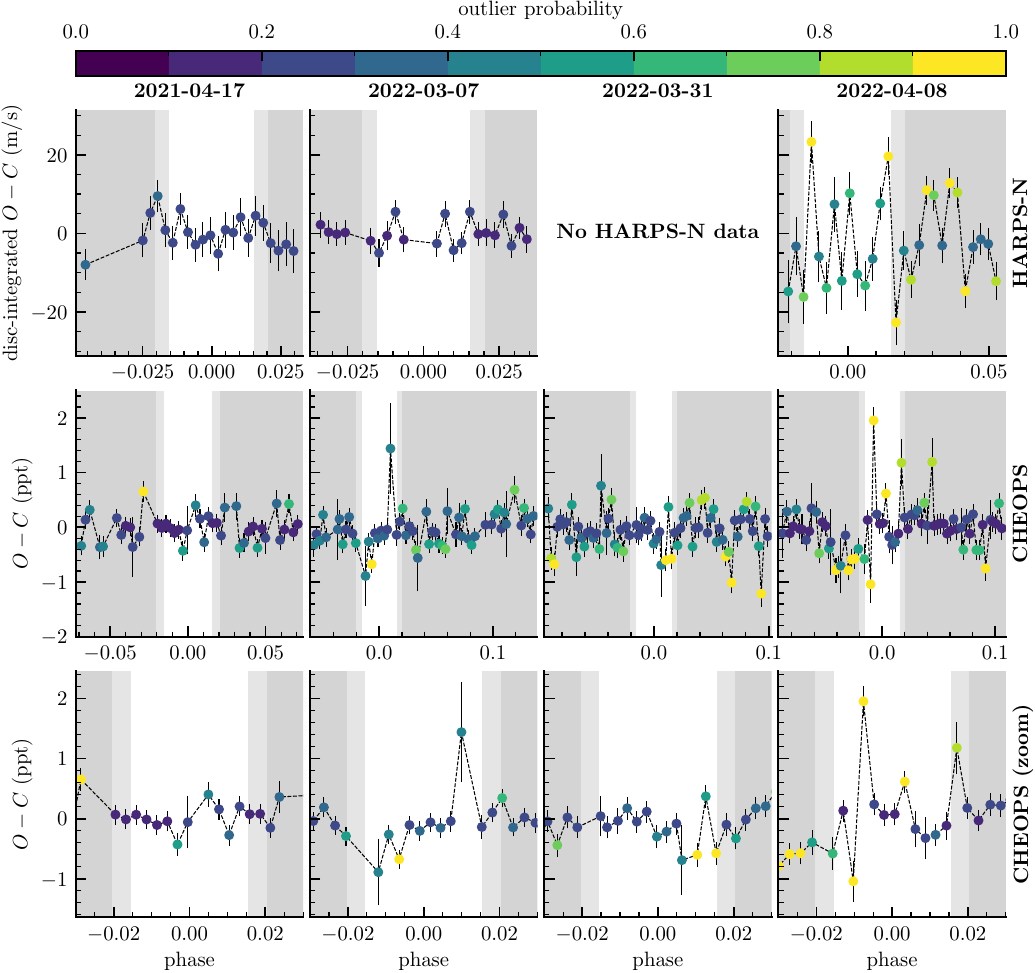}
    \caption{Transit observations from Campaign B (2021 April 17, 2022 March 7 and 31, and 2022 April 8) using HARPS-N and CHEOPS. The data points are binned to the same exposure time as individual HARPS-N exposures (if simultaneously observed), and their colour refers to their probability of being an outlier point from the regression analysis in Section~\ref{sec:outlier_identification}. Out-of-transit regions are shaded dark grey, transit ingress/egress regions are shaded light grey, while the highlighted central regions are transit phases between the second and third contact points. Each column is an observation date, while the rows are in descending order HARPS-N, CHEOPS, and a zoomed in view of the CHEOPS data around the transit.}
    \label{fig:prob_campaignB}
\end{figure*}

\begin{table*}
    \sisetup{separate-uncertainty=true}
    \renewcommand{\arraystretch}{1.2}
    \small
    % \sisetup{round-mode=places}
    \centering
    \label{table:priors_table}
    \caption{Results from the MCMC analysis for our full list of model parameters with their chosen priors.}
    \begin{tabular}{
    l
    % S[table-format=-4.8(8), round-mode=uncertainty]%, round-precision = 2, scientific-notation = false, drop-zero-decimal=false]
    >{$}c<{$}
    l
    % S[table-format=-1.0(1), round-mode=uncertainty, round-precision = 1, scientific-notation = false, drop-zero-decimal=false]
    c
    % S[table-format=1.2(2)]
    % S[table-format=-2.0(2)]
    % S[table-format=1.2(2)]
    % S[table-format=-1.0(1), round-mode=uncertainty, round-precision = 1, scientific-notation = false, drop-zero-decimal=false]
    }
        \toprule
        \toprule
        Parameter & \text{Value} & Unit & Prior \\
        \midrule
        % $P$ & & \unit{\day} & $\mathcal{G}()$ \\
        % $T_0$ & & BJD$_\mathrm{TDB}$ & $\mathcal{G}()$ \\
        % $r/R_\star$ & & \unit{\Rstar} & $\mathcal{G}()$ \\
        % $T_{14}$ & & \unit{\day} & $\mathcal{G}()$ \\
        % $b$ & & \unit{\Rstar} & $\mathcal{G}()$ \\[2pt]
        $\ln{P}$      & 0.97669862 \pm 0.00000047 & \unit{\day} & $\mathcal{G}(0.97670, 0.00038)$ \\
        $T_0$            & 2460258.825019 \pm 0.000093 & BJD$_\mathrm{TDB}$ & $\mathcal{G}(2460258.8254, 0.01)$ \\
        $\ln{T_{14}}$    & -2.2126 \pm 0.0039 & \unit{\day} & $\mathcal{G}(-2.22,0.46)$ \\
        $b$              & 0.135 \pm 0.079 & $R_\star$ & $\mathcal{U}(0,1.13)$ \\
        $\ln{(r/R_\star)}$         & -1.98960 \pm 0.00055 &  & $\mathcal{G}(-1.98960,0.00053)$ \\
        $R_\star$           & 0.990 \pm 0.020 & \unit{\Rsun} & $\mathcal{G}(0.99,0.02)$ \\
        $u_{1,\mathrm{EulerCam}}$           & 0.42 \pm 0.19 &  & $\mathcal{U}(-1,1)$ \\
        $u_{2,\mathrm{EulerCam}}$          & 0.23 \pm 0.27 &  & $\mathcal{U}(-1,1)$ \\
        $u_{1,\mathrm{SPECULOOS}}$           & 0.55 \pm 0.17 &  & $\mathcal{U}(-1,1)$ \\
        $u_{2,\mathrm{SPECULOOS}}$           & 0.26 \pm 0.25 &  & $\mathcal{U}(-1,1)$ \\
        $u_{1,\mathrm{NGTS}}$ & 0.312 \pm 0.085 &  & $\mathcal{U}(-1,1)$ \\
        $u_{2,\mathrm{NGTS}}$           & 0.19 \pm 0.15 &  & $\mathcal{U}(-1,1)$ \\
        $u_{1,\mathrm{CHEOPS}}$           & 0.434 \pm 0.063 &  & $\mathcal{U}(-1,1)$ \\
        $u_{2,\mathrm{CHEOPS}}$           & 0.18 \pm 0.12 &  & $\mathcal{U}(-1,1)$ \\
        $\ln{D_\mathrm{EulerCam}}$ & -0.787 \pm 0.055 &  & $\mathcal{G}(-0.787, 0.055)$ \\
        $\ln{D_\mathrm{SPECULOOS}}$ & -1.00 \pm 0.17 &  & $\mathcal{G}(-0.98, 0.18)$ \\
        $\ln{D_\mathrm{NGTS}}$ & -0.82 \pm 0.12 &  & $\mathcal{G}(-0.82, 0.13)$ \\
        $\ln{D_\mathrm{CHEOPS}}$ & -0.80 \pm 0.18 &  & $\mathcal{G}(-0.84, 0.21)$ \\
        % $\ln{f_\mathrm{spot,2021-02-12}}$ & 0.002 \pm 0.079 &  & $\mathcal{G}(0, 0.2)$ \\
        % $\ln{f_\mathrm{spot,2021-02-20}}$ & -0.010 \pm 0.078 &  & $\mathcal{G}(0, 0.2)$ \\
        % $\ln{f_\mathrm{spot,2021-03-16}}$ & -0.000 \pm 0.080 &  & $\mathcal{G}(0, 0.2)$ \\
        % $\ln{f_\mathrm{spot,EulerCam}}$ & -0.053 \pm 0.083 &  & $\mathcal{G}(0, 0.2)$ \\
        % $\ln{f_\mathrm{spot,SPECULOOS}}$ & 0.065 \pm 0.097 &  & $\mathcal{G}(0, 0.2)$ \\
        % $\ln{f_\mathrm{spot,NGTS}}$ & -0.073 \pm 0.083 &  & $\mathcal{G}(0, 0.2)$ \\
        % $\ln{f_\mathrm{spot,CHEOPS,2021-04-17}}$ & -0.030 \pm 0.057 &  & $\mathcal{G}(0, 0.2)$ \\
        % $\ln{f_\mathrm{spot,CHEOPS,2022-03-07}}$ & -0.014 \pm 0.057 &  & $\mathcal{G}(0, 0.2)$ \\
        % $\ln{f_\mathrm{spot,CHEOPS,2022-03-31}}$ & -0.034 \pm 0.058 &  & $\mathcal{G}(0, 0.2)$ \\
        % $\ln{f_\mathrm{spot,CHEOPS,2022-04-08}}$ & -0.032 \pm 0.057 &  & $\mathcal{G}(0, 0.2)$ \\
        $\ln{\sigma_\mathrm{EulerCam,2021-02-12}}$    & -9.333 \pm 3.033 &  & $\mathcal{G}(-9.32, 5)$ \\
        $\ln{\sigma_\mathrm{Speculoos,2021-02-12}}$    & -8.98 \pm 3.23 &  & $\mathcal{G}(-7.77, 5)$ \\
        $\ln{\sigma_\mathrm{NGTS,2021-02-12}}$    & -8.924 \pm 0.010 &  & $\mathcal{G}(-11.86, 5)$ \\
        $\ln{\sigma_\mathrm{EulerCam,2021-02-20}}$     & -7.61 \pm 0.38 &  & $\mathcal{G}(-9.38, 5)$ \\
        $\ln{\sigma_\mathrm{Speculoos,2021-02-20}}$    & -11.71 \pm 3.17 &  & $\mathcal{G}(-7.83, 5)$ \\
        $\ln{\sigma_\mathrm{NGTS,2021-02-20}}$    & -8.930 \pm 0.011 &  & $\mathcal{G}(-11.86, 5)$ \\
        $\ln{\sigma_\mathrm{EulerCam,2021-03-16}}$   & -7.29 \pm 0.13 &  & $\mathcal{G}(-9.45, 5)$ \\
        $\ln{\sigma_\mathrm{Speculoos,2021-03-16}}$    & -12.05 \pm 2.98 &  & $\mathcal{G}(-7.66, 5)$ \\
        $\ln{\sigma_\mathrm{NGTS,2021-03-16}}$    & -8.471 \pm 0.010 &  & $\mathcal{G}(-11.84, 5)$ \\
        $\ln{\sigma_\mathrm{CHEOPS,2021-04-17}}$     & -7.824 \pm 0.062 &  & $\mathcal{G}(-10.25, 5)$ \\
        $\ln{\sigma_\mathrm{CHEOPS,2022-03-07}}$   & -7.939 \pm 0.063 &  & $\mathcal{G}(-10.24, 5)$ \\
        $\ln{\sigma_\mathrm{CHEOPS,2021-03-31}}$   & -7.283 \pm 0.044 &  & $\mathcal{G}(-10.24, 5)$ \\
        $\ln{\sigma_\mathrm{CHEOPS,2021-04-08}}$   & -7.405 \pm 0.040 &  & $\mathcal{G}(-10.24, 5)$ \\
        $c_{1,\mathrm{FWHM,EulerCam,2021-03-16}}$           & -0.065 \pm 0.012 &  & $\mathcal{G}(0, 1)$ \\
        $c_{1,\mathrm{FWHM,Speculoos,2021-02-20}}$       & -0.073 \pm 0.014 &  & $\mathcal{G}(0,1)$ \\
        $c_{1,\delta y,\mathrm{Speculoos,2021-02-20}}$         & -0.57 \pm 0.38 &  & $\mathcal{G}(0, 10)$ \\
        $c_{2,\delta y,\mathrm{Speculoos,2021-02-20}}$         & -0.016 \pm 0.017 &  & $\mathcal{G}(0, 10)$ \\
        $c_{1,\delta y,\mathrm{Speculoos,2021-03-16}}$             & -0.226 \pm 0.023 &  & $\mathcal{G}(0, 10)$ \\
        $c_{1,\sin \phi,\mathrm{CHEOPS,2021-04-17}}$    & 0.000316 \pm 0.000054 &  & $\mathcal{G}(0, 10)$ \\
        $c_{1,\cos \phi,\mathrm{CHEOPS,2021-04-17}}$    & -0.000017 \pm 0.000044 &  & $\mathcal{G}(0, 10)$ \\
        $c_{1,\sin \phi,\mathrm{CHEOPS,2022-03-07}}$    & -0.000053 \pm 0.000034 &  & $\mathcal{G}(0, 10)$ \\
       $c_{1,\cos \phi,\mathrm{CHEOPS,2022-03-07}}$    & -0.000067 \pm 0.000046 &  & $\mathcal{G}(0, 10)$ \\
        $c_{1,\sin \phi,\mathrm{CHEOPS,2021-03-31}}$    & 0.000211 \pm 0.000051 &  & $\mathcal{G}(0, 10)$ \\
        $c_{1,\cos \phi,\mathrm{CHEOPS,2021-03-31}}$    & -0.000209 \pm 0.000050 &  & $\mathcal{G}(0, 10)$ \\
        $c_{1,\sin \phi,\mathrm{CHEOPS,2021-04-08}}$    & 0.000111 \pm 0.000041 &  & $\mathcal{G}(0, 10)$ \\
        $c_{1,\cos \phi,\mathrm{CHEOPS,2021-04-08}}$    & -0.000097 \pm 0.000048 &  & $\mathcal{G}(0, 10)$ \\
        $c_{1,\mathrm{sky,Speculoos,2021-02-12}}$ & 3.78 \pm 0.55  &  & $\mathcal{G}(0, 10)$ \\
        $c_{2,\mathrm{sky,Speculoos,2021-02-12}}$ & -0.141 \pm 0.037 &  & $\mathcal{G}(0, 10)$ \\
        $c_{1,\mathrm{sky,Speculoos,2021-02-20}}$ & -10.56 \pm 0.83 &  & $\mathcal{G}(0, 10)$\\
        $c_{2,\mathrm{sky,Speculoos,2021-02-20}}$ & 1.013 \pm 0.063 &  & $\mathcal{G}(0, 10)$ \\
        $c_{1,\mathrm{sky,EulerCam,2021-02-12}}$ & -24.33 \pm 0.81 &  & $\mathcal{G}(0, 10)$\\
        $c_{2,\mathrm{sky,EulerCam,2021-02-12}}$ & 1.820 \pm 0.046 &  & $\mathcal{G}(0, 10)$ \\
        \bottomrule
    \end{tabular}
\end{table*}
\begin{table*}
    \sisetup{separate-uncertainty=true}
    \renewcommand{\arraystretch}{1.2}
    \small
    % \sisetup{round-mode=places}
    \centering
    % \label{table:fitted_params}
    % \caption{Table A1 continued...}
    \begin{tabular}{
    l
    % S[table-format=-4.8(8), round-mode=uncertainty]%, round-precision = 2, scientific-notation = false, drop-zero-decimal=false]
    >{$}c<{$}
    l
    % S[table-format=-1.0(1), round-mode=uncertainty, round-precision = 1, scientific-notation = false, drop-zero-decimal=false]
    c
    % S[table-format=1.2(2)]
    % S[table-format=-2.0(2)]
    % S[table-format=1.2(2)]
    % S[table-format=-1.0(1), round-mode=uncertainty, round-precision = 1, scientific-notation = false, drop-zero-decimal=false]
    }
        \toprule
        \toprule
        Parameter & \text{Value} & Unit & Prior \\
        \midrule
        % \multicolumn{4}{c}{\textit{Table A1 continued...}} \\
        % $\cdots$ \\
        \textit{\textbf{Table A1 continued...}} \\
        $c_{1,t,\mathrm{EulerCam,2021-02-12}}$ & 0.20 \pm 0.10 & \unit{\per\day} & $\mathcal{G}(0, 1)$ \\
        $c_{2,t,\mathrm{EulerCam,2021-02-12}}$     & 0.0008 \pm 0.0048 & \unit{\per\day\squared} & $\mathcal{G}(0, 1)$ \\
        $c_{1,t,\mathrm{SPECULOOS,2021-02-12}}$ & -0.37 \pm 0.15 & \unit{\per\day} & $\mathcal{G}(0, 1)$ \\
        $c_{2,t,\mathrm{SPECULOOS,2021-02-12}}$ & -0.0061 \pm 0.0043 & \unit{\per\day\squared} & $\mathcal{G}(0, 1)$ \\
        $c_{1,t,\mathrm{NGTS(801),2021-02-12}}$ & -0.00019 \pm 0.00073 & \unit{\per\day} & $\mathcal{G}(0, 1)$ \\
        $c_{2,t,\mathrm{NGTS(801),2021-02-12}}$ & 0.000027 \pm 0.000054 & \unit{\per\day\squared} & $\mathcal{G}(0, 1)$ \\
        $c_{1,t,\mathrm{NGTS(804),2021-02-12}}$ & -0.00215 \pm 0.00073 & \unit{\per\day} & $\mathcal{G}(0, 1)$\\
        $c_{2,t,\mathrm{NGTS(804),2021-02-12}}$ & -0.000672 \pm 0.000050 & \unit{\per\day\squared} & $\mathcal{G}(0, 1)$ \\
        $c_{1,t,\mathrm{NGTS(809),2021-02-12}}$ & -0.00021 \pm 0.00075 & \unit{\per\day} & $\mathcal{G}(0, 1)$ \\
        $c_{2,t,\mathrm{NGTS(809),2021-02-12}}$ & -0.000202 \pm 0.000054 & \unit{\per\day\squared} & $\mathcal{G}(0, 1)$ \\
        $c_{1,t,\mathrm{NGTS(810),2021-02-12}}$ & 0.00063 \pm 0.00075 & \unit{\per\day} & $\mathcal{G}(0, 1)$ \\
        $c_{2,t,\mathrm{NGTS(810),2021-02-12}}$ & 0.000010 \pm 0.000051 & \unit{\per\day\squared} & $\mathcal{G}(0, 1)$ \\
        $c_{1,t,\mathrm{NGTS(811),2021-02-12}}$  & 0.0022 \pm 0.0021 & \unit{\per\day} & $\mathcal{G}(0, 1)$ \\
        $c_{2,t,\mathrm{NGTS(811),2021-02-12}}$  & -0.000145 \pm 0.000095 & \unit{\per\day\squared} & $\mathcal{G}(0, 1)$ \\
        $c_{1,t,\mathrm{EulerCam,2021-02-20}}$ & 0.147 \pm 0.043 & \unit{\per\day} & $\mathcal{G}(0, 1)$ \\
        $c_{2,t,\mathrm{EulerCam,2021-02-20}}$ & -0.0084 \pm 0.0013 & \unit{\per\day\squared} & $\mathcal{G}(0, 1)$ \\
        $c_{1,t,\mathrm{SPECULOOS,2021-02-20}}$  & -0.321 \pm 0.040 & \unit{\per\day} & $\mathcal{G}(0, 1)$ \\
        $c_{2,t,\mathrm{SPECULOOS,2021-02-20}}$& 0.0104 \pm 0.0032 & \unit{\per\day\squared} & $\mathcal{G}(0, 1)$ \\
        $c_{1,t,\mathrm{NGTS(801),2021-02-20}}$ & -0.00166 \pm 0.00062 & \unit{\per\day} & $\mathcal{G}(0, 1)$ \\
        $c_{2,t,\mathrm{NGTS(801),2021-02-20}}$ & 0.000064 \pm 0.000042 & \unit{\per\day\squared} & $\mathcal{G}(0, 1)$ \\
        $c_{1,t,\mathrm{NGTS(804),2021-02-20}}$ & 0.00202 \pm 0.00063 & \unit{\per\day} & $\mathcal{G}(0, 1)$ \\
        $c_{2,t,\mathrm{NGTS(804),2021-02-20}}$ & -0.000490 \pm 0.000045 & \unit{\per\day\squared} & $\mathcal{G}(0, 1)$ \\
        $c_{1,t,\mathrm{NGTS(809),2021-02-20}}$ & -0.00003 \pm 0.00061 & \unit{\per\day} & $\mathcal{G}(0, 1)$ \\
        $c_{2,t,\mathrm{NGTS(809),2021-02-20}}$ & 0.000028 \pm 0.000044 & \unit{\per\day\squared} & $\mathcal{G}(0, 1)$ \\
        $c_{1,t,\mathrm{NGTS(810),2021-02-20}}$ & -0.00078 \pm 0.00060 & \unit{\per\day} & $\mathcal{G}(0, 1)$ \\
        $c_{2,t,\mathrm{NGTS(810),2021-02-20}}$ & 0.000005 \pm 0.000044 & \unit{\per\day\squared} &$\mathcal{G}(0, 1)$ \\
        $c_{1,t,\mathrm{EulerCam,2021-03-16}}$ & 0.015 \pm 0.062 & \unit{\per\day} & $\mathcal{G}(0, 1)$ \\
        $c_{2,t,\mathrm{EulerCam,2021-03-16}}$ & 0.0100 \pm 0.0016 & \unit{\per\day\squared} & $\mathcal{G}(0, 1)$ \\
        $c_{1,t,\mathrm{SPECULOOS,2021-03-16}}$ & -0.007 \pm 0.036 & \unit{\per\day} & $\mathcal{G}(0, 1)$ \\
        $c_{2,t,\mathrm{SPECULOOS,2021-03-16}}$ & -0.0382 \pm 0.0022 & \unit{\per\day\squared} & $\mathcal{G}(0, 1)$ \\
        $c_{1,t,\mathrm{NGTS(801),2021-03-16}}$ & -0.0012 \pm 0.0020 & \unit{\per\day} & $\mathcal{G}(0, 1)$ \\
        $c_{2,t,\mathrm{NGTS(801),2021-03-16}}$ & 0.00012 \pm 0.00010 & \unit{\per\day\squared} & $\mathcal{G}(0, 1)$ \\
        $c_{1,t,\mathrm{NGTS(804),2021-03-16}}$ & -0.0016 \pm 0.0021 & \unit{\per\day} & $\mathcal{G}(0, 1)$ \\
        $c_{2,t,\mathrm{NGTS(804),2021-03-16}}$ & -0.00068 \pm 0.00010 & \unit{\per\day\squared} & $\mathcal{G}(0, 1)$ \\
        $c_{1,t,\mathrm{NGTS(809),2021-03-16}}$ & 0.0006 \pm 0.0020 & \unit{\per\day} & $\mathcal{G}(0, 1)$ \\
        $c_{2,t,\mathrm{NGTS(809),2021-03-16}}$ & 0.00009 \pm 0.00010 & \unit{\per\day\squared} & $\mathcal{G}(0, 1)$ \\
        $c_{1,t,\mathrm{NGTS(810),2021-03-16}}$ & -0.0003 \pm 0.0020 & \unit{\per\day} & $\mathcal{G}(0, 1)$ \\
        $c_{2,t,\mathrm{NGTS(810),2021-03-16}}$    & 0.00018 \pm 0.00010 & \unit{\per\day\squared} & $\mathcal{G}(0, 1)$ \\
        $c_{1,t,\mathrm{CHEOPS,2022-03-31}}$ & 0.00190 \pm 0.00019 & \unit{\per\day} & $\mathcal{G}(0, 1)$ \\
        $\ln \bar{y}_{\mathrm{EulerCam,2022-02-12}}$  & 0.00588 \pm 0.00043 & & $\mathcal{G}(0, 1)$ \\
        $\ln \bar{y}_{\mathrm{SPECULOOS,2022-02-12}}$& 0.01121 \pm 0.00090 & & $\mathcal{G}(0, 1)$ \\
        $\ln \bar{y}_{\mathrm{NGTS(801),2022-02-12}}$ & -3.35704 \pm 0.00017 & & $\mathcal{G}(-3.37, 1)$ \\
        $\ln \bar{y}_{\mathrm{NGTS(804),2022-02-12}}$ & -3.36365 \pm 0.00017 & & $\mathcal{G}(-3.37, 1)$\\
        $\ln \bar{y}_{\mathrm{NGTS(809),2022-02-12}}$ & -3.36753 \pm 0.00018 & & $\mathcal{G}(-3.37, 1)$ \\
        $\ln \bar{y}_{\mathrm{NGTS(810),2022-02-12}}$ & -3.36024 \pm 0.00018 & & $\mathcal{G}(-3.37, 1)$ \\
        $\ln \bar{y}_{\mathrm{NGTS(811),2022-02-12}}$ & -3.37839 \pm 0.00020 & & $\mathcal{G}(-3.37, 1)$ \\
        \bottomrule
    \end{tabular}
\end{table*}

\begin{table*}
    \sisetup{separate-uncertainty=true}
    \renewcommand{\arraystretch}{1.2}
    \small
    % \sisetup{round-mode=places}
    \centering
    % \label{table:fitted_params}
    % \caption{Table A1 continued...}
    \begin{tabular}{
    l
    % S[table-format=-4.8(8), round-mode=uncertainty]%, round-precision = 2, scientific-notation = false, drop-zero-decimal=false]
    >{$}c<{$}
    l
    % S[table-format=-1.0(1), round-mode=uncertainty, round-precision = 1, scientific-notation = false, drop-zero-decimal=false]
    c
    % S[table-format=1.2(2)]
    % S[table-format=-2.0(2)]
    % S[table-format=1.2(2)]
    % S[table-format=-1.0(1), round-mode=uncertainty, round-precision = 1, scientific-notation = false, drop-zero-decimal=false]
    }
        \toprule
        \toprule
        Parameter & \text{Value} & Unit & Prior \\
        \midrule
        % \multicolumn{4}{c}{\textit{Table A1 continued...}} \\
        \textit{\textbf{Table A1 continued...}} \\
        $\ln \bar{y}_{\mathrm{EulerCam,2022-02-20}}$ & 0.00528 \pm 0.00028 & & $\mathcal{G}(0, 1)$ \\
        $\ln \bar{y}_{\mathrm{SPECULOOS,2022-02-20}}$ & 0.01076 \pm 0.00042 & & $\mathcal{G}(0, 1)$ \\
        $\ln \bar{y}_{\mathrm{NGTS(801),2022-02-20}}$ & -3.35752 \pm 0.00018 & & $\mathcal{G}(-3.37, 1)$ \\
        $\ln \bar{y}_{\mathrm{NGTS(804),2022-02-20}}$ & -3.36348 \pm 0.00018 & & $\mathcal{G}(-3.37, 1)$ \\
        $\ln \bar{y}_{\mathrm{NGTS(809),2022-02-20}}$ & -3.37090 \pm 0.00018 & & $\mathcal{G}(-3.37, 1)$ \\
        $\ln \bar{y}_{\mathrm{NGTS(810),2022-02-20}}$ & -3.36003 \pm 0.00017 & & $\mathcal{G}(-3.37, 1)$ \\
        $\ln \bar{y}_{\mathrm{EulerCam,2022-03-16}}$ & 0.00652 \pm 0.00035 & & $\mathcal{G}(0, 1)$ \\
        $\ln \bar{y}_{\mathrm{SPECULOOS,2022-03-16}}$ & 0.00510 \pm 0.00039 & & $\mathcal{G}(0, 1)$ \\
        $\ln \bar{y}_{\mathrm{NGTS(801),2022-03-16}}$ & -3.35700 \pm 0.00033 & & $\mathcal{G}(-3.37, 1)$ \\
        $\ln \bar{y}_{\mathrm{NGTS(804),2022-03-16}}$ & -3.36474 \pm 0.00034 & & $\mathcal{G}(-3.37, 1)$ \\
        $\ln \bar{y}_{\mathrm{NGTS(809),2022-03-16}}$ & -3.36943 \pm 0.00033 & & $\mathcal{G}(-3.37, 1)$ \\
        $\ln \bar{y}_{\mathrm{NGTS(810),2022-03-16}}$ & -3.35752 \pm 0.00033 & & $\mathcal{G}(-3.37, 1)$ \\
        $\ln \bar{y}_{\mathrm{CHEOPS,2021-04-17}}$ & 0.000427 \pm 0.000035 & & $\mathcal{G}(0, 1)$ \\
        $\ln \bar{y}_{\mathrm{CHEOPS,2022-03-07}}$ & 0.000260 \pm 0.000032 & & $\mathcal{G}(0, 1)$ \\
        $\ln \bar{y}_{\mathrm{CHEOPS,2022-03-31}}$ & 0.000273 \pm 0.000037 & & $\mathcal{G}(0, 1)$ \\
        $\ln \bar{y}_{\mathrm{CHEOPS,2022-04-08}}$ & 0.000230 \pm 0.000032 & & $\mathcal{G}(0, 1)$ \\
        $\ln{\sigma_\mathrm{GP,EulerCam,2021-02-12}}$ & -6.76 \pm 0.32 & & $\mathcal{G}(-9.32, 10)$ \\
        $\ln{\sigma_\mathrm{GP,SPECULOOS,2021-02-12}}$ & -10.65 \pm 6.39 & & $\mathcal{G}(-7.77, 10)$ \\
        $\ln{\sigma_\mathrm{GP,NGTS,2021-02-12}}$ & -17.79 \pm 6.35 & & $\mathcal{G}(-11.86, 10)$ \\
        $\ln{\sigma_\mathrm{GP,EulerCam,2021-02-20}}$ & -6.81 \pm 0.35 & & $\mathcal{G}(-9.38, 10)$ \\
        $\ln{\sigma_\mathrm{GP,SPECULOOS,2021-02-20}}$ & -6.20 \pm 0.63 & & $\mathcal{G}(-7.83, 10)$ \\
        $\ln{\sigma_\mathrm{GP,NGTS,2021-02-20}}$ & -19.044 \pm 6.077 & & $\mathcal{G}(-11.86, 10)$ \\
        $\ln{\sigma_\mathrm{GP,EulerCam,2021-03-16}}$& -6.87 \pm 0.63 & & $\mathcal{G}(-9.45, 10)$ \\
        $\ln{\sigma_\mathrm{GP,SPECULOOS,2021-0=3-16}}$& -6.29 \pm 0.62 & & $\mathcal{G}(-7.66, 10)$ \\
        $\ln{\sigma_\mathrm{GP,NGTS,2021-03-16}}$ & -18.09 \pm 6.64 & & $\mathcal{G}(-11.84, 10)$ \\
        $\ln{\sigma_\mathrm{GP,CHEOPS,2021-04-17}}$& -17.16 \pm 6.30 & & $\mathcal{G}(-10.25, 10)$ \\
        $\ln{\sigma_\mathrm{GP,CHEOPS,2022-03-07}}$ & -14.93 \pm 6.93 & & $\mathcal{G}(-10.24, 10)$ \\
        $\ln{\sigma_\mathrm{GP,CHEOPS,2022-03-31}}$& -10.40 \pm 5.75 & & $\mathcal{G}(-10.24, 10)$ \\
        $\ln{\sigma_\mathrm{GP,CHEOPS,2022-04-08}}$ & -13.82 \pm 6.96 & & $\mathcal{G}(-10.24, 10)$ \\
        $\ln{\rho_\mathrm{GP,EulerCam,2021-02-12}}$& -4.81 \pm 0.43 & \unit{\day} & $\mathcal{G}(-3.22, 5), \mathcal{U}(-6.58, -1.23)$ \\
        $\ln{\rho_\mathrm{GP,SPECULOOS,2021-02-12}}$ & -2.97 \pm 1.46 & \unit{\day} & $\mathcal{G}(-3.22, 5), \mathcal{U}(-6.58, -1.23)$ \\
        $\ln{\rho_\mathrm{GP,NGTS,2021-02-12}}$ & -3.61 \pm 1.56 & \unit{\day} & $\mathcal{G}(-3.22, 5), \mathcal{U}(-6.58, -1.23)$ \\
        $\ln{\rho_\mathrm{GP,EulerCam,2021-02-20}}$ & -4.68 \pm 0.42 & \unit{\day} & $\mathcal{G}(-3.22, 5), \mathcal{U}(-6.58, -1.23)$ \\
        $\ln{\rho_\mathrm{GP,SPECULOOS,2021-02-20}}$ & -3.19 \pm 0.72 & \unit{\day} & $\mathcal{G}(-3.22, 5), \mathcal{U}(-6.58, -1.23)$ \\
        $\ln{\rho_\mathrm{GP,NGTS,2021-02-20}}$ & -3.86 \pm 1.54 & \unit{\day} & $\mathcal{G}(-3.22, 5), \mathcal{U}(-6.58, -1.23)$ \\
        $\ln{\rho_\mathrm{GP,EulerCam,2021-03-16}}$ & -3.64 \pm 0.69 & \unit{\day} & $\mathcal{G}(-3.22, 5), \mathcal{U}(-6.58, -1.23)$ \\
        $\ln{\rho_\mathrm{GP,SPECULOOS,2021-03-16}}$& -3.25 \pm 0.86 & \unit{\day} & $\mathcal{G}(-3.22, 5), \mathcal{U}(-6.58, -1.23)$ \\
        $\ln{\rho_\mathrm{GP,NGTS,2021-03-16}}$ & -3.96 \pm 1.41 & \unit{\day} & $\mathcal{G}(-3.22, 5), \mathcal{U}(-6.58, -1.23)$ \\
        $\ln{\rho_\mathrm{GP,CHEOPS,2021-04-17}}$ & -3.80 \pm 1.52 & \unit{\day} & $\mathcal{G}(-3.22, 5), \mathcal{U}(-6.58, -1.23)$ \\
        $\ln{\rho_\mathrm{GP,CHEOPS,2022-03-07}}$ & -3.70 \pm 1.47 & \unit{\day} & $\mathcal{G}(-3.22, 5), \mathcal{U}(-6.58, -1.23)$ \\
        $\ln{\rho_\mathrm{GP,CHEOPS,2022-03-31}}$ & -5.42 \pm 1.34 & \unit{\day} & $\mathcal{G}(-3.22, 5), \mathcal{U}(-6.58, -1.23)$ \\
        $\ln{\rho_\mathrm{GP,CHEOPS,2022-04-08}}$ & -4.48 \pm 1.62 & \unit{\day} & $\mathcal{G}(-3.22, 5), \mathcal{U}(-6.58, -1.23)$ \\
        \bottomrule
    \end{tabular}
\end{table*}

% If you want to present additional material which would interrupt the flow of the main paper,
% it can be placed in an Appendix which appears after the list of references.

%%%%%%%%%%%%%%%%%%%%%%%%%%%%%%%%%%%%%%%%%%%%%%%%%%

% Don't change these lines
\bsp	% typesetting comment
\label{lastpage}
\end{document}